\documentclass[a4paper,11pt,headings=big,DIV=12]{scrartcl}
\pdfoutput=1

\usepackage[a4paper, top=1.2in, bottom=1.2in, left=1.1in, right=1.1in]{geometry}
\usepackage[utf8]{inputenc}
\usepackage{amsmath, amssymb}
\usepackage{color}
\usepackage[dvipsnames, table]{xcolor}
\definecolor{blue}{rgb}{0,0,0.5}
\usepackage[colorlinks,linkcolor=blue,urlcolor=blue,citecolor=blue]{hyperref}
\usepackage{graphicx}
\graphicspath{{./Figures/}}
\usepackage{slashed}
\usepackage{cite}
\usepackage{enumitem}
\usepackage{bm}
\usepackage{physics}
\usepackage{xspace}
\usepackage[T1]{fontenc}
\usepackage{lmodern}
\usepackage{grffile}
\usepackage{multirow}
\usepackage{booktabs}
\def\parenbar#1{{\null\!      
   \mathop#1\limits^{\hbox{%
   \tiny{{\fontsize{3.5pt}{0em}\selectfont (}%
   \raisebox{-0.4pt}{--}%
   {\fontsize{3.5pt}{0em}\selectfont )}}}} 
   \!\null}} 
   
\usepackage{amsbsy}
\newcommand{\gcl}{\cellcolor[gray]{0.85}}
\newcommand{\R}{{\rm I\!R}}
\newcommand{\Cplx}{{\rm l\!\!\!C}}

\usepackage{float} 
\newcommand{\be}{\begin{equation}}
\newcommand{\ee}{\end{equation}}
\newcommand{\bea}{\begin{eqnarray}}
\newcommand{\eea}{\end{eqnarray}}
\newcommand{\<}{\langle}
\renewcommand{\>}{\rangle}
\newcommand{\mc}{\mathcal}

\newcommand{\nn}{\nonumber}

\newcommand{\mmy}{\mu \mu \gamma}

\newcommand{\ga}{\gamma}
\newcommand{\Ga}{\Gamma}

\newcommand{\De}{\Delta}
\newcommand{\eps}{\epsilon}
\newcommand{\la}{\lambda}
\newcommand{\eff}{\rm eff}
\newcommand{\Bs}{\ensuremath{B^0_s}\xspace}
\newcommand{\bBs}{\bar B^0_s}
\newcommand{\re}{{\rm Re}}
\newcommand{\im}{{\rm Im}}
\newcommand{\ADGf}{A_{\De\Ga_s}^{f}}
\newcommand{\ADGmmy}{\ensuremath{A_{\De\Ga_s}^{\mmy}}\xspace}

\newcommand{\DE}{\rm DE}
\newcommand{\Brems}{\rm Brems}
\newcommand{\GeV}{{\rm GeV}}
\newcommand{\PS}{{\rm PS}}
\newcommand{\Af}{\mc A_f}
\newcommand{\bAf}{\bar{\mc A}_f}
\newcommand{\Afsq}{\vert \Af \vert^2}
\newcommand{\bAfsq}{\vert \bAf \vert^2}

\newcommand{\A}{\mc A}
\newcommand{\bA}{\bar{\mc A}}
\newcommand{\Bsmmy}{B_s \to \mmy}
\newcommand{\bBsmmy}{\bar B_s \to \mmy}
\newcommand{\SM}{{\rm SM}}
\newcommand{\NP}{{\rm NP}}
\newcommand{\CP}{{\rm CP}}
\newcommand{\BBW}{{\rm BBW}}
\newcommand{\LP}{{\rm LP}}
\newcommand{\cut}{\rm cut}
\newcommand{\sh}{\hat{s}}
\newcommand{\Bsmumu}{\Bs \to \mu^{+} \mu^{-}}
\newcommand{\Bsmumugamma}{\Bs \to \mu^{+} \mu^{-} \gamma}
\newcommand{\mumu}{\mu^{+} \mu^{-}}

\newcommand{\FTVAb}{\overline{T}_{\perp,\parallel}}

\newcommand{\APP}{\mathfrak{a}}
\newcommand{\APPb}{\APP_{\perp,\parallel}}

\newcommand{\MB}{M_{B_s}}
\newcommand{\fB}{f_{B_s}}
\newcommand{\mlh}{\hat m_\mu}
\newcommand{\mbh}{\hat m_b}
\newcommand{\FA}{F_A}
\newcommand{\Fa}[1]{F_{A,#1}}
\newcommand{\FV}{F_V}
\newcommand{\Fv}[1]{F_{V,#1}}
\newcommand{\FTA}{F_{TA}}
\newcommand{\FTV}{F_{TV}}
\newcommand{\bFTA}{\bar{F}_{TA}}
\newcommand{\bFTV}{\bar{F}_{TV}}
\newcommand{\FTVA}{T_{\perp,\parallel}}
\newcommand{\C}{C}
\def\qsquare{\ensuremath{q^2}\xspace}
\def\bsmumugamma{\ensuremath{\Bs \to \mu^+ \mu^- \gamma}\xspace}
\def\bsmumu{\ensuremath{\Bs \to \mu^+ \mu^-}\xspace}
\def\bskmunu{\ensuremath{B^0_s \to K^- \mu^+ \nu_{\mu}}\xspace}

\DeclareOldFontCommand{\rm}{\normalfont\rmfamily}{\mathrm}
\DeclareOldFontCommand{\sf}{\normalfont\sffamily}{\mathsf}
\DeclareOldFontCommand{\tt}{\normalfont\ttfamily}{\mathtt}
\DeclareOldFontCommand{\bf}{\normalfont\bfseries}{\mathbf}
\DeclareOldFontCommand{\it}{\normalfont\itshape}{\mathit}
\DeclareOldFontCommand{\sl}{\normalfont\slshape}{\@nomath\sl}
\DeclareOldFontCommand{\sc}{\normalfont\scshape}{\@nomath\sc}

\begin{document}

\begin{flushright}
\small
CERN-TH-2021-026\\
LAPTH-010/21
\end{flushright}
\vskip0.5cm

\begin{center}
{\sffamily \bfseries \LARGE \boldmath
On the effective lifetime of $B_s \to \mu \mu \gamma$
}\\[0.8 cm]
{\normalsize \sffamily \bfseries Alexandre Carvunis$^1$, Francesco Dettori$^2$, Shireen Gangal$^1$,\\Diego Guadagnoli$^{3,1}$, Camille Normand$^1$} \\[0.5 cm]
\small
$^1${\em LAPTh, Universit\'{e} Savoie Mont-Blanc et CNRS, Annecy, France}\\
[0.1cm]
$^2${\em Dipartimento di Fisica, Universit\`a di Cagliari and INFN Sezione di Cagliari,  Cagliari, Italy}\\
[0.1cm]
$^3${\em Theoretical Physics Department, CERN, CH-1211 Geneva 23, Switzerland}
\end{center}

\medskip

\begin{abstract}

\noindent
We consider the $\Bsmumugamma$ effective lifetime, and the related CP-phase sensitive quantity $\ADGmmy$, as a way to obtain qualitatively new insights on the current $B$-decay discrepancies. Through a fit comparing pre- to post-Moriond-2021 data we identify a few theory benchmark scenarios addressing these discrepancies, and featuring large CP violation in addition.
We then explore the possibility of telling apart these scenarios with $\ADGmmy$, once resonance-modeling and form-factor uncertainties are taken into account. We do so in both regions of low and high invariant di-lepton mass-squared $q^2$.
For low $q^2$, we show how to shape the integration range in order to reduce the impact of the $\phi$-resonance modelling on the $\ADGmmy$ prediction. For high $q^2$, we find that the corresponding pollution from broad-charmonium resonances has a surprisingly small effect on $\ADGmmy$.
This is due to a number of cancellations, that can be traced back to the complete dominance of semi-leptonic operator contributions for high $q^2$---at variance with low $q^2$---and to $\ADGmmy$ behaving like a ratio-of-amplitudes observable.
Our study suggests that $\ADGmmy$ is---especially at high $q^2$---a potentially valuable probe of short-distance CP-violating effects in the very same Wilson coefficients that are associated to current $b \to s$ discrepancies. Its discriminating power, however, relies on progress in form-factor uncertainties. Interestingly, high $q^2$ is the region where $\Bsmumugamma$ is already being accessed experimentally, and the region where form factors are more accessible through non-perturbative QCD methods.

\end{abstract}


\section{Introduction} \label{sec:intro}

The decay $\Bs \to \mu \mu \gamma$ has recently attracted new attention in connection with, but also independently from, the discrepancies currently observed in $b \to s$ transitions \cite{Aaij:2017vbb, Aaij:2019wad,Aaij:2015esa,Aaij:2015oid, Aaij:2020nrf}. One reason is the fact that, once sizeable new-physics effects on $\Bs \to \mu \mu$ have been excluded by recent measurements of its branching fraction, the decay $\Bs \to \mu \mu \gamma$ becomes interesting, as it allows to probe a larger set of Wilson coefficients than $\Bs \to \mu \mu$, in particular all those of current interest in connection with the mentioned discrepancies \cite{Aebischer:2019mlg,Alguero:2019ptt,Arbey:2019duh,Ciuchini:2019usw,Kowalska:2019ley,Bhattacharya:2019dot,Alok:2019ufo, Biswas:2020uaq,Alok:2017jgr}. Moreover, $\Bs \to \mu \mu \gamma$ enjoys an enhancement due to the lifting of the chiral suppression by the photon coupling, which translates into a branching ratio of the order of $10^{-8}$ \cite{Eilam:1996vg,Aliev:1996ud,Guadagnoli:2017quo,Kozachuk:2017mdk}.

Admittedly, from a theory point of view $\Bs \to \mu \mu \gamma$ is not nearly as clean as $\Bs \to \mu \mu$ \cite{Bobeth:2013uxa,Beneke:2017vpq,Beneke:2019slt} because of the required $\Bs \to \gamma$ form factors (f.f.'s), and the limited knowledge thereof \cite{Beneke:2020fot, Kozachuk:2017mdk, Melikhov:2004mk, Kruger:1996cv, Aliev:1996ud, Dubnicka:2018gqg, Albrecht:2019zul}. However, this difficulty can be circumvented in selected regions of the measurable phase space \cite{Dettori:2016zff}, and/or by defining {\em ratio} observables whereby the f.f. uncertainties cancel to a large extent \cite{Guadagnoli:2017quo, Kozachuk:2017mdk}.
Besides, from an experimental point of view the high-$q^2$ $\Bs \to \mu \mu \gamma$ spectrum may be accessed from the very same, abundant dataset as $\Bs \to \mu \mu$ \cite{Dettori:2016zff}, i.e. without direct detection of the photon. 
In fact, a search with this method has been performed recently by the LHCb experiment~\cite{LHCb:2021vsc,LHCb:2021awg}, with the full statistics collected so far, for dimuon masses above 4.9~\GeV, 
yielding the first experimental limit on the branching fraction of this decay of $2.0\times 10^{-9}$ at 95\% CL for the mass region considered. 
This confirms that prospects of detection or of improved limits at LHCb are favourable with future datasets. 

One under-explored feature of $\Bs \to \mu \mu \gamma$ is its capacity to probe new CP violation, i.e. complex Wilson coefficients. Actually, and as well-known \cite{Altmannshofer:2012az}, in most constraining $b \to s$-sector branching ratios and CP-averaged angular observables, only NP contributions aligned in phase with the SM can interfere with the SM contributions. As a consequence, NP with non-standard CP violation is in fact {\em constrained more weakly} than NP where CP violation stems only from the CKM phase. As a further consequence, for the coefficients present in the SM, i.e. $C_{7}$, $C_9$ and $C_{10}$, the constraints on the imaginary part of the NP contributions end up being {\em looser} than on the real part, as we will discuss later.

The effective lifetime of $\Bs \to \mu \mu \gamma$ allows to access the quantity known as $\ADGmmy$, which offers a sensitive probe of new CP-violating effects in $b \to s$-sector Wilson coefficients. We introduce and calculate explicitly this quantity.%
\footnote{The effective lifetime was first introduced in Refs. \cite{Fleischer:2011cw}, and it was specifically discussed as a $\Bs \to \mu \mu$ observable in \cite{DeBruyn:2012wj} as a probe of new CP violation in scalar operators.}
$\ADGmmy$ is naturally a ratio-of-amplitudes observable, so that sensitivity to hadronic uncertainties may be accordingly reduced. We explore this question in detail. Specifically, this quantity may be studied in two separate regions, to be indicated as low- and high-$q^2$, located beneath and respectively above the $J/\psi$ and $\psi(2S)$ resonances. In the low-$q^2$ region, a calculation of the long-distance dynamics based on factorisation methods was recently made available \cite{Beneke:2020fot}. Besides, for both low and high $q^2$, a further recent parameterization based on light-cone sum rules (LCSRs) has recently appeared in Ref. \cite{Janowski:2021yvz}.\footnote{Before this parameterization was made public, it was applied in Ref. \cite{Albrecht:2019zul}. For a related discussion see \cite{Pullin:2021ebn}. A further f.f. determination, based on a phenomenological model, is due to Refs. \cite{Kozachuk:2017mdk,Melikhov:2004mk,Melikhov:2000yu,Kruger:1996cv}.}
For low $q^2$ we are thus able to perform an explicit comparison between the two parameterizations, and thereby explore the sensitivity of $\ADGmmy$ to the f.f. choice. Although a similar comparison is not yet possible for high $q^2$, we are however able to address the question of the $\ADGmmy$ sensitivity to broad-charmonium resonances, which we find to be reassuringly small. This region is the most sensitive to the operators $\mc O_9$ and $\mc O_{10}$ allegedly responsible for $b \to s$ discrepancies, and is the region where $\Bsmumugamma$ may be accessible in the short/medium term with the method in Ref. \cite{Dettori:2016zff}.

The plan of the paper is as follows. In Sec. \ref{sec:notation} we collect the basic facts about effective lifetimes and rederive the related CP-violating observable for the case of $\Bs \to \mu \mu \gamma$, as an example of a decay with more than two particles in the final state. We specifically discuss the necessity of including phase-space integrals, separately in the numerator and denominator of the \ADGmmy definition.
In Sec. \ref{sec:parenthesis} we discuss to what extent $b \to s$ and related data as of Moriond 2021 allow for complex contributions to the Wilson coefficients $C_{7,9,10}$. We identify several benchmark new-physics scenarios. Our aim is to compare their effect on $\ADGmmy$ with the effect of hadronic uncertainties, including f.f. parameterisations, as well as resonant effects from the $\phi$ and from broad charmonium. We perform such comparison in Sec. \ref{sec:low_q2}, devoted to low $q^2$, and Sec. \ref{sec:high_q2}, on high $q^2$.
Finally, in Sec. \ref{sec:conclusions} we provide a discussion of the experimental prospects as well as our conclusions. We relegate to appendices: notational details on the necessary amplitudes and f.f.'s (Appendix \ref{app:ffs}); a discussion of f.f.'s within the BBW approach \cite{Beneke:2020fot} and a comparison with the LCSR parameterisation in Ref. \cite{Janowski:2021yvz} (Appendix \ref{app:BBW}); explicit $\ADGmmy$ formul\ae{} (Appendix \ref{app:ADG_formula}); a detailed analytic argument on why the uncertainty induced by broad-charmonium resonances is small (Appendix \ref{app:why_cc_small}); our input table (Appendix \ref{app:input_table}).

\section{Basics and main formul\ae} \label{sec:notation}

Given a final state $f$ common to both the $\Bs$ and the $\bBs$, the most `natural' experimental observable assuming equal production rates for a $\Bs$ and a $\bBs$ is the `untagged' rate \cite{Dunietz:2000cr,DeBruyn:2012wj}
\be
\label{eq:untagged_rate}
\< \Ga(B_s(t) \to f) \> ~\equiv~ \Gamma(\Bs(t) \to f) + \Gamma(\bBs(t) \to f) ~=~ \int_{\PS} \left(  \vert \Af(t) \vert^2 + \vert \bAf(t) \vert^2 \right)\,,
\ee
where
\be
\int_{\PS} ~\equiv~ \int \frac{(2\pi)^4}{2 \MB} d \Phi_f\,,
\ee
with $d \Phi_f$ an element of $n$-body phase space for the final state $f$\cite{Zyla:2020zbs}, and where $\parenbar{{\mc A}}_f(t)$ denote the amplitudes of decay to $f$ for a $B_s$-meson that was a $\parenbar{B}^0_s$ at $t = 0$.

The explicit time dependence of the two amplitudes squared appearing on the r.h.s. of eq.~(\ref{eq:untagged_rate}) is well-known. One introduces
\bea
\label{eq:B-defs}
&&\vert B_{L,H} \> ~=~ p \vert \Bs \> \pm q \vert \bBs \>\,,\nn \\
&&\left( \frac{q}{p} \right)^2 ~=~ e^{-2i\phi_M} \left( 1 + a \right)\,,\nn \\
[0.2cm]
&&\Delta M_s = M_H - M_L\,,~~\Ga_s = \frac{\Ga_H + \Ga_L}{2}\,,~~\Delta \Ga_s = \Ga_L - \Ga_H\,,
\eea
where the deviation of $|(q/p)^2|$ from unity measures CP violation in $\Bs - \bBs$ oscillations, and is quantified by $a \approx A_{\rm SL} \simeq -3.5 \times 10^{-3}$ \cite{Zyla:2020zbs}, with $A_{\rm SL}$ the `wrong-charge' asymmetry for $\parenbar{B}^0_s \to \ell^{\mp} + X$ decays. From eqs. (\ref{eq:B-defs}), the amplitudes squared $\vert \parenbar{{\mc A}}_f(t) \vert^2$ can be calculated as follows \cite{Dunietz:2000cr}
\bea
\label{eq:Asq}
\vert \parenbar{{\mc A}}_f(t) \vert^2 &=& \frac{e^{- \Ga_s t}}{2} \Bigl[ \left( \Afsq + \vert q/p \vert^2 \bAfsq \right) \cosh(\De \Ga_s \, t / 2) \pm \left( \Afsq - \vert q/p \vert^2 \bAfsq \right) \cos(\De M_s \, t) \nn \\
&-&2 \, \re \left( q / p \, \bAf \Af^* \right) \sinh(\De \Ga_s \, t / 2) 
\mp 2 \, \im \left( q / p \, \bAf \Af^* \right) \sin(\De M_s \, t)\,,
\eea
where we omitted terms of $O(a)$ in the $\vert \bAf (t)\vert^2$ case. Clearly, the $\sin$ and $\cos$ terms in eq.~(\ref{eq:Asq}) will cancel in the sum on the r.h.s. of eq.~(\ref{eq:untagged_rate}), and one gets \cite{DeBruyn:2012wj}
\be
\label{eq:untagged_rate_2}
\< \Gamma(B_s(t) \to f)\> ~=~ \left( R_H^f + R_L^f \right) e^{- \Ga_s t} \left[ \cosh \left( \frac{y_s t}{\tau_s} \right) + A_{\De \Ga_s}^f \sinh \left( \frac{y_s t}{\tau_s} \right) \right]\,,
\ee
with $R_H^f + R_L^f  = \int_{\PS} \left( \Afsq + \vert q/p \vert^2 \bAfsq \right)$. For $\ADGf$ one then gets
\bea
\label{eq:ADGf}
&&\ADGf = \frac{-2 \int_{\PS} \re \left( q / p \, \bAf \Af^* \right)}{\int_{\PS} \left( \Afsq + \vert q/p \vert^2 \bAfsq \right)}~.
\eea%
The $\ADGf$ expression usually found in the literature is the r.h.s. of eq. (\ref{eq:ADGf}), {\em without} the phase-space integrals. We note explicitly that these phase-space integrals are {\em separate} in the numerator {\em and} the denominator of $\ADGf$, see eq. (\ref{eq:ADGf}). Therefore, for decays to more than two particles in the final state, they cannot be accounted as the overall normalization factor that is usually understood in the literature.
In fact, the products $\bAf \Af^*$, $\Afsq$ and $\bAfsq$ will depend on kinematic invariants if $f$ consists of $\ge 3$ particles, and this dependence is different for the different Wilson-coefficient combinations that these amplitude products depend on. The integrals allow to integrate out such kinematic dependence, in accord with the l.h.s. of eq. (\ref{eq:untagged_rate_2}), which depends on $t$ only.

From eq. (\ref{eq:untagged_rate}) one may define the general relation between the experimental and the theoretical branching ratio as \cite{Dunietz:2000cr,DeBruyn:2012wj,Descotes-Genon:2015hea}
\be
\label{eq:BRexp_vs_th}
\mc B(B_s \to f)_{\exp} \equiv \frac{1}{2} \int_0^\infty \< \Ga(B_s(t) \to f) \> dt = \left( \frac{1 + \ADGf y_s}{1-y_s^2} \right) \mc B(B_s \to f)_{\rm th}\,,
\ee
where
\be
\label{eq:BRth}
\mc B(B_s \to f)_{\rm th} ~\equiv~ \frac{\tau_s}{2} \< \Ga(B_s(0) \to f) \>~.
\ee
The parameter $\ADGf$ relating the two branching ratios is final-state as well as model dependent. Interestingly however, $\ADGf$ can be extracted directly from another observable, the effective lifetime \cite{Fleischer:2011cw,DeBruyn:2012wj}, defined as
\be
\tau_{\eff}^f ~\equiv~ \frac{\int_0^\infty t \< \Ga(B_s(t) \to f)\> dt}{\int_0^\infty \< \Ga(B_s(t) \to f)\> dt} ~=~ \frac{\tau_s}{1-y_s^2}\left( \frac{1+2 A_{\De\Ga_s}^f y_s + y_s^2}{1+ A_{\De\Ga_s}^f y_s} \right)~.
\ee
In the above relations we introduced, as customary, the average $B_s$-system lifetime $\tau_s$
and the fractional lifetime difference $y_s \equiv \De \Ga_s / (2 \Ga_s)$.

\subsection{\boldmath$A_{\De\Ga_s}^{\mmy}$ calculation} \label{sec:ADG_calculation}

The calculation of $A_{\De\Ga_s}^{\mmy}$ is a special case of eq. (\ref{eq:ADGf}).
The integral over the 3-body phase space reads
\be
\label{eq:intPS}
\int_{\PS, \mmy} = \frac{\MB}{2^8 \pi^3} \int d\hat{s} \, d\hat{t}\,,
\ee
where, after defining the decay kinematics $\parenbar{B}^0_s(p) \to \mu^+(p_1) + \mu^-(p_2) + \gamma(k)$, we introduce \cite{Melikhov:2004mk} the kinematic invariants $s = (p_1 + p_2)^2$, $t = (p - p_1)^2$, and a hatted quantity denotes that it has been normalized to the appropriate power of $\MB$ to make it dimensionless. Further using
\be
\hat t = {\hat m}_\mu^2 + \frac{1- \hat s}{2} \left( 1 - \sqrt{1 - \frac{4 {\hat m}^2_\mu}{\hat s}} \cos \theta \right)\,,
\ee
with $\theta = \arccos 
\frac{\vec p_2 \cdot \vec k}{|\vec p_2| |\vec k|}$, 
one has
\be
\label{eq:PS_def}
\int_{\PS, \mmy} = 
\frac{\MB}{2^8 \pi^3} \int f(\hat s, {\hat m}^2_\mu)\,\,
d\hat{s} \, d \cos \theta\,,
~~~~\mbox{with } f(\hat s, {\hat m}^2_\mu) \equiv 
\frac{1- \hat s}{2} 
\sqrt{1 - \frac{4 {\hat m}^2_\mu}{\hat s}}~.
\ee
We obtain
\bea
\label{eq:ADGmmy}
\ADGmmy &=& - \frac{
\int_{\PS, \mmy} \re \left( q / p \, \bA \A^* \right)
}{
\int_{\PS, \mmy} \left \vert \mc A \right \vert^2
} \nn \\
[0.3cm]
&=& 
- \frac{
\int d\hat{s} \, d \cos \theta \, \, f(\hat s, {\hat m}^2_\mu) \,\, \re \left( q / p \, \bA \A^* \right)
}{
\int d\hat{s} \, d \cos \theta \, \, f(\hat s, {\hat m}^2_\mu) \,\,
\left \vert \mc A \right \vert^2
}\,,
\eea
where, to ease notation, we used the abbreviation
\be
\parenbar{\A} ~\equiv~ \parenbar{\A}_{\mmy}\,,
\ee
took into account that $|\bA| = |\A|$, and again neglected terms of $O(a)$ (see eq.~(\ref{eq:B-defs})). An explicit formula for $\A$ and the relation between $\A$ and $\bA$ are reported in Appendix \ref{app:ffs}. The corresponding formula for $\ADGmmy$ is provided in App. \ref{app:ADG_formula}. 

A few comments are in order. First, the CKM phase in the weak-Hamiltonian coupling in $\bA \A^*$ exactly cancels the $q/p$ phase, as expected. More specifically, after introducing the CP transformation
\be
\label{eq:CPBs}
\CP |\Bs \> = e^{i \phi_{\CP}} | \bBs \>\,,
\ee
with $\phi_{\CP}$ a convention-dependent phase ($\phi_{\CP} = 0$ in FLAG \cite{Aoki:2019cca} and $= \pi$ in \cite{Dunietz:2000cr}), the phase $\phi_M$ appearing in the $q/p$ ratio (see eq. (\ref{eq:B-defs})) is given by (see e.g. \cite{Fleischer:2002ys})
\be
\phi_M = \pi + 2 \arg (V_{tb} V_{ts}^*) - \phi_{\CP}~.
\ee
Making the flavour of the initial state explicit, we then have $\bAf / \Af \equiv \bAf^{(q)} / \Af^{(q)} = e^{+i \phi_M} \xi_f^{(q)}$, with $\xi_f^{(q)}$, $(q = d,s)$, a convention-independent quantity that depends on the initial state $B^0_q$ as well as on the final state $f$. The quantity $\xi_f^{(q)}$ can be determined in terms of two of the three observables
\bea
&&A^{f}_{\Delta \Gamma_q} \equiv \frac{-2 \int_{\PS} |\mc A_f|^2 \, \re \, \xi_f^{(q)}}{\int_{\PS} |\mc A_f|^2 \, (1 + |\xi_f^{(q)}|^2)}~,~~~
A^{f, \rm mix}_{\CP} \equiv \frac{-2 \int_{\PS} |\mc A_f|^2 \, \im \, \xi_f^{(q)}}{\int_{\PS} |\mc A_f|^2 \, (1 + |\xi_f^{(q)}|^2)}~,~~~\nn \\
[0.2cm]
&&\hspace{2.5cm}A^{f, \rm dir}_{\CP} \equiv \frac{\int_{\PS} |\mc A_f|^2 \, (1 - |\xi_f^{(q)}|^2)}{\int_{\PS} |\mc A_f|^2 \, (1 + |\xi_f^{(q)}|^2)}~,
\eea
where, for $q = s$, the first relation coincides with eq. (\ref{eq:ADGf}).
Since the $\phi_M$ dependence is the same in ${\bAf}/{\Af}$ and in $\bAf \Af^*$, the latter appearing in eq. (\ref{eq:ADGmmy}), we see that $\ADGf$ is phase-convention independent, as well-known~\cite{Dunietz:2000cr,Fleischer:2002ys}.

We observe that, given the very $\ADGmmy$ definition, complex phases in any of the Wilson-coefficient combinations contribute to `misaligning' numerator and denominator in eq. (\ref{eq:ADGmmy}), in turn causing $\ADGmmy$ to depart from unity.%
\footnote{We also note that kinematic structures are different for different Wilson-coefficient combinations.
 In eq.~(\ref{eq:ADGmmy}), the phase-space integrals take care of integrating over the kinematic variables---see also comment made below eq.~ (\ref{eq:ADGf}). Concretely, we integrate $\cos \theta$ in the full range $[-1, 1]$. We then integrate $\hat s$ in appropriate ranges, chosen to minimise pollution from resonant regions while maximising event rates.} This makes $\ADGmmy$ a very sensitive probe of new, short-distance, CP-violating effects; besides, the fact that $\ADGmmy$ is a `ratio-of-amplitudes-squared' quantity helps reducing its sensitivity to certain hadronic effects for high $q^2$, as we will see in detail in Sec. \ref{sec:broad-c} and Appendix \ref{app:why_cc_small}. Of course, the above mentioned `misalignment' depends on the theory scenario assumed, e.g. the SM vs. a given new-physics shift to the Wilson coefficients, in particular on the imaginary component of such shifts. Hence, to address the hadronic sensitivity of $\ADGmmy$ in detail, we have to first establish theory scenarios to use as benchmarks. We discuss the latter in the next section.

\section{Parenthesis: a CPV Global Fit in the Light of Recent Data} \label{sec:parenthesis}

This section lies somewhat outside the main line of discussion of this paper, which is centered around $\ADGmmy$. This section, however, emphasises that large CP-violating effects on certain $b \to s$ Wilson coefficients \cite{Altmannshofer:2021qrr,Altmannshofer:2012az,Alok:2017jgr} are still possible with updated data, and that $\ADGmmy$ offers a new, theoretically clean observable to put them to the test.

To make our point, we consider the $b \to s \mu \mu$ effective Hamiltonian (for notation see Appendix \ref{app:ffs}), and we compare real vs. complex deviations on one-Wilson-coefficient combinations including $C_{7,9,10}^{(\prime)}$, as well as on the chiral-basis combinations $C_9^{(\prime)} - C_{10}^{(\prime)} = C_{LL (RL)}$ and $C_9^{(\prime)} + C_{10}^{(\prime)} \equiv C_{LR (RR)}$. These scenarios include a few that are well-known to describe the persistent set of deviations in $b \to s$ data remarkably well, in particular $C_9^{\NP}$ and $C_9^{\NP} = -C_{10}^{\NP}$. Blissfully, this is still the case after the most recent LHCb analyses including the full Run-2 dataset \cite{Aaij:2021vac,LHCb:2021vsc,LHCb:2021awg}. These updates, recently presented at Moriond EW 2021, are included in our analysis, which is summarised in table \ref{tab:the_big_picture}. Existing studies discussing the complex-Wilson-coefficient case include \cite{Altmannshofer:2012az,Alok:2017jgr}, as well as the very recent \cite{Altmannshofer:2021qrr}. Below we add comments of comparison with these studies.

\begin{table}[h]
\renewcommand{\arraystretch}{1.2} 
  \centering
  \begin{tabular}{|c c|c|c|c||c|c|c|}
    \cline{3-8}
    \multicolumn{2}{c|}{} & \multicolumn{3}{c||}{Pre-Moriond 2021} & \multicolumn{3}{c|}{Post-Moriond 2021} \\
  \hline
  \multicolumn{2}{|c|}{Scenario} & Best-fit & Pull & $p$-value & Best-fit & Pull & $p$-value \\
  \hline
  \multirow{2}{*}{$C_7$} & \gcl $\R$ & \gcl \small{$-0.0079$} & \gcl $0.58 \sigma $ & \gcl $0.11\% $ & \gcl \small{$-0.0079$} & \gcl $0.57 \sigma $ & \gcl $0.12\% $ \\
    & $\Cplx$ & \small{$-0.0045 -0.056 \, i $} & $0.61 \sigma $ & $0.11 \% $& \small{$-0.0044-0.056 \, i $} & $0.61 \sigma $ & $0.11 \% $ \\
  \hline
  \multirow{2}{*}{$C_9$} & \gcl $\R$ & \gcl \small{$-0.97$} & \gcl $\boldsymbol{6.4 \sigma}$ & \gcl $10.0\% $ & 
  \gcl \small{$-0.93$} & \gcl $\boldsymbol{6.7 \sigma}$ & \gcl $12.0\% $\\
    & $\Cplx$ & \small{$-0.98-0.22 \, i $} & $\boldsymbol{6.1 \sigma} $ & $9.4 \% $& \small{$-0.93-0.25 \, i $} & $\boldsymbol{6.4 \sigma}$ & $12.0 \% $ \\
  \hline
  \multirow{2}{*}{$C_{10}$} & \gcl $\R$  & \gcl \small{$0.72$} & \gcl $\boldsymbol{5.8 \sigma} $ & \gcl $6.1\% $ & \gcl \small{$0.68$} & \gcl $\boldsymbol{6.0 \sigma}$ & \gcl $5.7\% $ \\
  & $\Cplx$ & \small{$0.80 +0.74 \, i $} & $\boldsymbol{5.6 \sigma}$ & $6.0 \% $ & \small{$0.76+0.75 \, i $} & $\boldsymbol{5.8 \sigma}$ & $5.6 \% $\\
  \hline
  \multirow{2}{*}{$C_{LL}$} &  \gcl $\R$ & \gcl \small{$-1.1$} & \gcl $\boldsymbol{6.9 \sigma }$ & \gcl $18.0\% $ & \gcl \small{$-0.96$} & \gcl $\boldsymbol{7.0 \sigma}$ & \gcl $16.0\% $  \\
  & $\Cplx$ & \small{$-1.2 - 1.5 \, i $} & $\boldsymbol{6.7 \sigma}$ & $18.0 \% $ &\small{$-1.1-1.4 \, i $} & $\boldsymbol{6.8 \sigma}$ & $16.0 \% $ \\
  \hline
  \multirow{2}{*}{$C_{LR}$} &  \gcl $\R$ & \gcl \small{$0.34$} & \gcl $1.2 \sigma $ & \gcl $0.13\% $  & \gcl \small{$0.28$} & \gcl $1.1 \sigma $ & \gcl $0.09\% $ \\
  & $\Cplx$ & \small{$0.34+0.032 \, i $} & $0.74 \sigma $ & $0.11 \% $& \small{$0.28+0.017 \, i $} & $0.59 \sigma $ & $0.08 \% $\\
  \hline
  \multirow{2}{*}{$C_7^{\prime}$} & \gcl $\R$ & \gcl \small{$0.004$} & \gcl $0.28 \sigma $ & \gcl $0.12\% $ & \gcl \small{$0.005$} & \gcl $0.29 \sigma $ & \gcl $0.07\% $\\
  & $\Cplx$ & \small{$0.004 - 0.001 \, i $} & $0.05 \sigma $ & $0.10 \% $ & \small{$0.005 -0.0003 \, i $} & $0.05 \sigma $ & $0.06 \% $   \\
  \hline
  \multirow{2}{*}{$C_9^{\prime}$}  & \gcl $\R$ & \gcl \small{$0.14$} & \gcl $0.74 \sigma $ & \gcl $0.13\% $ & \gcl \small{$0.0044$} & \gcl $0.06 \sigma $ & \gcl $0.09\% $ \\ 
   & $\Cplx$ & \small{$0.13+0.24 \, i $} & $0.54 \sigma $ & $0.12 \% $ &\small{$0.0012+0.2 \, i $} & $0.24 \sigma $ & $0.08 \% $ \\ 
  \hline
  \multirow{2}{*}{$C_{10}^{\prime}$}  & \gcl $\R $ & \gcl \small{$-0.18$} & \gcl $1.7 \sigma $ & \gcl $0.14\% $ & \gcl \small{$-0.09$} & \gcl $0.81 \sigma $ & \gcl $0.08\% $\\
  & $\Cplx$ & \small{$-0.20 - 0.14 \, i $} & $1.3 \sigma $ & $0.13 \% $ & \small{$-0.063 -0.11 \, i $} & $0.45 \sigma $ & $0.07 \% $ \\ 
  \hline
  \multirow{2}{*}{$C_{RL}$} & \gcl $\R$ & \gcl \small{$0.22$} & \gcl $1.5 \sigma $ & \gcl $0.17\% $ & \gcl \small{$0.088$} & \gcl $0.23 \sigma $ & \gcl $0.07\% $ \\
   & $\Cplx$ & \small{$0.24+0.40 \, i $} & $1.3 \sigma $ & $0.16 \% $ & \small{$0.085+0.32 \, i $} & $0.40 \sigma $ & $0.07 \% $ \\ 
  \hline
  \multirow{2}{*}{$C_{RR}$} & \gcl $\R$ & \gcl \small{$-0.37$} & \gcl $1.4 \sigma $ & \gcl $0.17\% $ & \gcl \small{$-0.28$} & \gcl $1.1 \sigma $ & \gcl $0.09\% $ \\
   & $\Cplx$ & \small{$-0.37-0.003 \, i $} & $0.93 \sigma $ & $0.15 \% $& \small{$-0.28 -0.004 \, i $} & $0.65 \sigma $ & $0.08 \% $ \\ 
  \hline
  \end{tabular}
  \caption{Comparison table of real vs. complex one-Wilson-coefficient scenarios for the $b \to s \mu \mu$ Hamiltonian. The best-fit columns refer to the NP contribution only. The pull is meant with respect to the SM likelihood. Pre- vs. post-Moriond 2021 results refer to the exclusion vs. inclusion of the updates in the last row of table \ref{tab:list_of_obs}. For reference, the $p$-value of the SM with the considered set of observables is $0.12 \%$ with pre-Moriond results and $0.09 \%$ after Moriond 2021, meaning that scenarios other than $C_9$, $C_{10}$ or $C_{LL}$ are no better than no NP at all.}
  \label{tab:the_big_picture}
\renewcommand{\arraystretch}{1.0} 
\end{table}

We use the common notation $C_i = C_i^{\SM} + C_i^{\NP}$, with the renormalization scale set to 4.8 GeV. We constrain $C_i^{\NP}$ with a maximum-likelihood approach, as implemented in the Python packages {\tt smelli} and {\tt flavio} \cite{flavio,smelli,wilson}. We scan the likelihood of each scenario in the $\Re(C_i^\text{\NP})/\Im(C_i^\text{\NP})$ plane, using all observables directly relevant to the $b \to s$ sector, which are summarised in table \ref{tab:list_of_obs}. In particular, we include the $R_K$ and $\Bs \to \mumu$ updates as of Moriond 2021.\footnote{The recent $\Bs \to \mumu$ update includes the first (ever) limit on $\Bs \to \mumu \gamma$ in a limited portion of the $q^2$ spectrum, following the method in Ref. \cite{Dettori:2016zff}. This decay is sensitive to all Wilson coefficients relevant to the current discrepancies \cite{Guadagnoli:2016erb,Melikhov:2004mk}. The new bound is not yet constraining for our analysis.}

The important qualitative message from table \ref{tab:the_big_picture} is that, for a given Wilson coefficient or Wilson-coefficient combination, sizeable imaginary contributions are entirely compatible with data. Representative shifts can be read off from the second column in table \ref{tab:the_big_picture}, which refers to the NP contribution only. These shifts show that the wealth of available $b \to s$ data is still relatively under-constraining for Wilson-coefficient shifts with a large phase not aligned to the corresponding one in the SM contribution. As we discuss in later sections, $\ADGmmy$ is a novel, efficient probe of precisely this case.

\begin{table}[h]
\renewcommand{\arraystretch}{1.2}
  \centering
  \begin{tabular}{|l|c||l|c|}
  \hline
  \textbf{$ \pmb{b \to s \mu \mu}$ CP-even obs.\phantom{xxxxx}}           & Ref.                                             & $\pmb{\Lambda_b \to \Lambda \mu \mu}$ \textbf{obs.\phantom{xxxxx}}                 & Ref.                          \\
  \hline
  $\left< \frac{d \mc B}{dq^2}\right> (B^+\to K^{(*)}\mu\mu )$    & \multirow{2}{*}{\cite{Aaij:2014pli}}                &$\left< \frac{d \mc B}{dq^2} \right> (\Lambda_b\to\Lambda\mu\mu)$ & \cite{Aaij:2015xza} \\
  $\left<\frac{d\mc B}{dq^2}\right>(B_0\to K\mu\mu)$          &                                                          & $\left<A_{{\rm FB}l,lh,h}\right>(\Lambda_b\to\Lambda\mu\mu)$           &  \cite{Aaij:2018gwm}            \\
  \cline{3-4}
   $\left< \frac{d \mc B}{dq^2} \right> (B_s\to \phi\mu\mu)$   & \cite{Aaij:2015esa,CDF:2012qwd}                       &   $\pmb{b \to s \gamma}$ \textbf{obs.} & Ref.                                                    \\
  \cline{3-4}
   $\left< \frac{d \mc B}{dq^2} \right> (B_0\to K^*\mu\mu)$ & \cite{Aaij:2016flj}                                      &  $\mc B(B^+\to K^*\gamma)$         & \multirow{4}{*}{\cite{Amhis:2014hma}}                                \\
   $\left<\mc B\right>(B\to X_s\mu\mu)$                           & \cite{Lees:2013nxa}                                 & $\mc B(B_0\to K^*\gamma)$ &                                                                            \\
   $\left< F_L\right>(B_0\to K^* \mu \mu)$   & \cite{Aaboud:2018krd,Aaij:2020nrf,CDF:2012qwd,Khachatryan:2015isa}      &  $S_{K^*\gamma}$   &                                                                                      \\
   $\left<P_1\right>(B_0\to K^* \mu \mu)$     & \multirow{2}{*}{\cite{Aaboud:2018krd,Aaij:2020nrf,CMS:2017ivg}}         &  $A_{\CP}(B\to X\gamma)$   &                                                                              \\  
   $\left<P'_5\right>(B_0\to K^* \mu \mu)$      &                                                                       & $\mc B(B\to X_s\gamma)$       & \cite{Misiak:2017bgg}                                                    \\
   $\left<P'_4\right>(B_0\to K^* \mu \mu)$      & \cite{Aaboud:2018krd,Aaij:2020nrf}                                    &  $\mc B(B_s\to \phi\gamma)$  & \cite{Dutta:2014sxo}                                                      \\
   $\left< A_{{\rm FB}}\right> (B_0\to K^* \mu \mu)$  & \cite{CDF:2012qwd,Khachatryan:2015isa}                          & \small{$\mc B(B_0\to K^* \gamma)/ \mc B(B_s \to \phi \gamma)$} & \cite{Aaij:2012ita}                     \\
   $\left<P_2\right>(B_0\to K^*\mu\mu)$         & \cite{Aaij:2020nrf}                                                   & $S_{\phi\gamma}$      & \multirow{2}{*}{\cite{Aaij:2019pnd}}                                              \\
   $\left< F_L\right>(B^+\to K^{*+} \mu \mu)$   & \multirow{3}{*}{\cite{Aaij:2020ruw}}                                  &   $A_{\Delta\Gamma}(B_s\to \phi\gamma)$ &                                                                 \\
   \cline{3-4}
   $\left<P_{1,2}\right>(B^+\to K^{*+}  \mu \mu)$     &                                                                 &  $R_{K/K^*}$     & \cite{Aaij:2017vbb,Abdesselam:2019wac,Abdesselam:2019lab,Aaij:2019wad}          \\
   $\left<P'_{4,5}\right>(B^+\to K^{*+}  \mu \mu)$       &                                                              & \gcl $R_{K/K^*}$     & \gcl \cite{Aaij:2021vac} $\leftarrow $\cite{Aaij:2019wad}                     \\
   \cline{1-4}
   $\pmb{B_0 \to K^* \mu \mu}$ \textbf{CPV obs.} & Ref.                                                                      & $\mc B(B_{d,s}\to \mu\mu)$  & \cite{Aaij:2017vad,Aaboud:2018mst,Sirunyan:2019xdu,LHCb-CONF-2020-002}              \\
   \cline{1-2}
   $A_{7,8,9}(B_0 \to K^* \mu \mu)$ & \cite{Aaij:2015oid}                                                              &  \gcl $\mc B(B_{d,s}\to \mu\mu)$    & \gcl \cite{LHCb:2021vsc,LHCb:2021awg} $\leftarrow$ \cite{Aaij:2017vad} \\
  \hline
  \end{tabular}
  \caption{List of the most constraining observables and their measurements implemented in the {\tt flavio} and {\tt smelli} Python packages at the date of publication. In the last table line (light gray) we specify the measurements that were updated as of Moriond 2021, and with the notation $\leftarrow [...]$ the measurements $[...]$ that these updates supersede.}
  \label{tab:list_of_obs}
\renewcommand{\arraystretch}{1.0} 
  \end{table}

\begin{figure}[!t]
  \centering
  \includegraphics[scale=0.56]{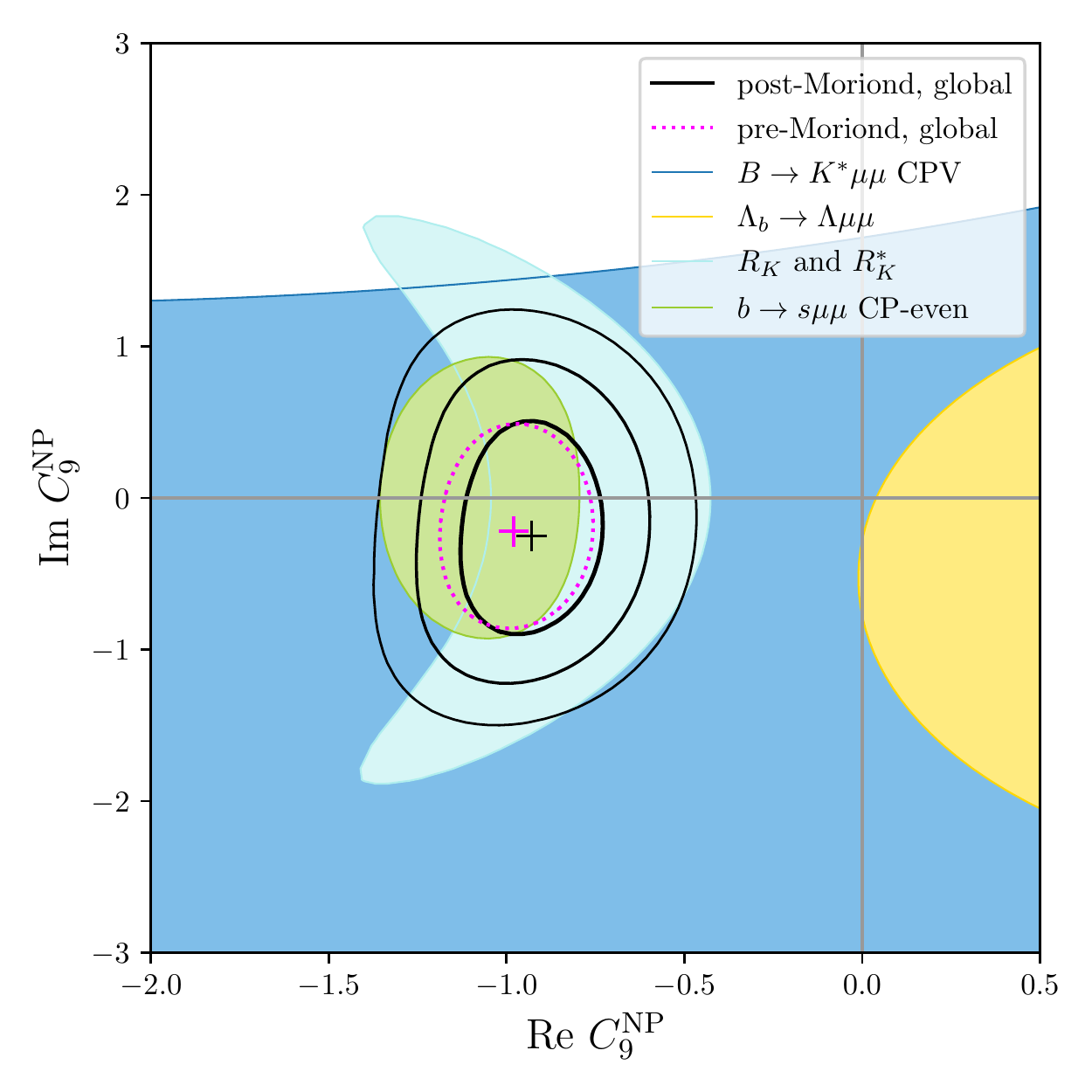}
  \includegraphics[scale=0.56]{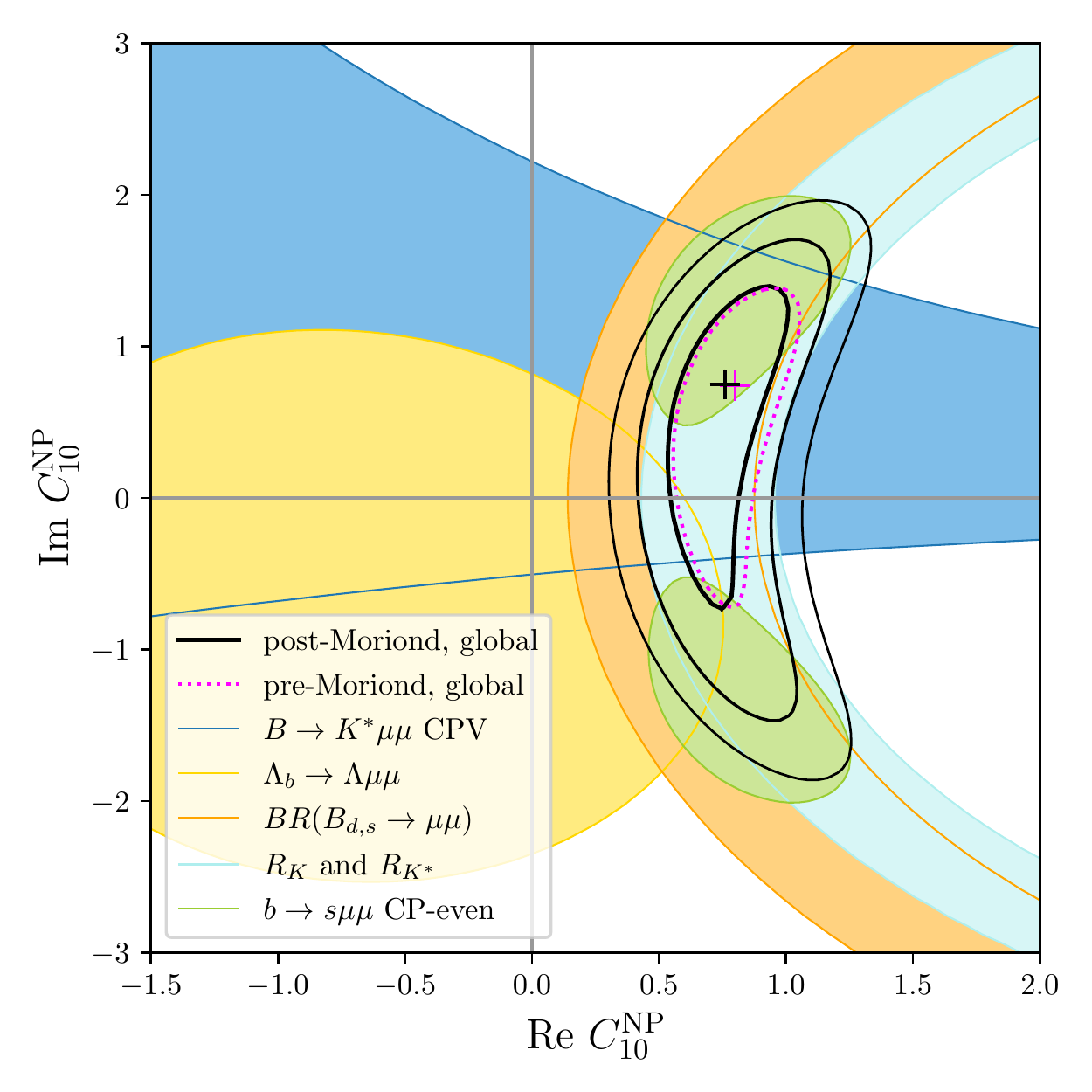} \\
  \includegraphics[scale=0.56]{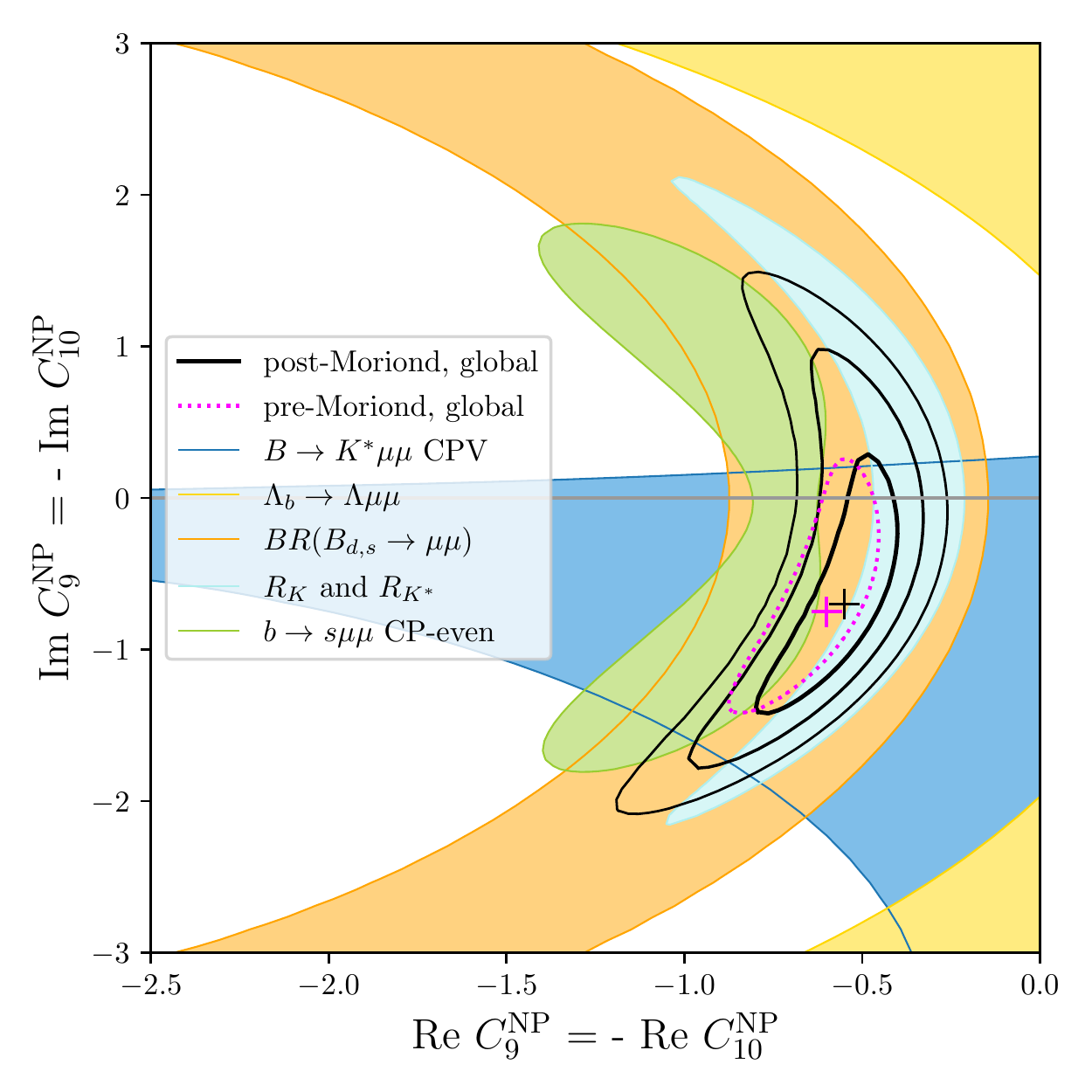}
  \includegraphics[scale=0.56]{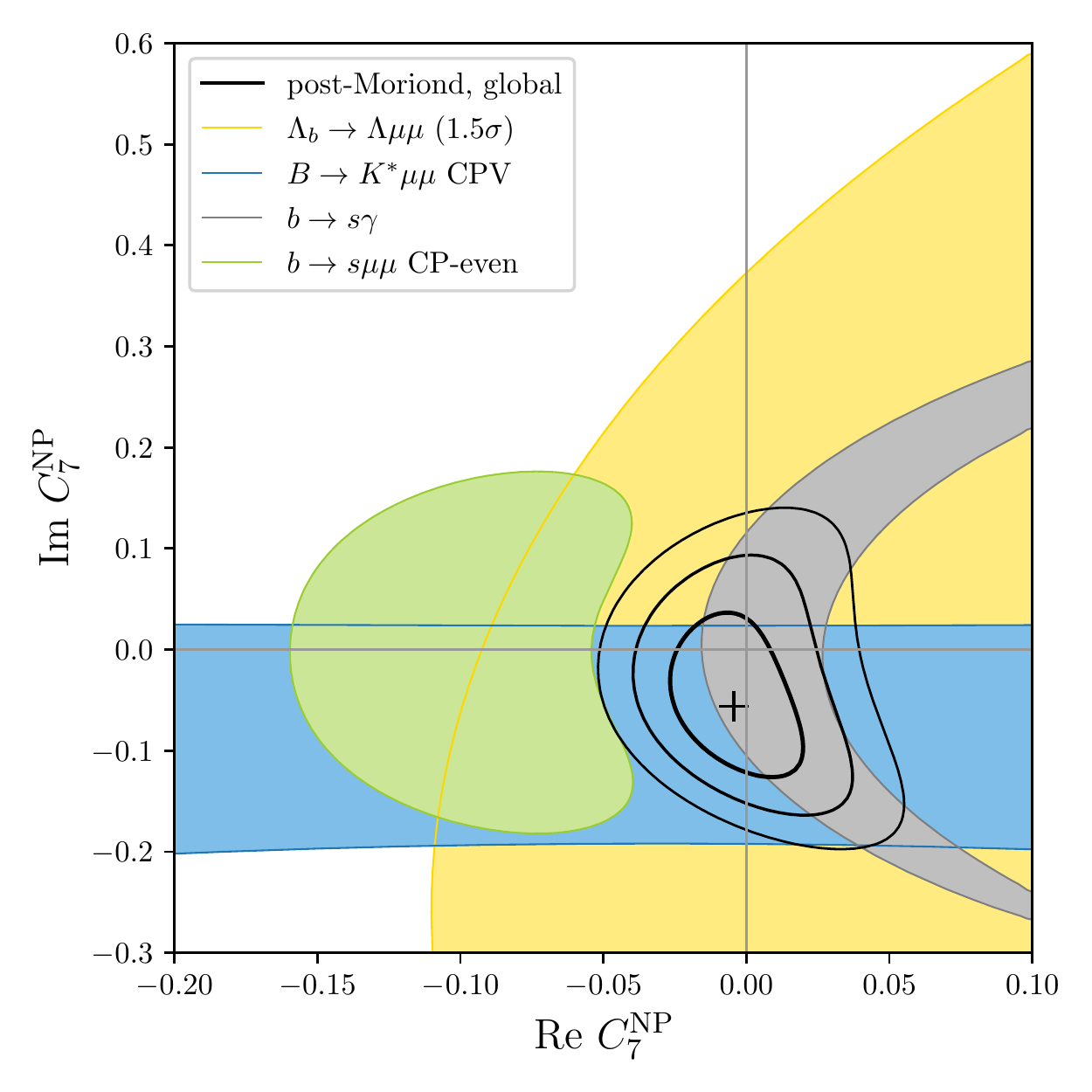}
  \caption{$1\sigma$ constraints on the NP Wilson coefficients from $\Lambda_b \to \Lambda \mu\mu$ (yellow), $b \to s \mu\mu$ (green), $b \to s\gamma$ (grey), $R_{K/K^*}$ (light blue), $B_s \to \mu\mu$ (orange) and $B \to K^* \mu \mu$ CPV (dark blue).
Pre- and post-Moriond 2021 global-fit regions are denoted in dotted purple (1$\sigma$ only) and solid black (1, 2 and $3\sigma$). The corresponding crosses indicate the best-fit point. Pre- and post-Moriond global-fit contours exactly overlap in the $C_7$ case (last panel).}
  \label{fig:global_fit}
\end{figure}

In fig.~\ref{fig:global_fit} we also show the 1, 2, 3$\sigma$ CL contours for the global fit in four selected scenarios, as solid black lines. These scenarios are $C_7^{\NP}, C_9^{\NP}$, $C_{10}^{\NP}$, and $C_9^{\NP} = -C_{10}^{\NP}$ $= C_{LL}^{\NP}/2$.\footnote{These scenarios will henceforth be referred to by dropping the $\NP$ superscript, i.e. as $C_7$, $C_9$, $C_{10}$ and $C_{LL}$, respectively. In addition, the `SM scenario' will denote the case with all $C_i^{\NP} = 0$.} We also plot individual constraints coming from the subsets of observables that are the most constraining among all those included in both {\tt smelli} and {\tt flavio}. These subsets are indicated in boldface in table \ref{tab:list_of_obs}. We make the following comments about these results:

\begin{itemize}

\item The fit improvement with respect to the purely SM solution ($\Re(C_i^\text{\NP}) = \Im(C_i^\text{\NP}) = 0$) is quite significant for the real part of the Wilson coefficient in the $C_9$, $C_{10}$ and $C_{LL}$ scenarios, as quantified by the pull, and as well-known \cite{Aebischer:2019mlg,Alguero:2019ptt,Arbey:2019duh,Ciuchini:2019usw,Kowalska:2019ley,Bhattacharya:2019dot,Alok:2019ufo}. On the other hand, in each considered scenario there is no significant pull on the sheer imaginary part of the NP Wilson coefficient considered: in each scenario, the SM solution is consistent with the available data. More CP-sensitive data would therefore be welcome in this respect.

\item The allowed region in the $C_9$ and $C_{LL}$ scenarios is smaller in our case as compared to the results of \cite{Alok:2017jgr}. The difference can be traced back to both updated measurements and the addition of new observables to the global fit. In particular, the recent measurements of the branching ratios and angular observables in $B \to K^*\mu\mu$ decays increases the constraining power along the imaginary axis; the addition of $\Lambda_b \to \Lambda \mu \mu$ branching ratio and angular observables pulls the allowed region towards the SM prediction; the updates in $R_{K/K^*}$ and in the $B_s\to \mu\mu$ branching ratio modifies the constraint on the real axis.

\item The imaginary part of $C_{10}$ and $C_{LL}$ is mostly constrained by the updated measurements in $B \to K^{(*)} \mu\mu$ observables. In particular, the addition of $B_0 \to K^* \mu \mu$ CP asymmetric angular observables \cite{Aaij:2015oid} (yielding the contours displayed in darker blue) shows a preference for $\Im(C_{10}^{\NP}) > 0$ and $\Im(C_{LL}^{\NP}) < 0$. On the other hand, $R_{K^{(*)}}$ and $\mc B(B_s \to \mu \mu)$ constrain more the real part.

\item $C_7$ is mostly constrained by $B \to K^*\mu\mu$ and $b\to s\gamma$ observables.\footnote{In the fit to $C_7$ we also included $B_0 \to K^* e^+ e^- $ decays angular observables \cite{Aaij:2020umj}. This constraint is at present very weak, and is not displayed in the $C_7$ plot. In fact, the $1 \sigma$ region covers the full area shown.} We do not find a significant deviation from the SM prediction for either the real or imaginary parts. Among the $b \to s \gamma$ observables implemented in {\tt flavio} and {\tt smelli}, the most constraining ones are the branching ratios of $B^{0,+} \to K^{* 0,+} \gamma$, $B_s \to \phi \gamma$ and $B \to X_s \gamma$. On the contrary, $A_{\Delta \Gamma}(B^0_{d,s}\to \phi \gamma)$, $S_{K^*\gamma}$ and $S_{\phi\gamma}$ do not yield competitive constraints on the real and imaginary parts of $C_7$.

\end{itemize}

From the above fits, we infer the following benchmark scenarios to be considered in the sections to follow:%
  \be
  \label{eq:NPbenchmarks}
  \def\arraystretch{1.2}
  \begin{tabular}{l||ccc}
  Scenario & $C_7^{\NP}$ & $C_9^{\NP}$ & $C_{10}^{\NP}$ \\
  \hline
  $C_7$ & $0.02 - 0.13 i$ & 0 & 0 \\
  $C_9$ & 0 & $-1.0 - 0.9 i$ & 0 \\
  $C_{10}$ & 0 & 0 & $1.0 + 1.4 i$ \\
  $C_{LL}$ & 0 & $-0.7 - 1.4 i$ & $0.7 + 1.4 i$ \\
  \end{tabular}
  \def\arraystretch{1.0}
  \ee%
These benchmark points are chosen to lie within the 1$\sigma$ region of the fit, and to have an imaginary part as large as possible.

\section{\boldmath $\ADGmmy$ at low $q^2$} \label{sec:low_q2}

The scenarios collected in eq. (\ref{eq:NPbenchmarks}) serve us to establish reference shifts to $\ADGmmy$ with respect to its SM value. We can then investigate to what extent these shifts are resolved once we keep into account hadronic uncertainties due to f.f.'s and to resonance modelling.

We first consider the region of low $q^2$, located between the lower kinematic limit $4 m_\mu^2$ and an upper bound around 6 GeV$^2$, determined in order to minimise possible pollution from the $J/\psi$-resonance sideband. The low-$q^2$ region is of interest because of the sizeable enhancement of the $C_7$ contribution due to the `nearby' photon pole, and because a determination of the f.f.'s using factorisation methods has recently become available \cite{Beneke:2020fot}. Away from resonant regions, this determination allows for a rigorous assessment of theory errors.

Besides the analysis in Ref. \cite{Beneke:2020fot} (BBW in what follows), another  $\bar B_s \to \gamma$ f.f. determination, based on Light-Cone Sum Rules (LCSR) has recently appeared in Ref. \cite{Janowski:2021yvz}, and will henceforth be denoted as JPZ. The JPZ parameterization also holds for the high-$q^2$ region to be discussed in Sec. \ref{sec:high_q2}. Before the BBW and JPZ parameterizations, the only other existing f.f. parameterisation, Ref. \cite{Kozachuk:2017mdk}, was based on a phenomenological model.

We are thus in a position to perform a direct comparison between the BBW and JPZ f.f.'s in $\ADGmmy$.
To this end, we first re-express the full amplitude calculated in \cite{Beneke:2020fot} in terms of the $F_{V,A,TV,TA}$ f.f.'s used in \cite{Kozachuk:2017mdk} as well as in related literature \cite{Kruger:1996cv,Melikhov:2000yu}. We thereby obtain `effective' BBW f.f.'s, $F_{V,A,TV,TA}^{\BBW}$, in the sense that the amplitudes in eqs. (\ref{eq:A_DE})-(\ref{eq:A_Brems}), with the effective f.f.'s $F_{V,A,TV,TA}^{\BBW}$, reproduce the BBW amplitude. A detailed discussion, along with explicit formul\ae, and a numerical comparison of $F_{V,A,TV,TA}^{\BBW}$ with the JPZ f.f.'s \cite{Janowski:2021yvz}, is presented in Appendix \ref{app:BBW}.

Here we would like to study the question how the f.f. choice, BBW vs. JPZ, impacts the prediction on $\ADGmmy$ in the different scenarios of eq. (\ref{eq:NPbenchmarks}) as well as in the SM.
Before presenting such comparison, an important qualification is in order. The low-$q^2$ region we are considering includes a narrow resonance, the $\phi$, which escapes at present a description in terms of fully rigorous QCD methods, and on the other hand accounts for a substantial fraction of the low-$q^2$ signal. One has therefore to establish a compromise between minimising the theoretical error and maximising the experimental statistics. In our numerical study, we use sliding $s_1$ and $s_2$ values such that $4 m_{\mu}^2 < s_1 < M_{\phi}^2 < s_2 < 6~\GeV^2$, and we identify the $\ADGmmy$ integration region $[4 m_{\mu}^2, s_1] \cup [s_2, 6~\GeV^2]$ with the above compromise in mind.

The $\phi$ region is at present accounted through phenomenological approaches, whereby the relevant amplitude is shifted by a Breit-Wigner(BW)-like shape, and the latter is suitably parameterised.
We follow the approach in Ref. \cite{Guadagnoli:2017quo}, that we very briefly summarise here. One first identifies the relevant reduced amplitude
\be
\label{eq:APPperp}
\APPb(q^2) = (C_7 \pm \frac{m_s}{m_b} C_7') \FTVAb(q^2) + 
(C_8 \pm \frac{m_s}{m_b} C_8') G_{\perp,\parallel}(q^2) + \sum_{i=1}^6 C_i L_{i \perp,\parallel}(q^2) \;,
\ee
where $\FTVAb(q^2)$, $G_{\perp,\parallel}(q^2)$ and $L_i(q^2)$ denote long-distance contributions, and
\be
\FTVAb(q^2) =  \FTVA(q^2,0) +  \FTVA(0,q^2)  \;,
\ee
takes into account diagrams where the e.m.-penguin photon emits the lepton pair and diagrams where the e.m.-penguin photon is the final-state one. Note that in more common notation $\bar T^{\bar{B}_s^0 \to \phi}_{\perp,\parallel}(0) = 2 T^{\bar{B}_s^0 \to \phi}_1(0) = 2 T^{\bar{B}_s^0 \to \phi}_2(0)$, see e.g. \cite{Straub:2015ica}. One then rewrites this amplitude as an $n$-times subtracted dispersion relation, such description being necessary because of the resonant behaviour. Taking $n = 1$ yields
\begin{alignat}{4}
\label{eq:FTVTA0q2}
\APP_\iota (q^2) =  
\APP_\iota (0) +   \frac{q^2}{M_\phi^2}   \, \frac{f_\phi M_\phi  {\cal \hat{A}}^{\bar{B}_s^0 \to \phi \gamma}_{\iota} }{q^2 - M_\phi^2 + i M_\phi \Gamma_\phi}~ + ~\dots\,.
\end{alignat}
This relation corresponds to the one also used in \cite{Melikhov:2004mk} in the approximation $ {\cal \hat{A}}^{\bar{B}_s^0 \to \phi \gamma}_{\perp,\parallel}   = 2 T^{\bar{B}_s^0 \to \phi}_1(0) \times C_7$ with the identification $T_1^{\bar{B}_s^0 \to \phi}(0) = -g_+^{\bar{B}_s^0 \to \phi}(0)$ \cite{Melikhov:2004mk} (see also \cite{Kruger:1996cv,Melikhov:2000yu}).
The f.f. is known most precisely from LCSR,\footnote{%
Another approach is based on relativistic quark models, with meson wave-functions constrained by leptonic decay constants. Predictions depend on the parameters used. The latest determination is provided in Ref. \cite{Kozachuk:2017mdk}.}
yielding \cite{Straub:2015ica,Ball:2004rg,Ball:2006eu}
\be
\label{eq:g+LCSR}
T_1^{\bar{B}^0_s \to \phi}(0)|_{\rm LCSR} =  0.309 \pm 0.027\,.
\ee
Finally, the $\mathcal O_8$ and four-quark contributions in eq. (\ref{eq:APPperp}) are known in the $1/m_b$-limit \cite{Beneke:2001at,Bosch:2001gv} and in LCSR \cite{Dimou:2012un,Lyon:2013gba}.

As pointed out in Ref. \cite{Guadagnoli:2017quo}, an alternative and possibly more effective strategy towards improving the prediction for the amplitudes ${\cal \hat{A}}^{\bar{B}_s^0 \to \phi \gamma}_{\perp,\parallel}$ appearing in eq. (\ref{eq:FTVTA0q2}) is to extract them from experiment. This approach is promising since the branching ratio \cite{Zyla:2020zbs}
\be
\label{eq:Bsphigamma_exp}
\mc B(\bar{B}^0_s \to \phi \gamma) = (3.52 \pm 0.34) \times 10^{-5}
\ee
is known to $10\%$ accuracy. An update including the entire dataset is in progress. Hence the statistical component of the error on this measurement---about half of the error quoted in eq. (\ref{eq:Bsphigamma_exp})---will decrease steadily.
As a consequence one can expect to extract the amplitudes at the $5\%$ level, which is well below a theory error above $10\%$.%
\footnote{%
As discussed in Ref. \cite{Guadagnoli:2017quo}, one may apply the above approach to other resonances, in particular the $\phi^\prime$. In spite of a potentially sizeable $T_1^{B_s^0 \to \phi^\prime}$ coupling, the large $\phi'$ width turns out to suppress the $\phi'$ contribution to the $\Bsmumugamma$ spectrum to be a below-1\% correction to the total branching ratio.
Finally, the above approach may be applied to narrow charmonium as well. However, the required radiative branching ratios are, again, not yet measured. Besides, short-distance dynamics in this region is dominated by the $\mc O_{9,10}^{(\prime)}$ operators, that one can more cleanly extract from the region $\sh > 0.55$.}
In other words, rather than taking $T_1^{\bar{B}^0_s \to \phi}(0)$ from eq.~(\ref{eq:g+LCSR})---which translates into an error around 15\% on the low-$q^2$ $\Bsmumugamma$ branching ratio---one may trade $T_1^{\bar{B}^0_s \to \phi}(0)$ for eq. (\ref{eq:Bsphigamma_exp}).\footnote{Of course, such approach overlooks systematic effects that go beyond f.f. dominance.} This translates into
\be
\label{eq:g+Bsphiy}
T_1^{\bar{B}^0_s \to \phi}(0)|_{\exp} =  0.375 \pm 0.018
\ee
which is over 1 standard deviation above eq. (\ref{eq:g+LCSR}).

The above discussion allows to identify the central value in eq. (\ref{eq:g+LCSR}) as reference, as well as the two values $(T_1)_{\min} = 0.282$ and $(T_1)_{\max} = 0.393$, which are respectively the $1\sigma$ lower bound from eq. (\ref{eq:g+LCSR}) and the $1\sigma$ upper bound from eq. (\ref{eq:g+Bsphiy}). These two values determine a realistic range for $T_1^{\bar{B}^0_s \to \phi}$, that in particular takes into account the `tension' between the determinations (\ref{eq:g+LCSR}) and (\ref{eq:g+Bsphiy}).

A first task is thus to identify the $\ADGmmy$ integration region $[4 m_{\mu}^2, s_1] \cup [s_2, 6~\GeV^2]$ such that a variation of $T_1^{\bar{B}^0_s \to \phi}$ in the $[(T_1)_{\min}, (T_1)_{\max}]$ range affects the $\ADGmmy$ prediction negligibly with respect to the rest of the error components. We display such dependence in fig. \ref{fig:JPZ_T1_dep} for the SM case as well as for all the NP scenarios in eq. (\ref{eq:NPbenchmarks}).
\newcommand{\w}{.4214}
\begin{figure}[ht!]
  \centering
  \includegraphics[width=\w\textwidth]{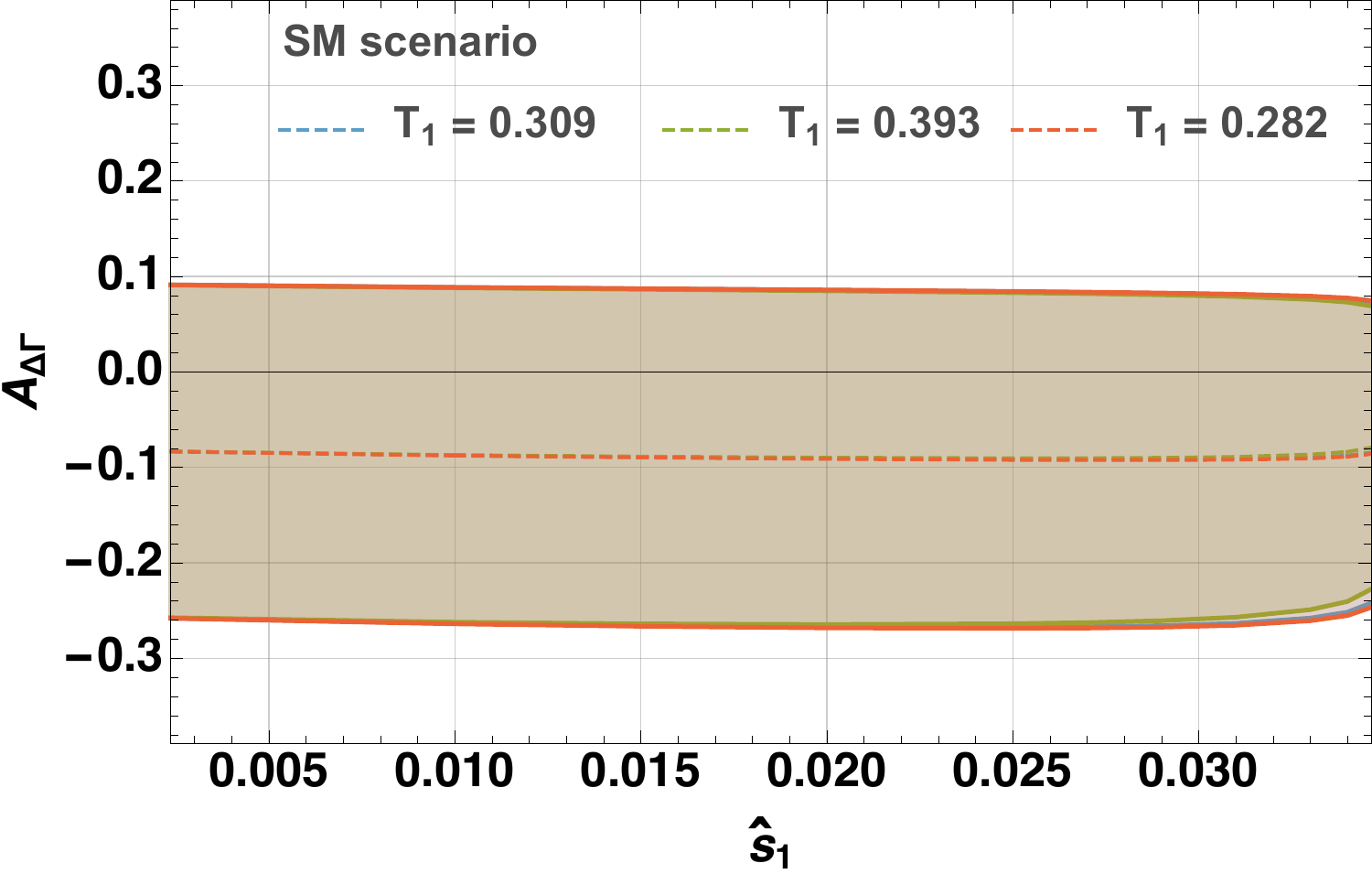}
  \hfill
  \includegraphics[width=\w\textwidth]{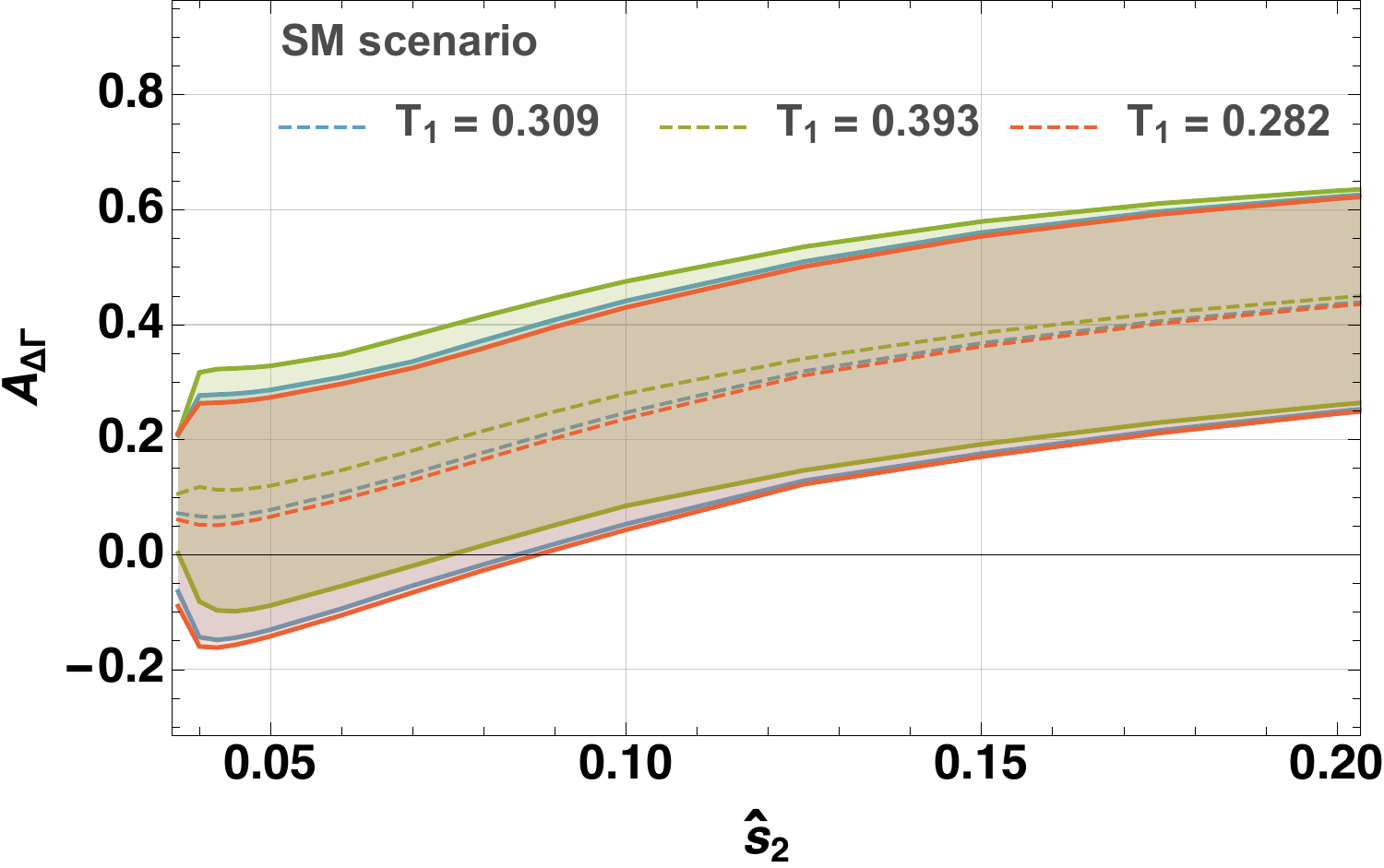}\\
  \includegraphics[width=\w\textwidth]{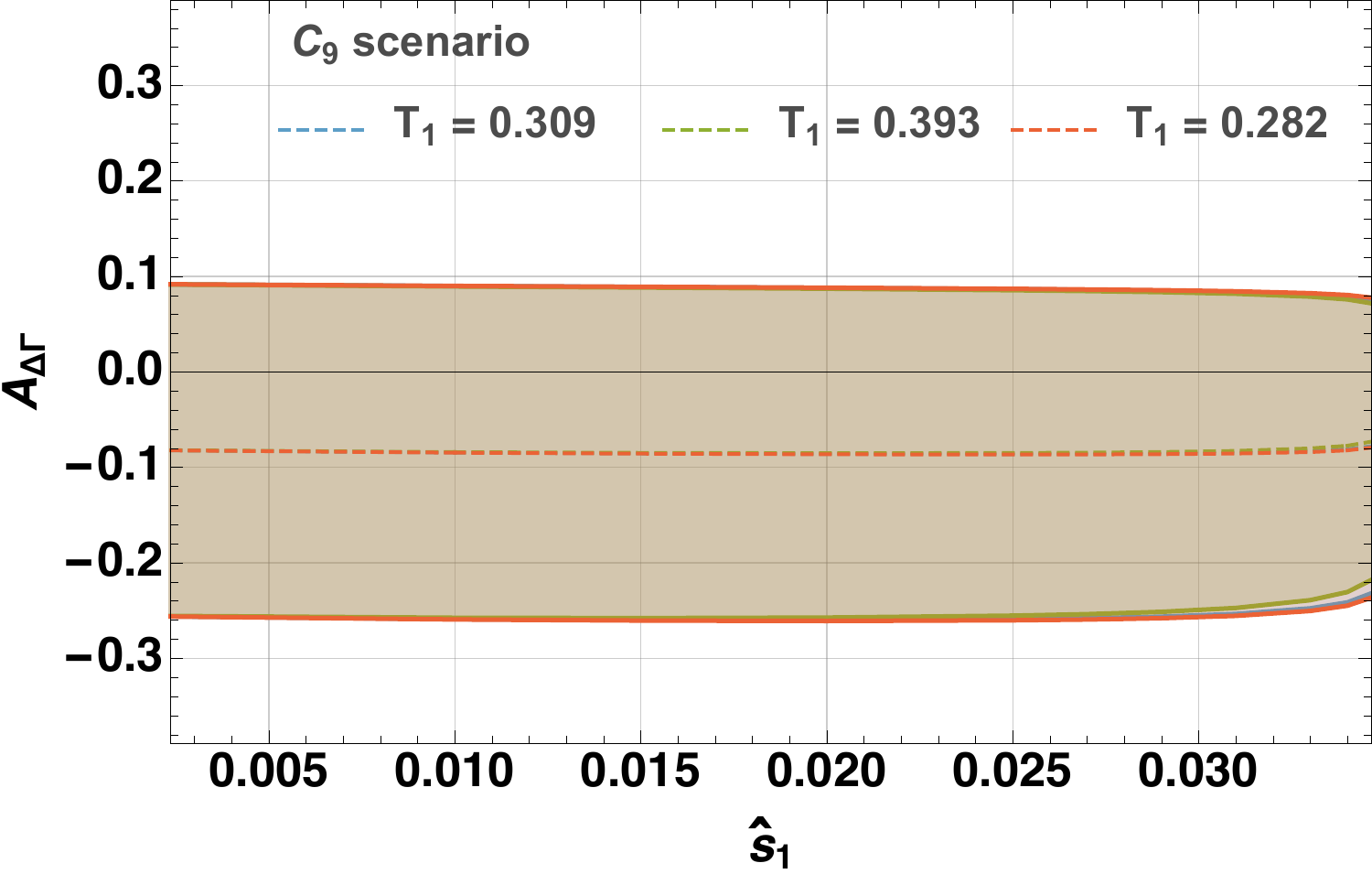}
  \hfill
  \includegraphics[width=\w\textwidth]{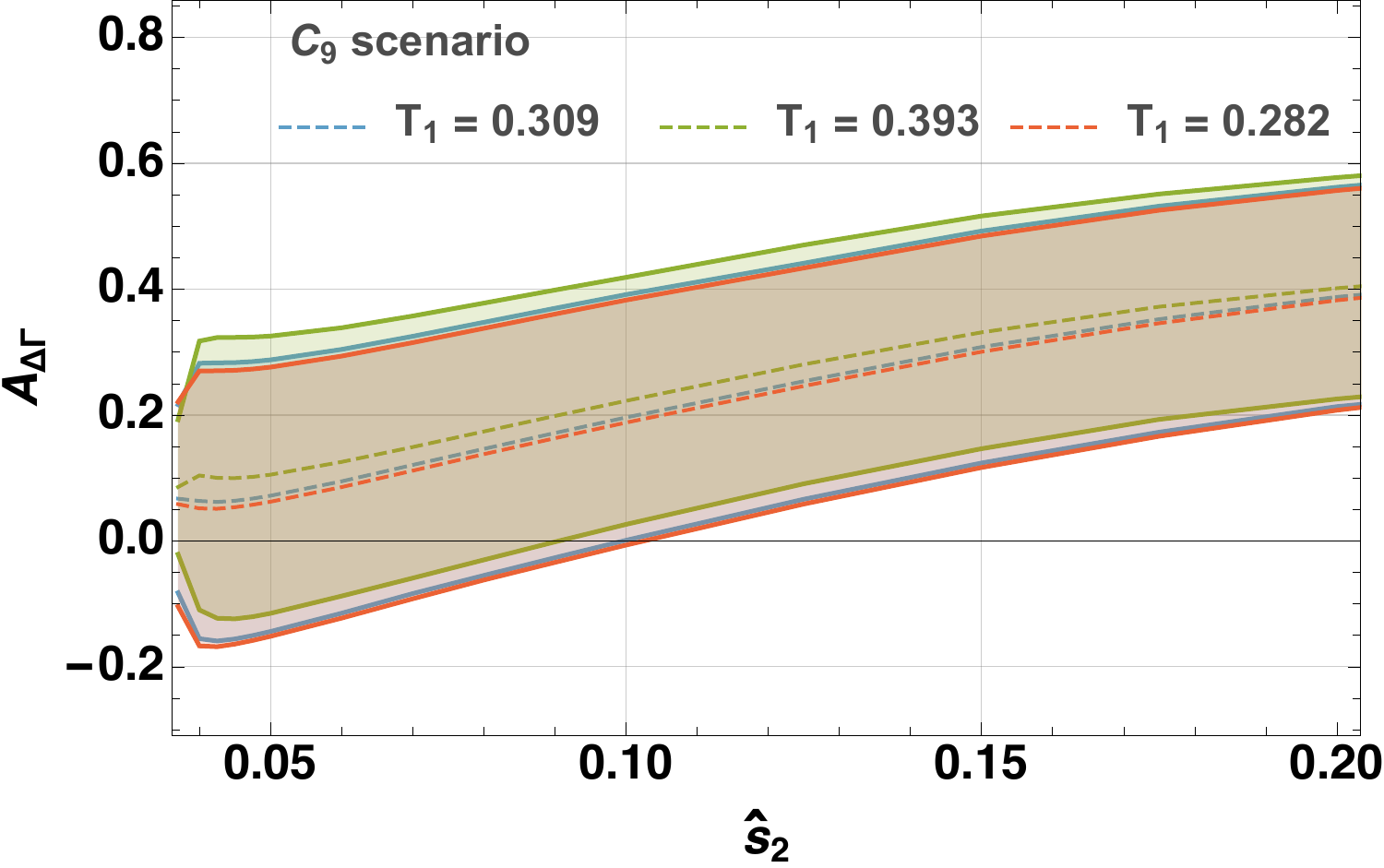}\\
  \includegraphics[width=\w\textwidth]{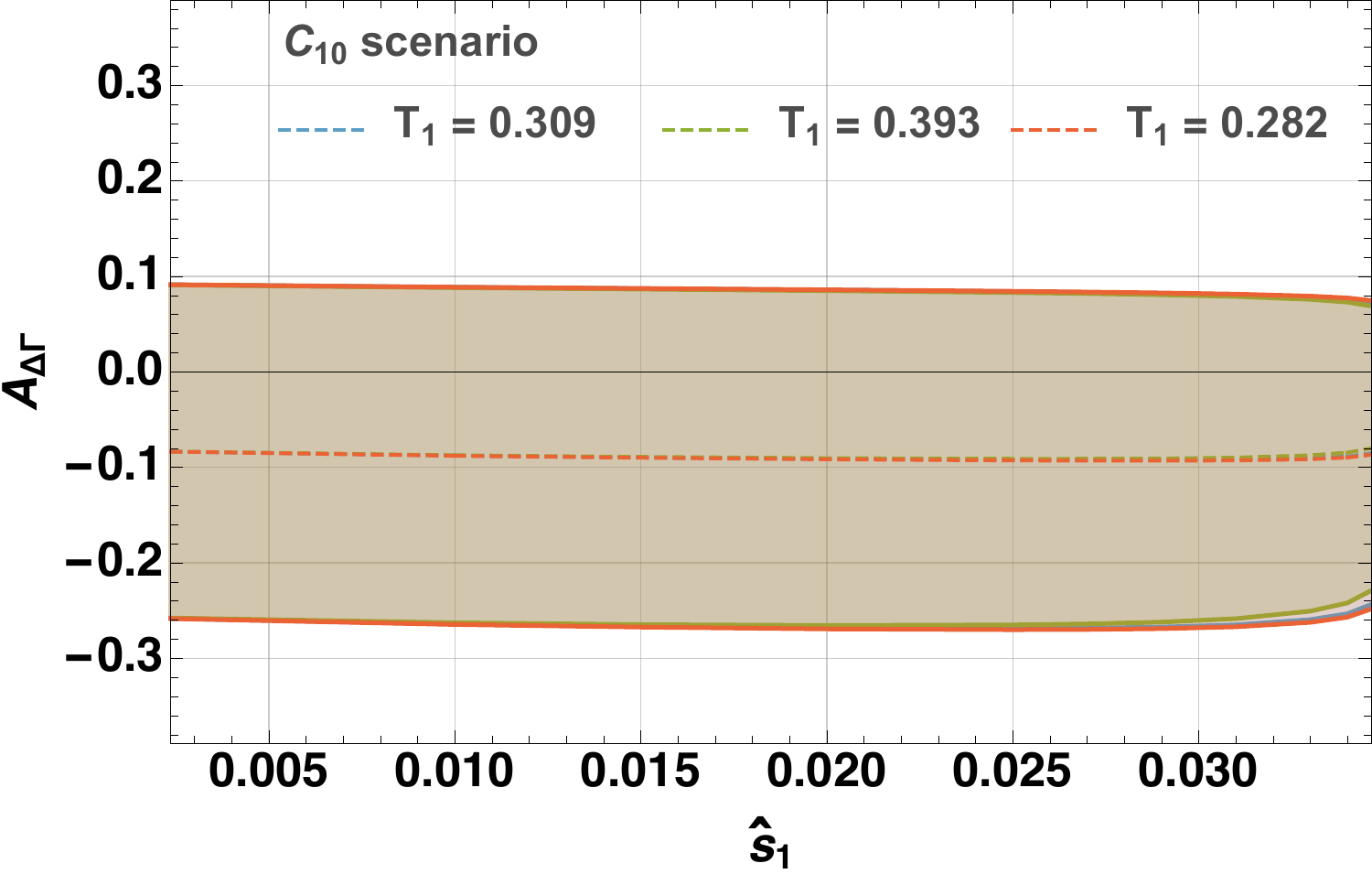}
  \hfill
  \includegraphics[width=\w\textwidth]{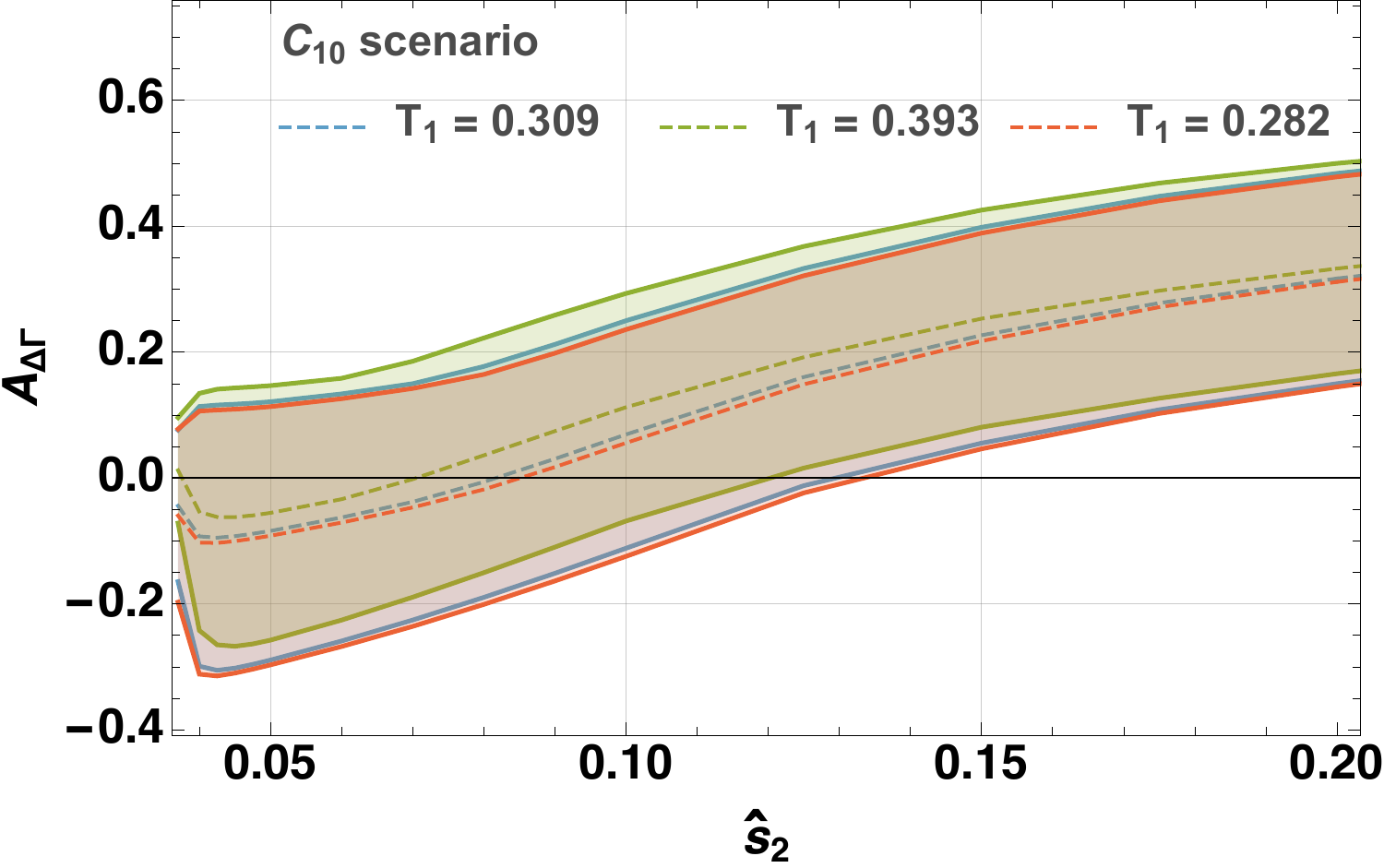}\\
  \includegraphics[width=\w\textwidth]{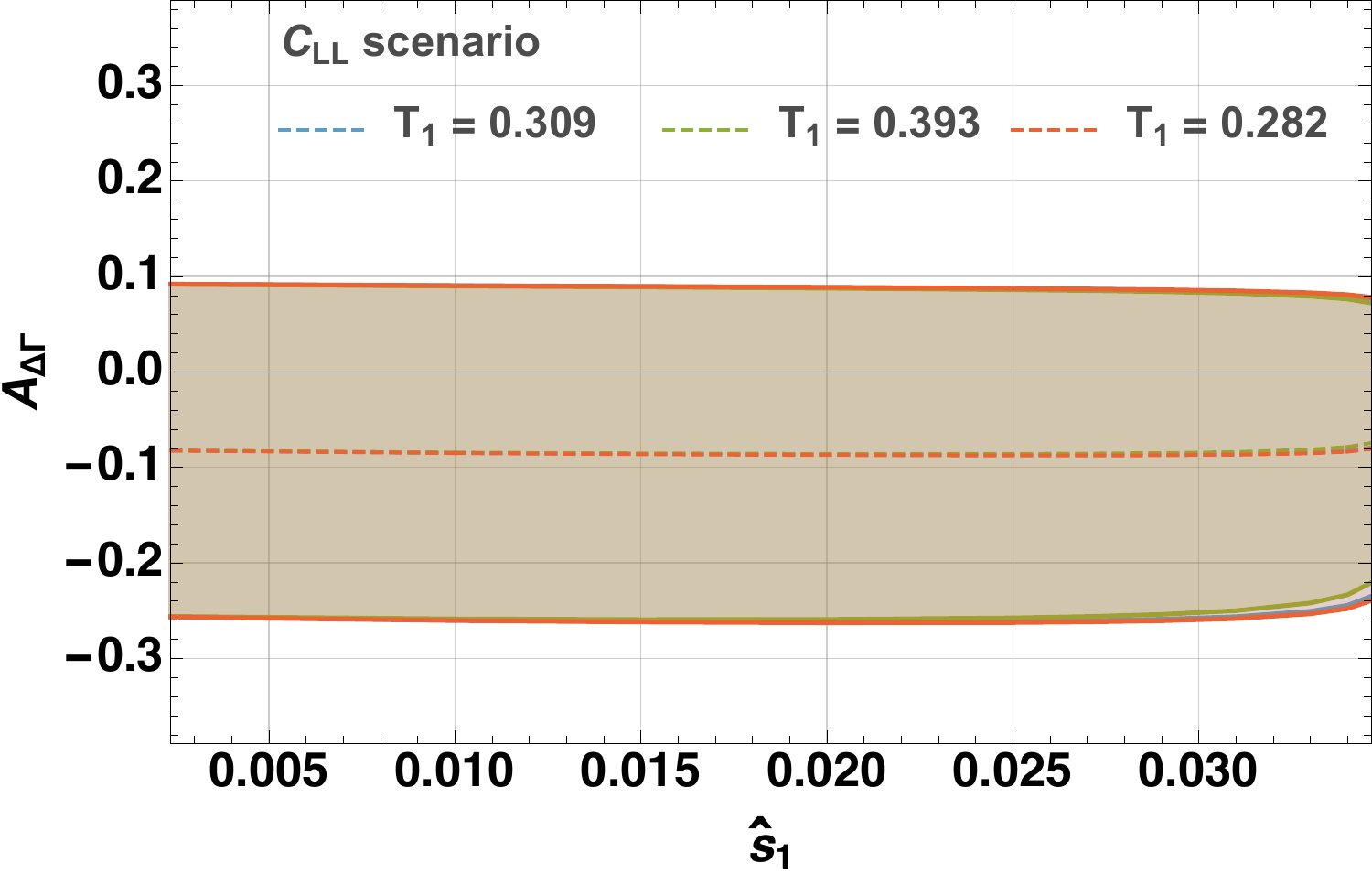}
  \hfill
  \includegraphics[width=\w\textwidth]{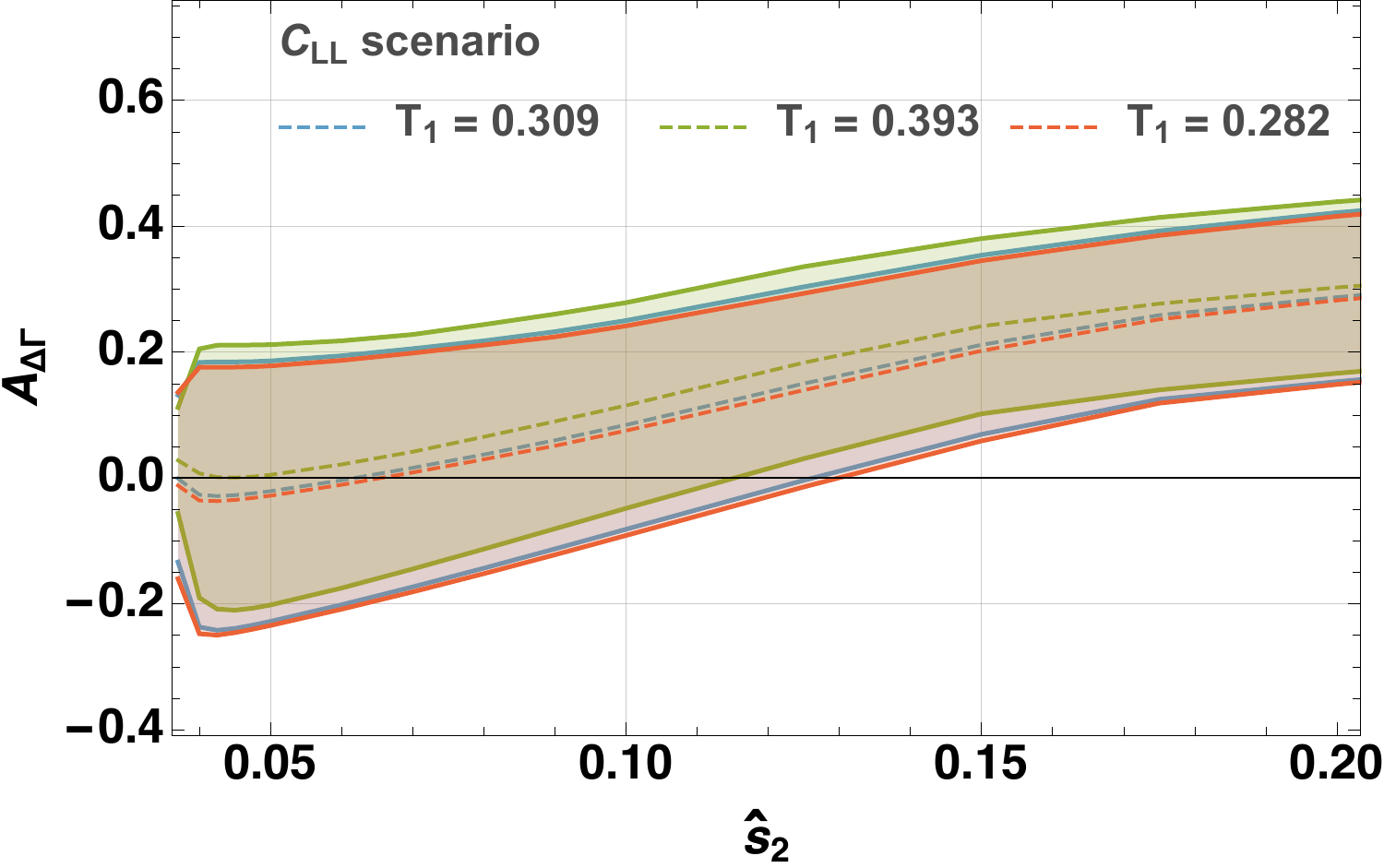}\\
  \includegraphics[width=\w\textwidth]{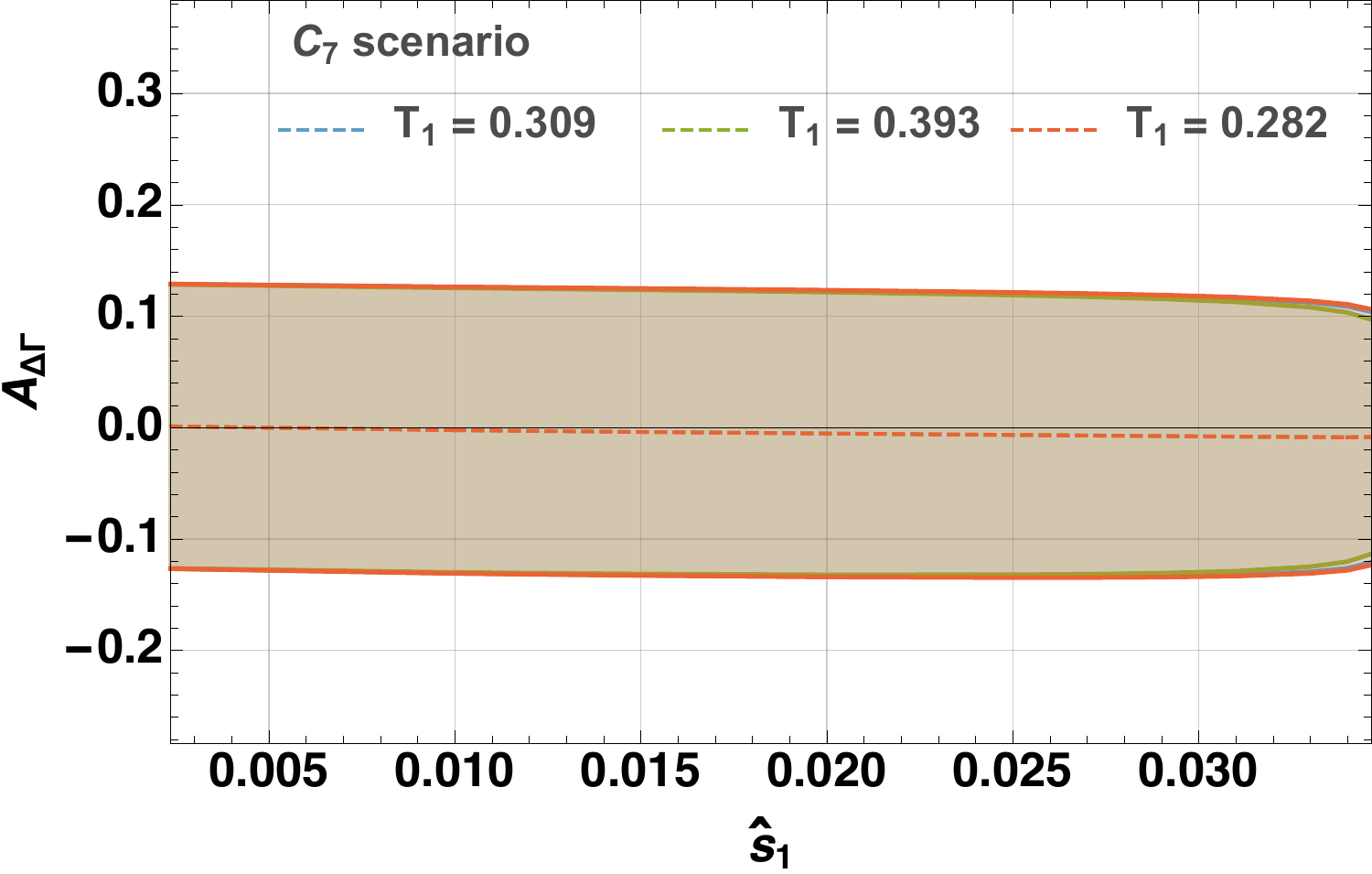}
  \hfill
  \includegraphics[width=\w\textwidth]{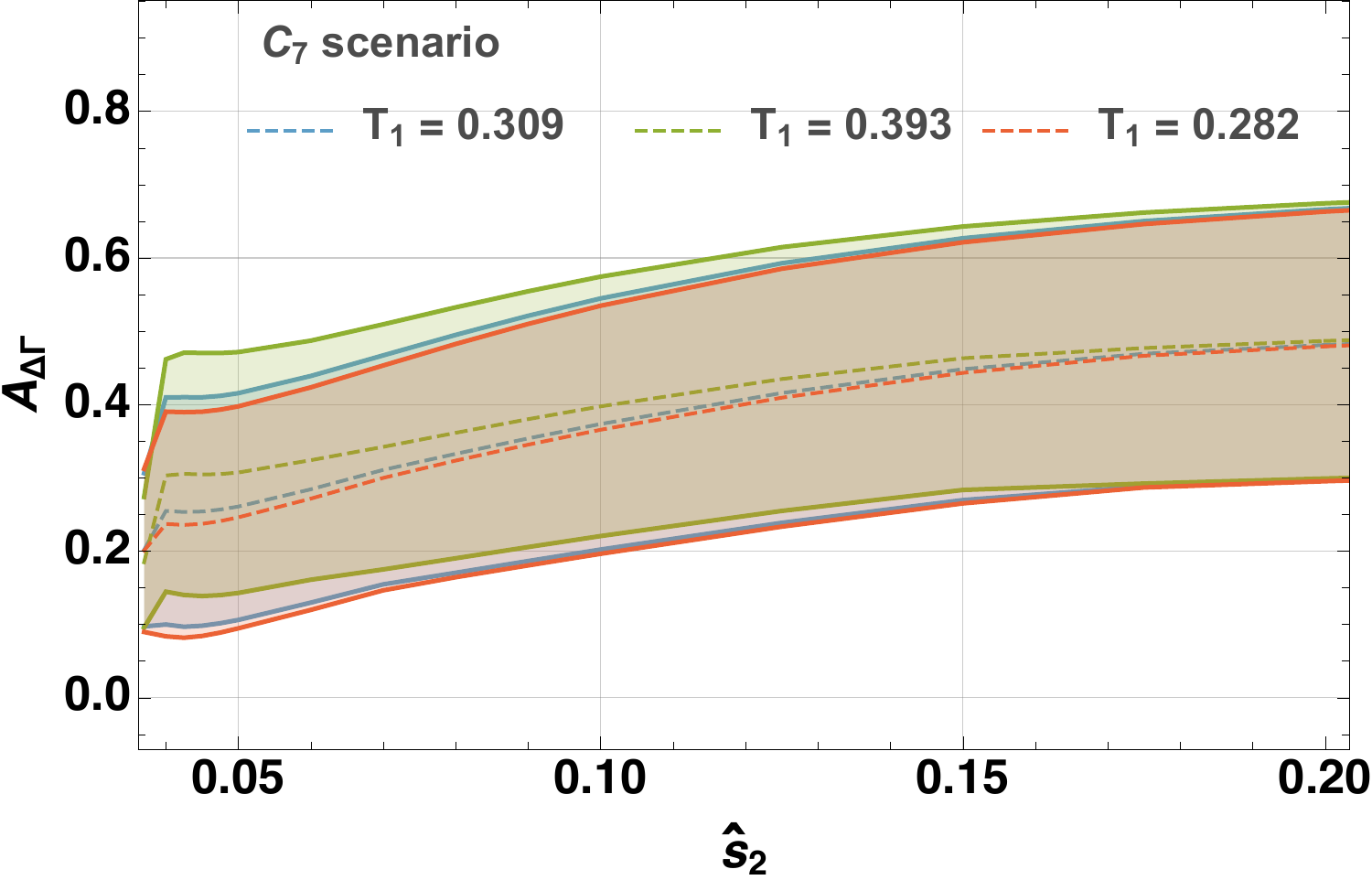}
  \caption{$\ADGmmy$ prediction, integrated in the interval $[4 m_{\mu}^2, s_1]$ (left panels) or $[s_2, 6~\GeV^2]$ (right panels), within the scenarios indicated. We use JPZ f.f.'s \cite{Janowski:2021yvz} as well as three choices of $T_1^{\bar{B}^0_s \to \phi}$---see below eq. (\ref{eq:g+Bsphiy}).}
  \label{fig:JPZ_T1_dep}
\end{figure}

The figure shows that the choice of $T_1^{\bar{B}^0_s \to \phi}$ in the quite generous range discussed has little impact on the prediction of $\ADGmmy$ even for $s_1$ as large as $0.035$ and $s_2$ as small as $0.037$---for reference, $q^2 = M_{\phi}^2$ corresponds to $s = 0.036$. This conclusion holds generally true in all considered theory scenarios. The figure suggests however to choose $\sh_1 \le$ 0.03 and $\sh_2 \ge 0.05$ ($\sh_i \equiv s_i / \MB^2$, see text below eq. (\ref{eq:intPS})), given the rapid variation of the $\ADGmmy$ prediction above and respectively below such values. With these $\sh_1$ and $\sh_2$ values in mind, we choose $T_1^{\bar{B}^0_s \to \phi}$ as the central value in eq. (\ref{eq:g+LCSR}) for definiteness.

We next discuss the comparison between the BBW and the JPZ parameterisation on the $\ADGmmy$ prediction in the low-$q^2$ region of integration. Of course, such comparison depends on the errors associated to either parameterisation.

The error on the BBW parameterisation is determined as the envelope of the errors discussed in detail in Appendix \ref{app:BBW}, and in most of the integration interval is entirely dominated by the uncertainty on the inverse moment of the $B_s$-meson distribution amplitude, $\lambda_{B_s}$. The JPZ parameterization comes with an estimation of the associated errors, as well as of the correlations between the different form factors, and we adhere to these estimations. In the range $\sh \in [0,1]$, f.f. errors are around 10\% for $F_V$ and $F_{TV}$, and comprised between 9\% and 17\% (28\%) for $F_{A}$ ($F_{TA}$).

The $\ADGmmy$ prediction integrated in the low-$q^2$ region $[4 m_{\mu}^2, s_1] \cup [s_2, 6~\GeV^2]$ with BBW vs. JPZ f.f.'s is displayed in fig. \ref{fig:BBW_vs_JPZ}, the two columns of panels denoting the region below and respectively above the $\phi$ peak, and the rows referring to the different scenarios, including the SM and those in eq. (\ref{eq:NPbenchmarks}). The figure shows that, whatever the choice of $s_1$ and $s_2$, the two parameterisations yield generally consistent results, within the large errors. There are a few notable exceptions though. The clearest is the $C_7$ scenario in the full $[4 m_\mu^2, \sh_1]$ range. The JPZ parameterization tends to predict lower values for $\re(F_{TA})$ and especially $\re(F_{TV})$ than the BBW counterparts, see fig. \ref{fig:Reff_BBW_vs_JPZ} in Appendix \ref{app:BBW}. This difference translates into lower $\ADGmmy$ predictions in the full $[4 m_\mu^2, \sh_1]$ range for any scenario (see left column of fig. \ref{fig:BBW_vs_JPZ}). These predictions become sizeably lower in the $C_7$ scenario, where the NP shift multiplies the $F_{TV,TA}$ f.f.'s. The other appreciable difference occurs in the $C_{LL}$ scenario for $\sh_2 \lesssim 0.10$ and in the $C_9$ scenario for $\sh_2 \lesssim 0.07$. We note in this respect that the JPZ shape visibly decreases for decreasing $\sh_2$, whereas the BBW shape is nearly constant throughout the $\sh_2$ range. This difference holds true in the full $\sh_2$ interval considered, and for any scenario, hence it is not attributable to features specifically related to the $\phi$-resonance region. On top of this difference, the BBW error in $\ADGmmy$ tends to squeeze for decreasing $\sh_2$ in the SM, $C_9$, and $C_{LL}$ scenarios, due to an accidental cancellation between $\lambda_{B_s}$- and $r_{\LP}$-induced components of the uncertainty, namely the two dominant ones.
The fact that, aside from these somewhat accidental instances, the two parameterizations give consistent results on $\ADGmmy$ is, we believe, a non-trivial finding.

\newcommand{\ww}{.422}
\begin{figure}[ht!]
  \centering
  \includegraphics[width=\ww\textwidth]{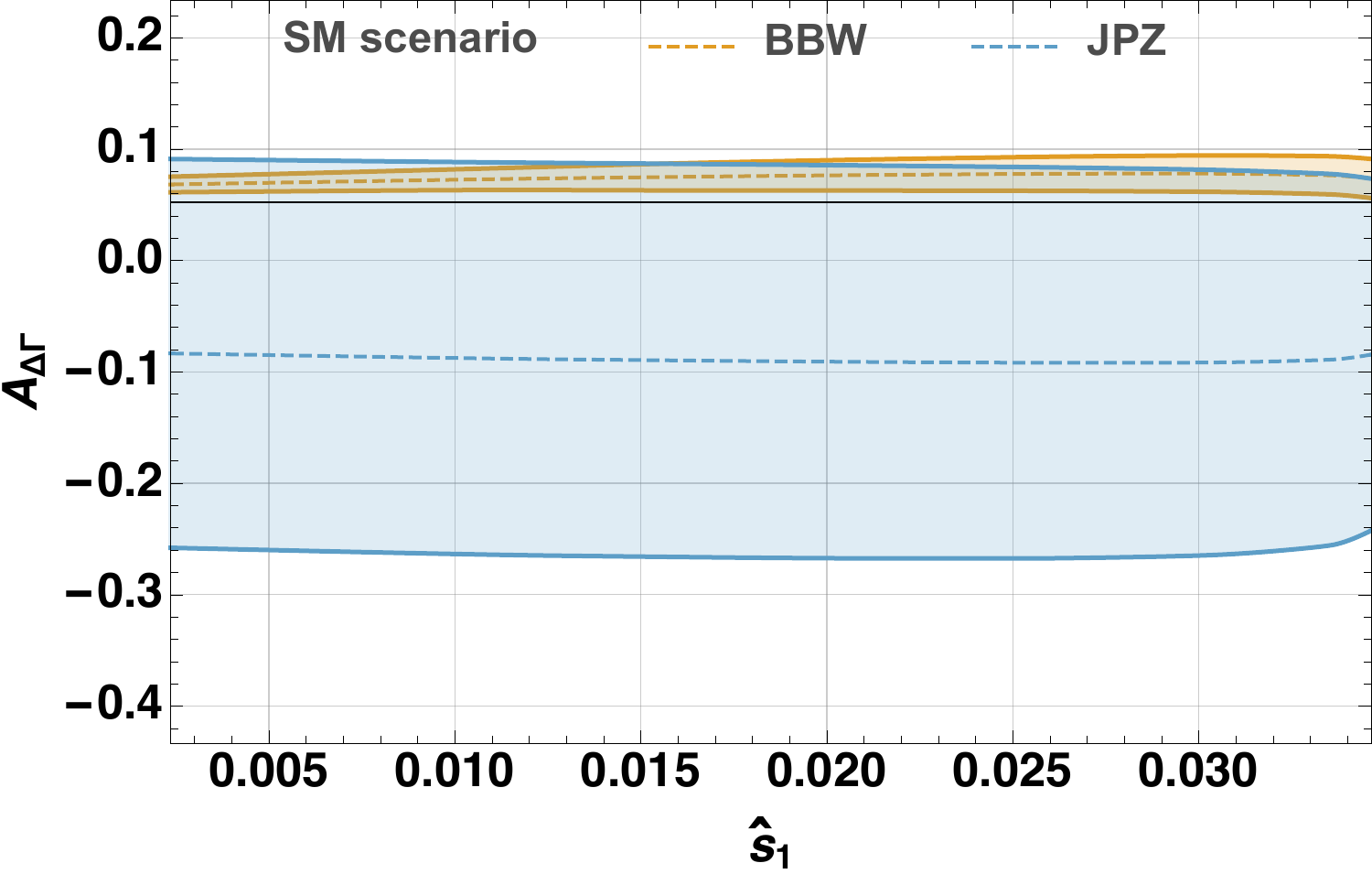}
  \hfill
  \includegraphics[width=\ww\textwidth]{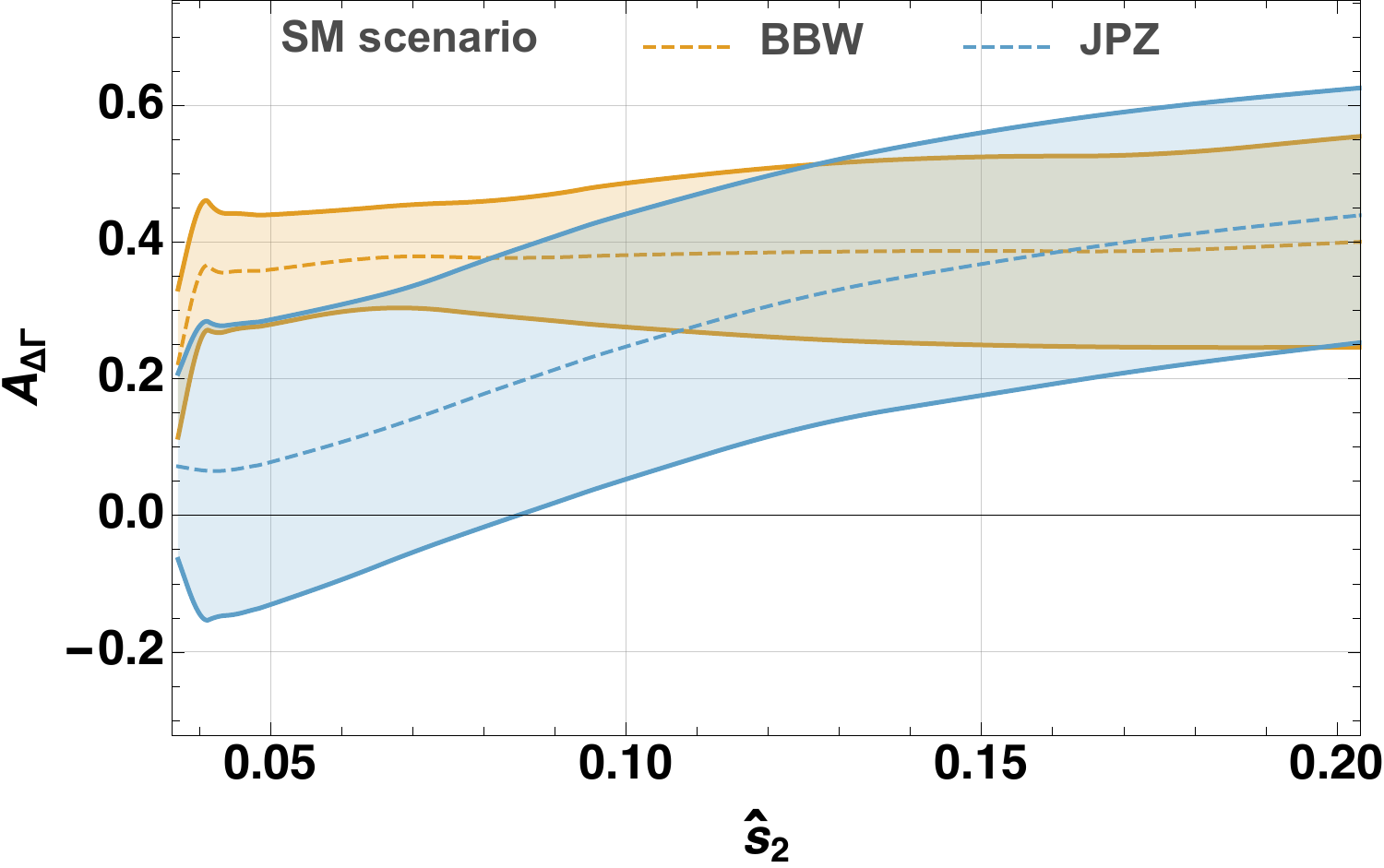}\\
  \includegraphics[width=\ww\textwidth]{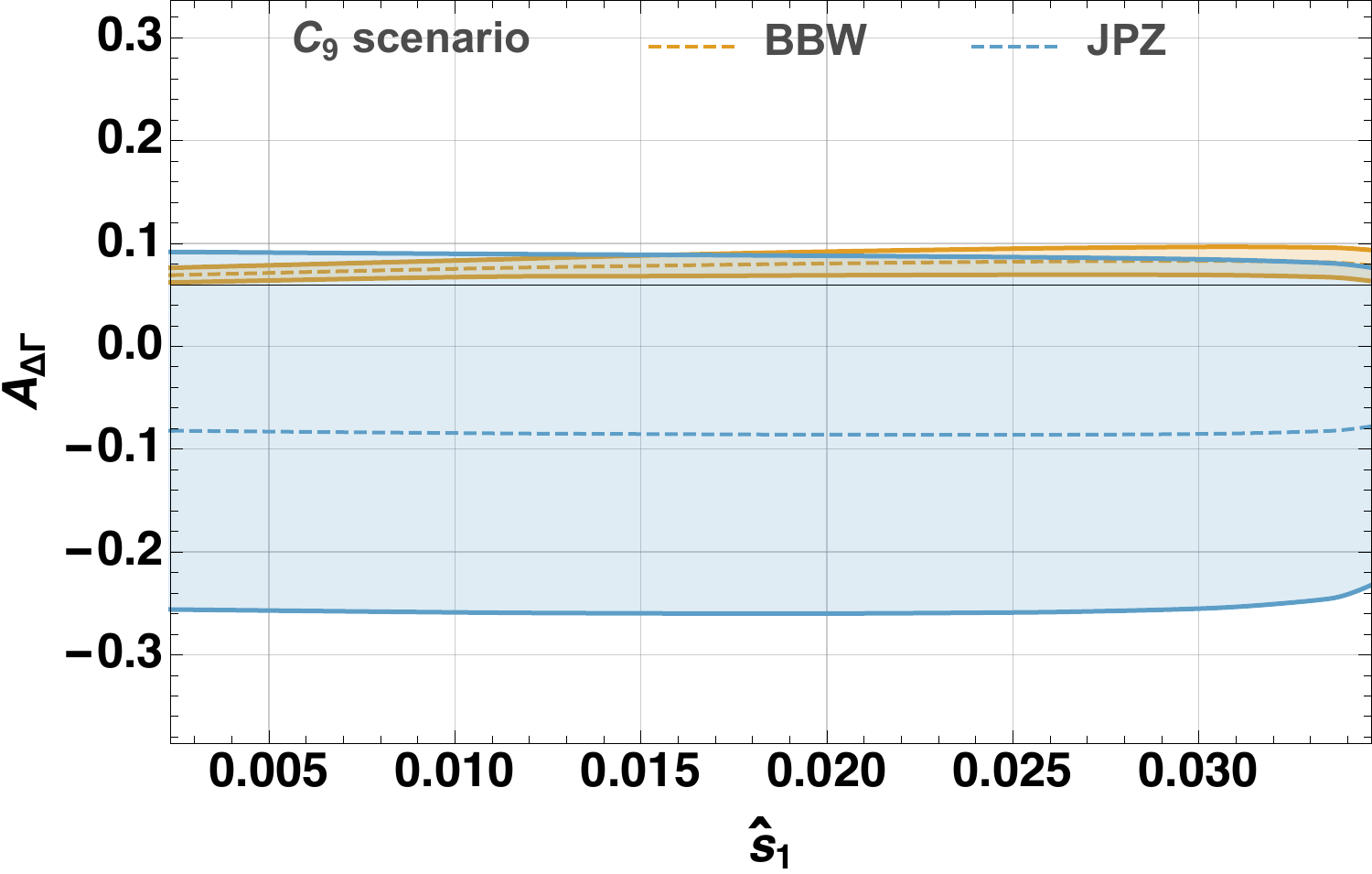}
  \hfill
  \includegraphics[width=\ww\textwidth]{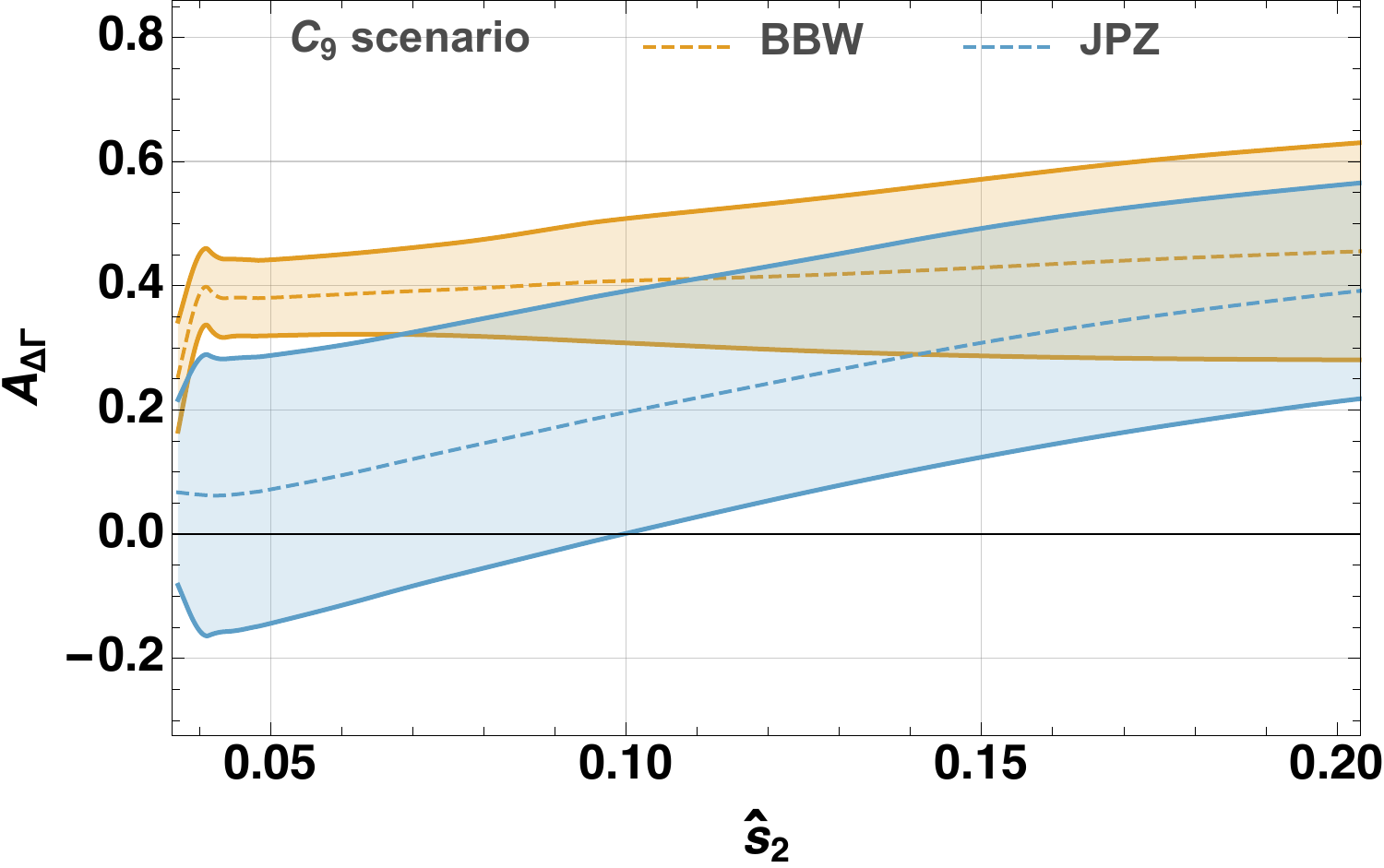}\\
  \includegraphics[width=\ww\textwidth]{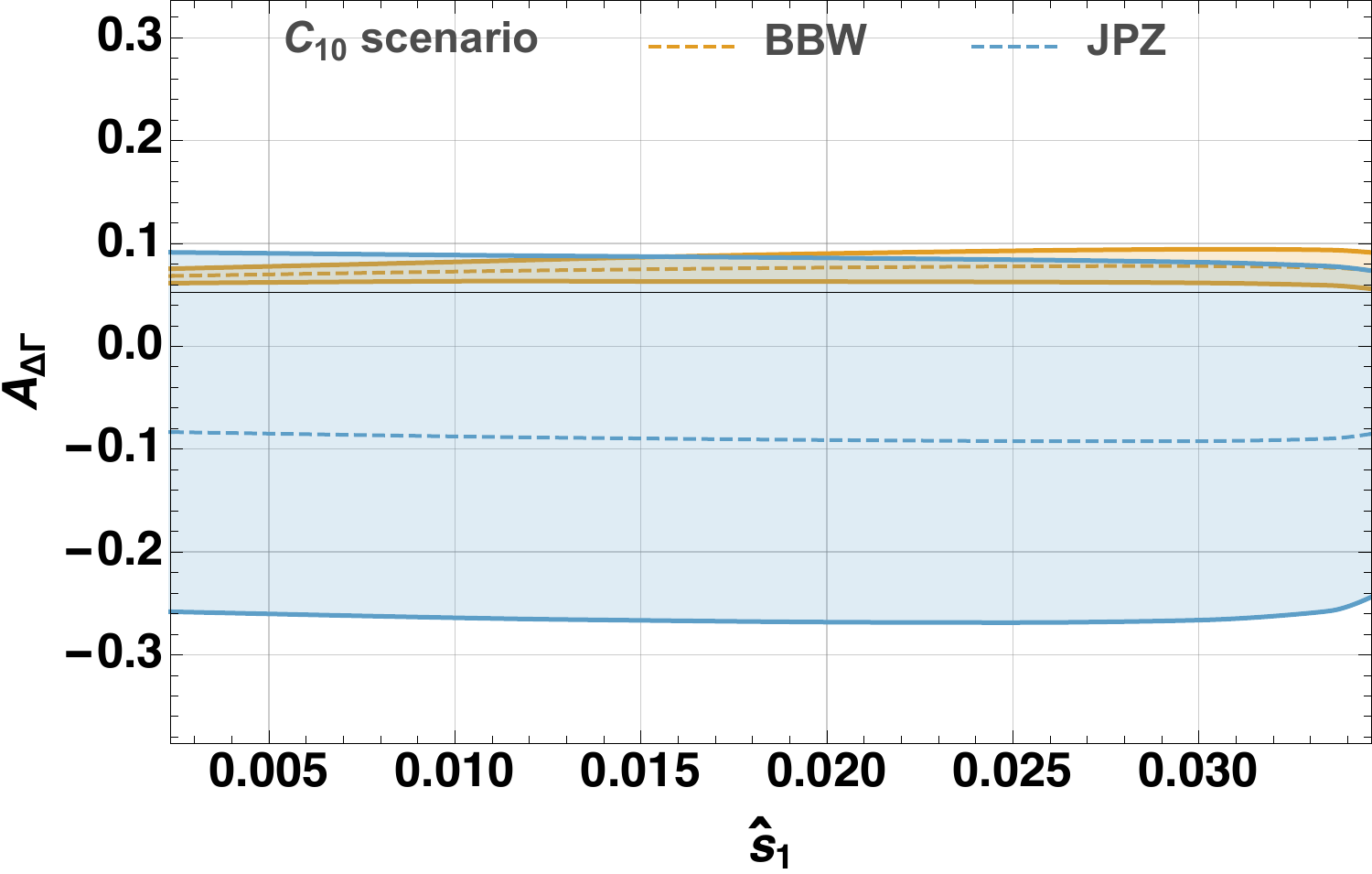}
  \hfill
  \includegraphics[width=\ww\textwidth]{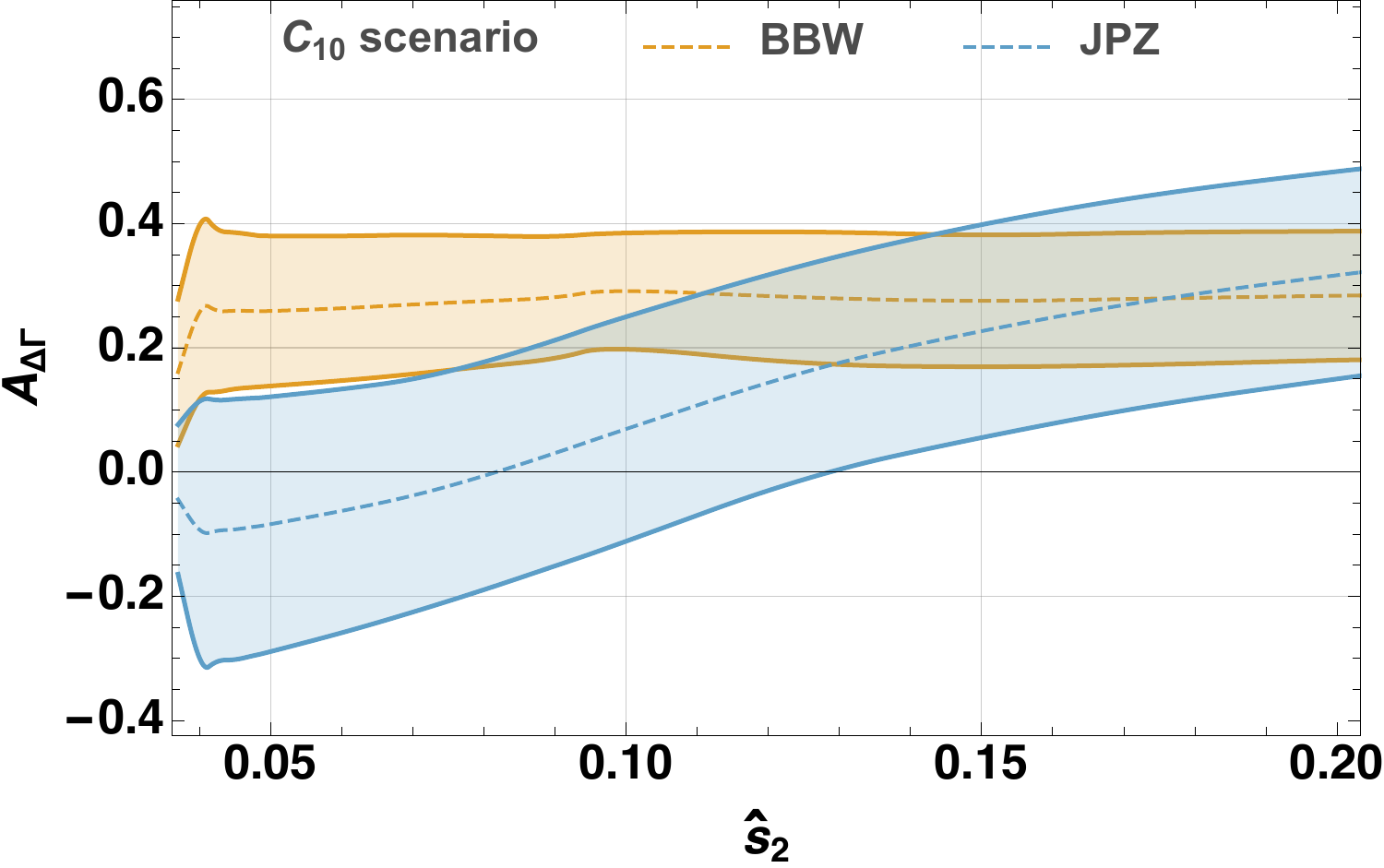}\\
  \includegraphics[width=\ww\textwidth]{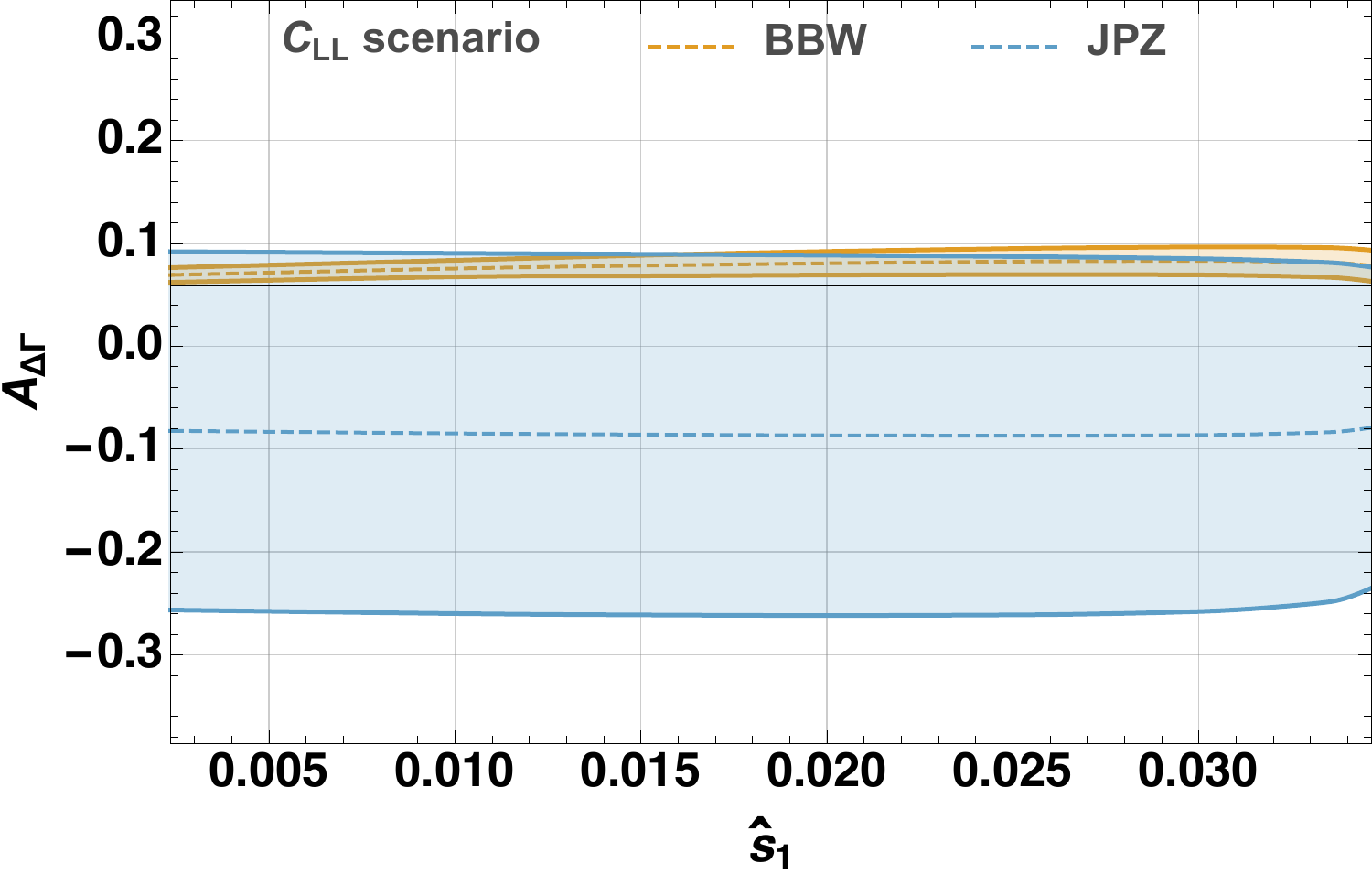}
  \hfill
  \includegraphics[width=\ww\textwidth]{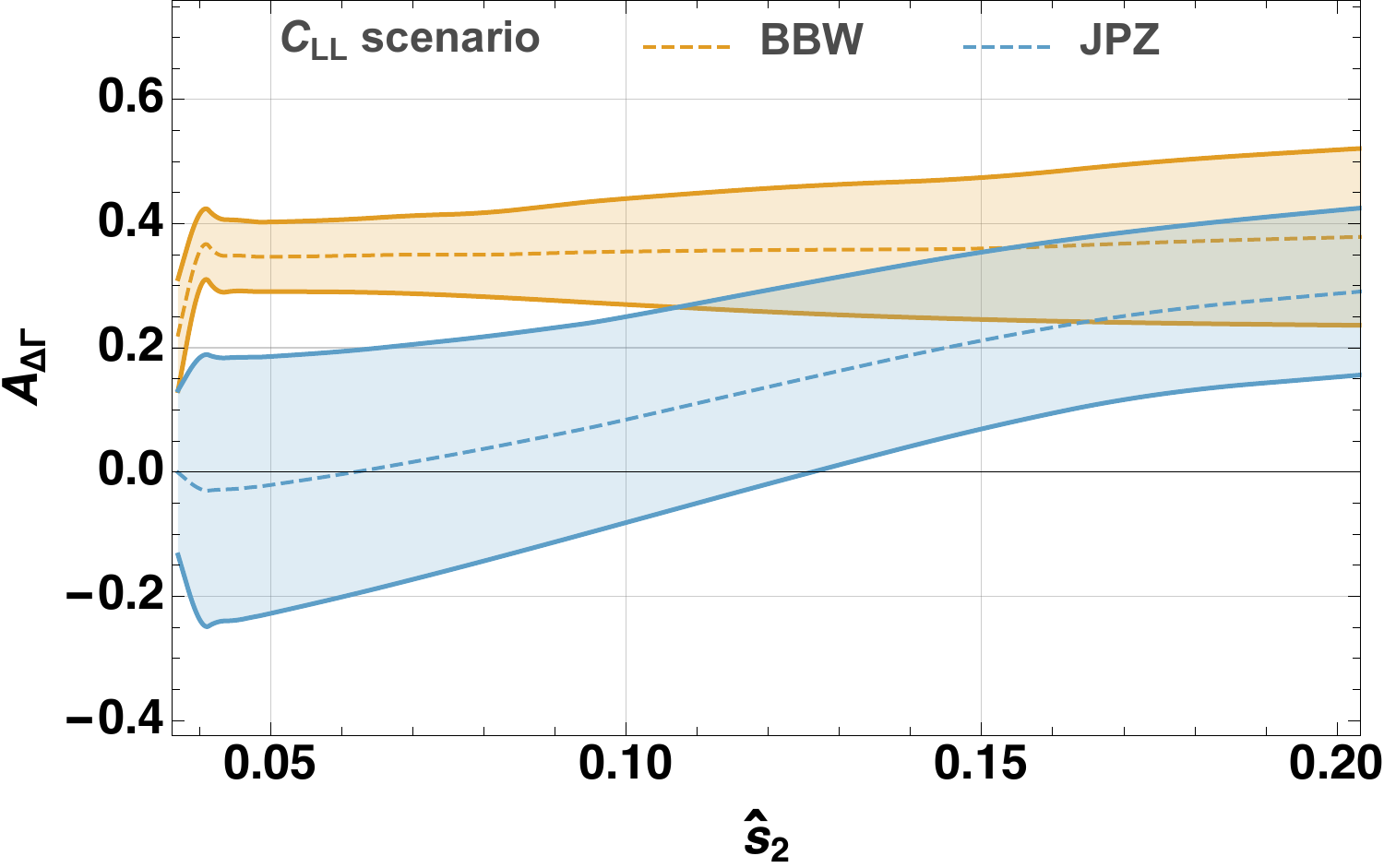}\\
  \includegraphics[width=\ww\textwidth]{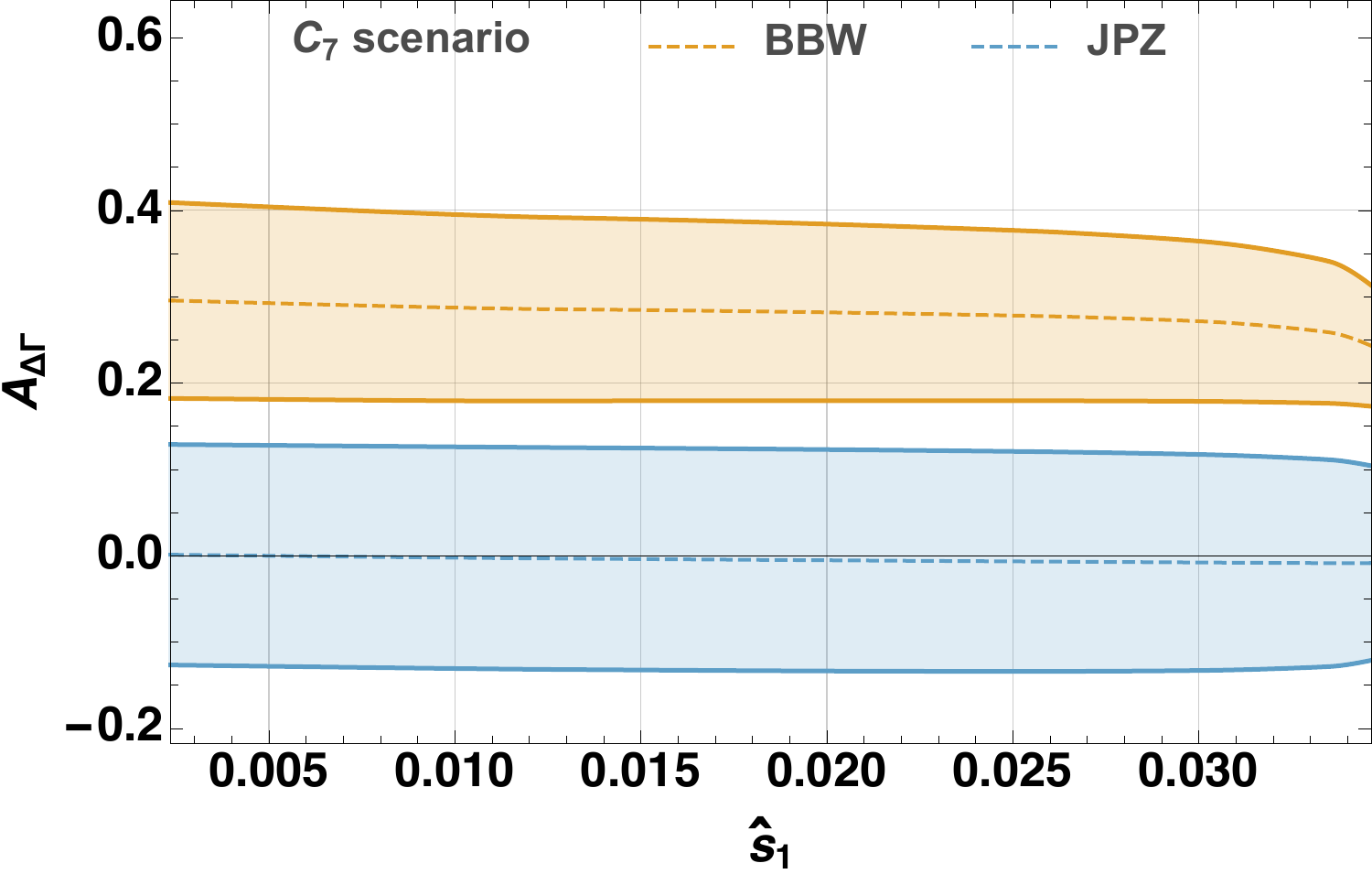}
  \hfill
  \includegraphics[width=\ww\textwidth]{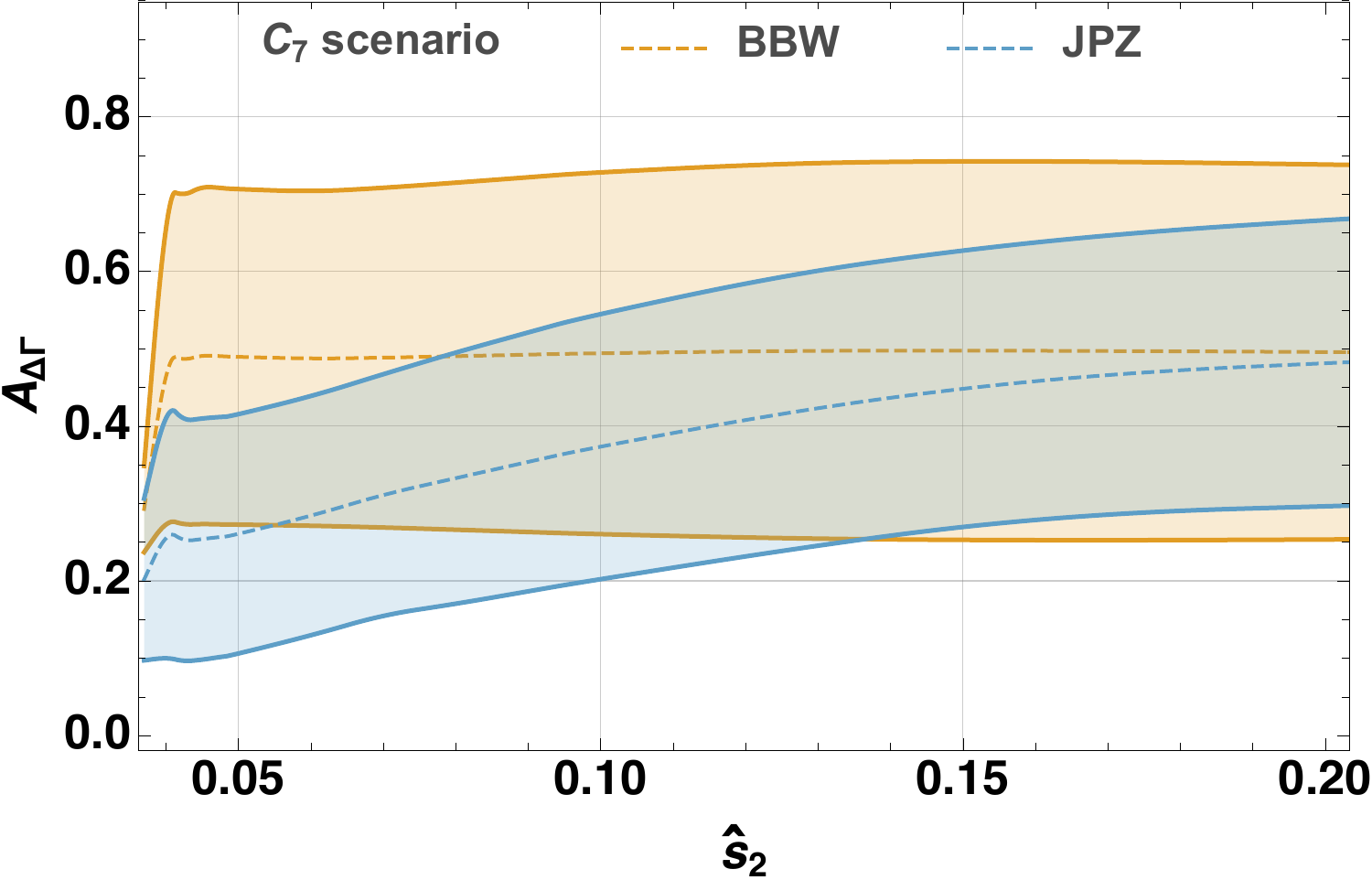}\\
  \caption{$\ADGmmy$ prediction, integrated in the interval $[4 m_{\mu}^2, s_1]$ (left panels) or $[s_2, 6~\GeV^2]$ (right panels), within the scenarios indicated. We superimpose BBW \cite{Beneke:2020fot} (yellow) and JPZ \cite{Janowski:2021yvz} (blue) f.f.'s. See text for further details.}
  \label{fig:BBW_vs_JPZ}
\end{figure}

\section{\boldmath $\ADGmmy$ at high $q^2$} \label{sec:high_q2}

In this section we discuss the high-$q^2$ region, the one most sensitive to $\mc O_{9,10}$, and the region where $\Bsmmy$ may be accessed in the short/medium term with the method in Ref. \cite{Dettori:2016zff}. Throughout the numerics in this section, we use JPZ f.f.'s \cite{Janowski:2021yvz} only, as BBW f.f.'s are deemed valid only for $q^2$ values below the narrow-charmonium threshold \cite{Beneke:2020fot}.

\subsection{\boldmath NP sensitivity at high $q^2$}

We would like to first display the sensitivity of $\ADGmmy$ to the new-physics scenarios summarised in (\ref{eq:NPbenchmarks}). To this end, we fix for definiteness the $\sh$ integration range as
\be
\label{eq:s_range}
\hat s \in [\hat s_{\rm min}, \hat s_{\rm max} ] = \left[0.59 , 1\right]\,.
\ee
The lower bound, corresponding to $\sqrt s \simeq 4.1$ GeV, has been chosen to minimise the possible contamination by broad-charmonium resonances, while maximising statistics. This value will be justified better in Sec. \ref{sec:broad-c}. Conversely, we fix the upper bound to be the endpoint $\sh = 1$. More generally, the upper bound may be expressed as $\sh_{\max} = 1 - 2 E_{\cut} / \MB$, where $E_{\cut}$ may be interpreted as the smallest photon energy detected by the experimental apparatus. This quantity is actually inferred from the apparatus' resolution in $s$, and is of the order of 50~MeV.\footnote{This figure may be estimated by noting that the experimental resolution in the muon momenta gives the $B_s \to \mu^+ \mu^-$ peak an approximately Gaussian shape, the width being for example of about 25~MeV for the LHCb experiment and ranges from 32 to 75~MeV for the CMS experiment~\cite{CMS:2014xfa}.} Concretely, the region of the $\Bsmmy$ spectrum very close to the endpoint (i.e. with $2 E_{\cut} / \MB \ll 1$) is completely dominated by bremsstrahlung (a.k.a. final-state radiation) contributions \cite{Buras:2012ru,Isidori:2007zt}, that experimentally are routinely subtracted by a MonteCarlo \cite{Davidson:2010ew}. Throughout this paper, we accordingly consider the $\Bsmmy$ spectrum with the bremsstrahlung contribution set to zero, whereas we keep interference terms, that the MonteCarlo does not subtract.

\begin{figure}[!ht]
  \centering
  \includegraphics[width=.49\textwidth]{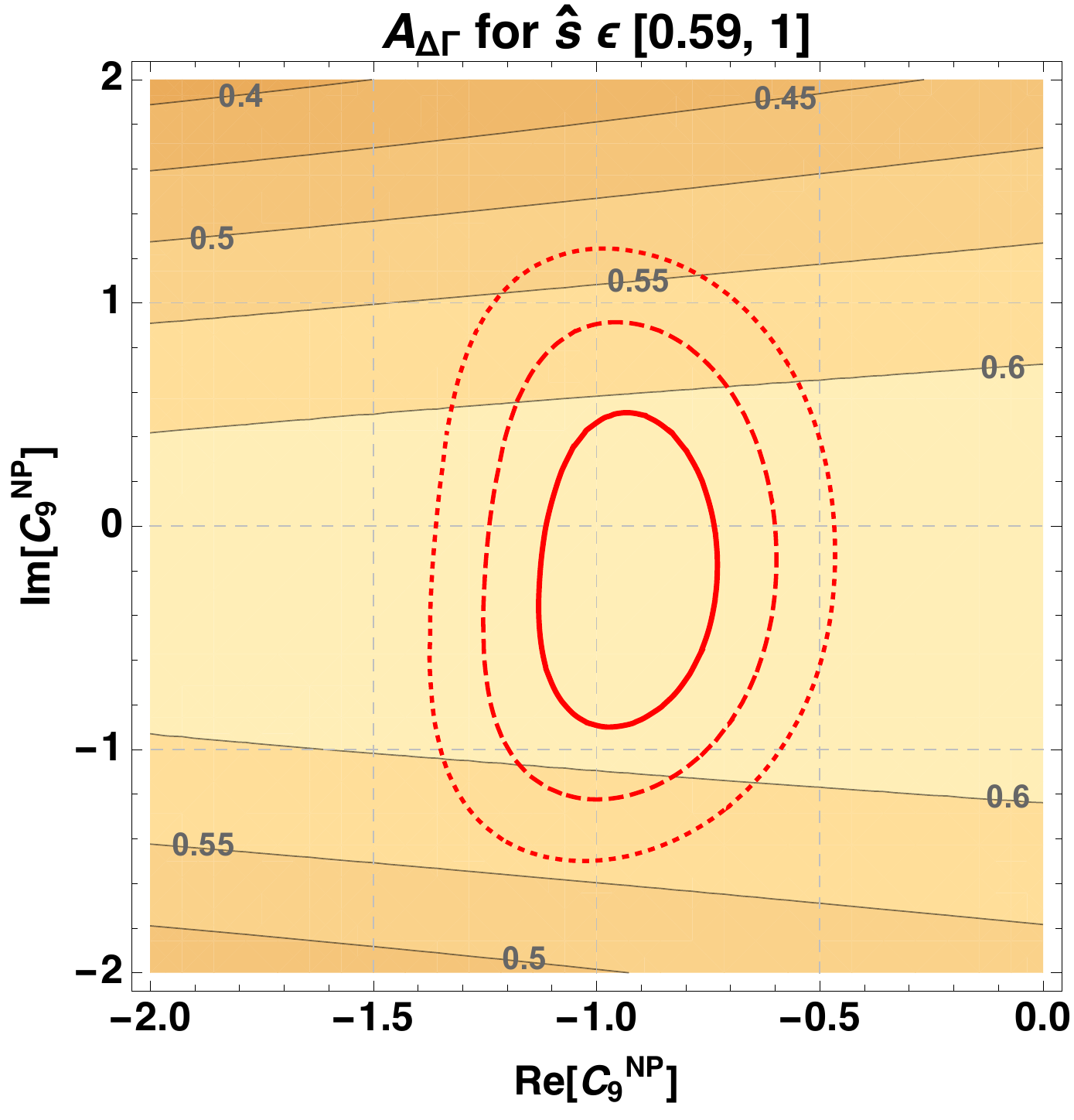}
  \hfill
  \includegraphics[width=.49\textwidth]{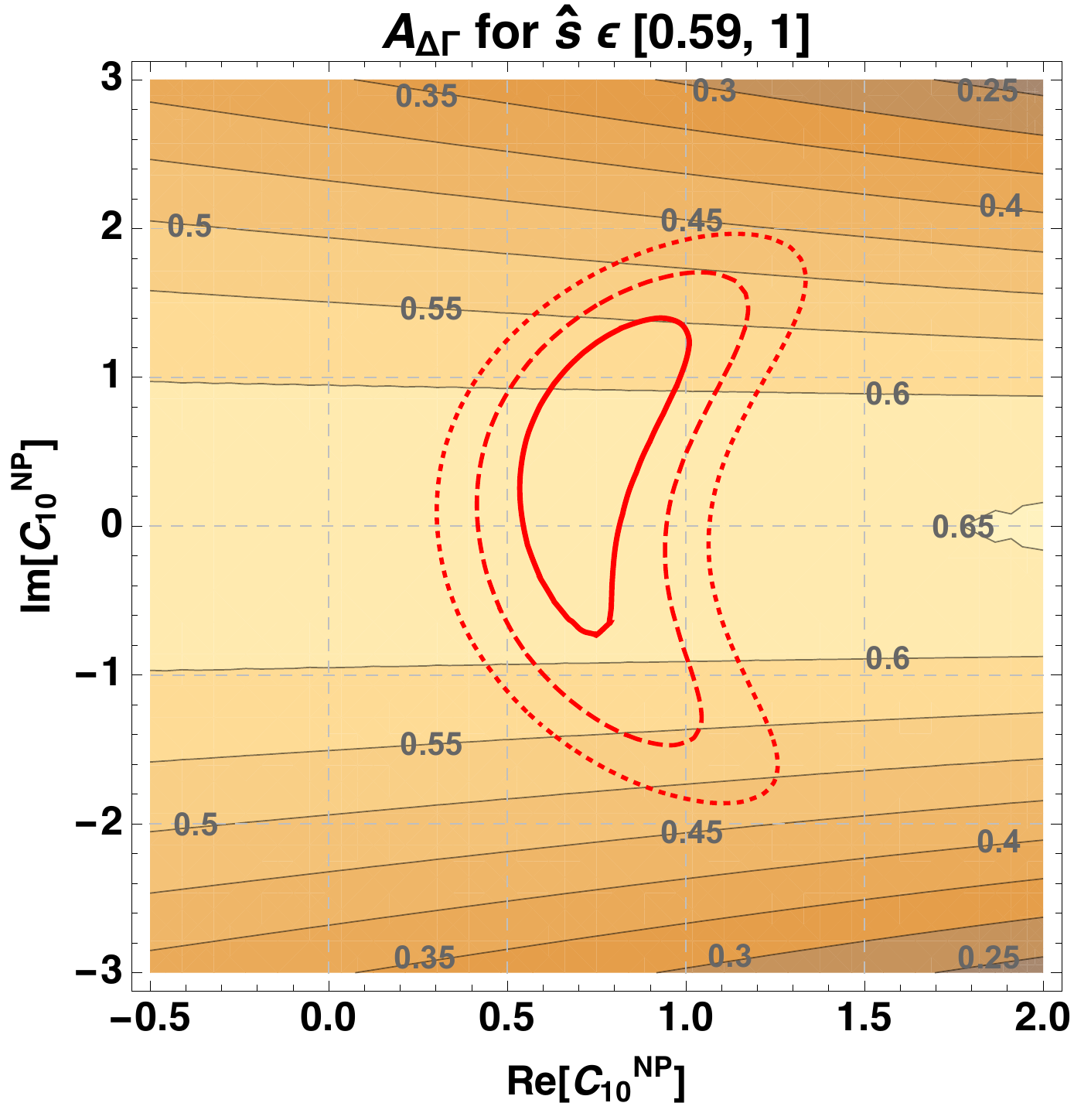}
  \hfill
  \includegraphics[width=.487\textwidth]{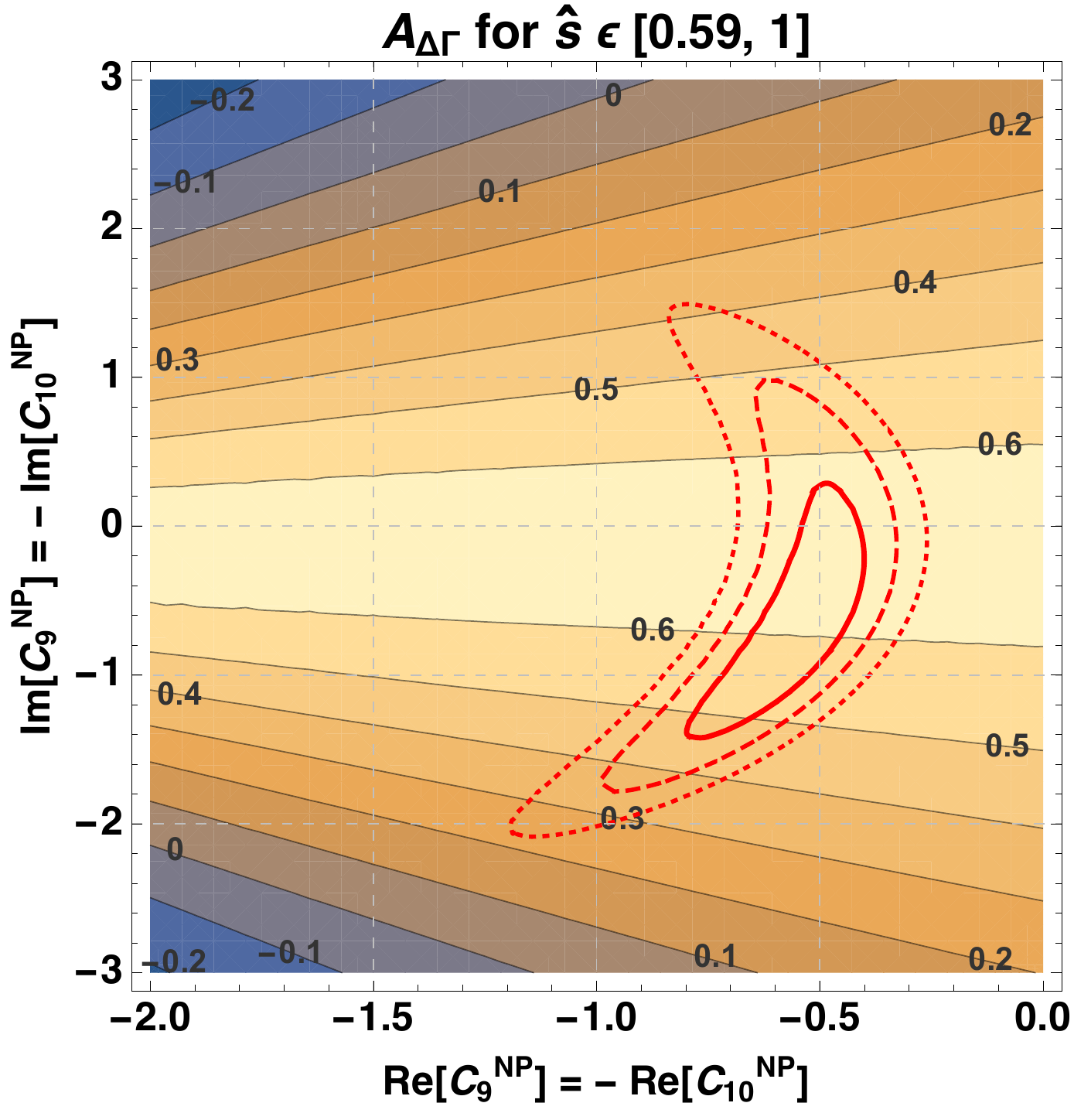}
  \hfill
  \includegraphics[width=.503\textwidth]{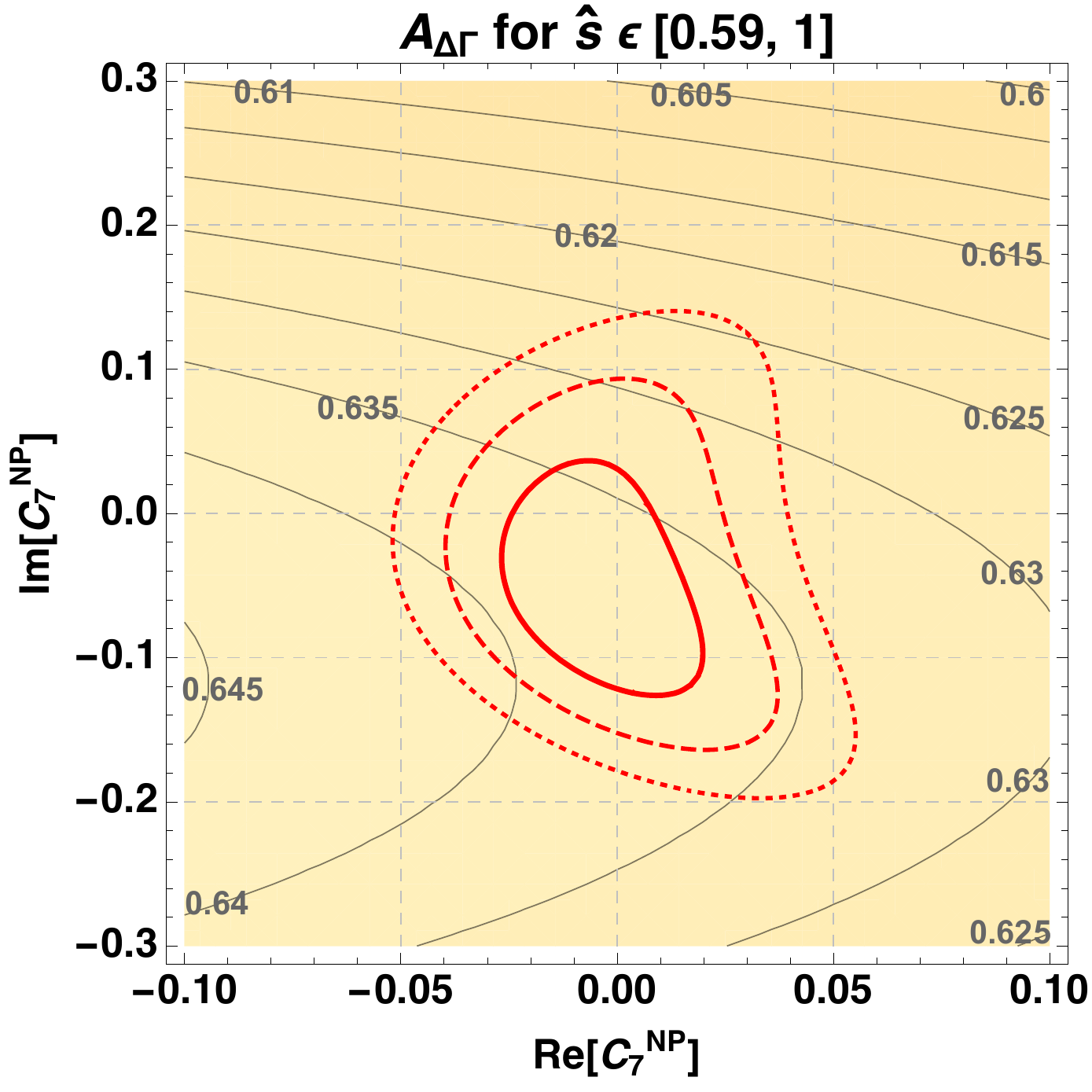}
  \hfill
  \caption{$\ADGmmy$ contours in the $\re(C_i^{\NP})$ vs. $\im(C_i^{\NP})$ plane, for $i = 7,9,10$, and for the scenario $\re$ vs. $\im$ of $C_9^{\NP} = - C_{10}^{\NP}$. The solid or dashed or dotted contours denote respectively the 1, 2 and 3$\sigma$ regions allowed by the global fit discussed in Sec. \ref{sec:parenthesis}. See text for further details.}
  \label{fig:ADG_vs_Ci_region1}
\end{figure}

With these qualifications, a first interesting question is the $\ADGmmy$ central-value prediction as a function of $\re(C_i^{\NP})$ vs. $\im(C_i^{\NP})$, with $i = 7,9,10,LL$, where we recall that the $i=LL$ case denotes $C_9^{\NP} = - C_{10}^{\NP}$. We display such dependence in the panels of fig. \ref{fig:ADG_vs_Ci_region1}. We also show as solid, dashed or dotted red contours the 1, 2 and 3$\sigma$ regions allowed by the global fit described in Sec. \ref{sec:parenthesis}.

As expected, in the chosen integration region the dependence of $\ADGmmy$ on $C_7^{\NP}$ is fairly weak, although shifts in $\im(C_7^{\NP})$ allowed by present global fits would lead to departures with respect to $({\ADGmmy})_{\rm SM}$ of O(10\%), thereby in principle detectable. Effects in the $C_9$, $C_{10}$ scenarios, and especially in the $C_{LL}$ one are more sizable, because in this region the dependence of $\ADGmmy$ on these Wilson-coefficient combinations is stronger than it is on $C_7$, similarly as for the $\Bsmmy$ differential width.

The effects just discussed have to be compared with the theory error associated to $\ADGmmy$, which we do next.

\subsection{\boldmath $\ADGmmy$ theory error at high $q^2$}

The two outstanding sources of theory error for $\ADGmmy$ in this region are the possible pollution from broad-charmonium resonances and the f.f. uncertainty. We address these two sources in turn.

\subsubsection{Impact of broad charmonium}\label{sec:broad-c}

In eq. (\ref{eq:s_range}) we fixed $\sh_{\min}$ somewhat arbitrarily. The lower the $\sh_{\min}$ value, the larger the potential contribution of broad-charmonium resonances, including in particular $\psi(2S)$, $\psi(3770)$, $\psi(4040)$, $\psi(4160)$ and $\psi(4415)$. For reference, their peaks correspond to $\hat s =\{ 0.47, 0.49, 0.57, 0.61, 0.68 \}$. Inclusion of these resonances within the $\hat s$ integration range may represent an important source of theory systematics \cite{Lyon:2014hpa} (see \cite{Aaij:2016cbx} for a dedicated LHCb study). Here we would like to address the question how this systematics impacts the $\ADGmmy$ prediction.

Following a standard approach to account for these effects, we note that they enter our decay of interest via the subprocess $B^0_s \to V (\to \ell \ell) \gamma$, where $V$ denotes any of the aforementioned resonances. The associated long-distance contributions may be modelled as a sum over Breit-Wigner (BW) poles \cite{Kruger:1996cv}, i.e.
\be
\label{eq:Vcc_shift}
C_9 ~\to~ C_9 ~-~ \frac{9 \pi}{\alpha^2} \, \bar C \,  \sum_V 
 |\eta_V| e^{i \delta_V} \frac{\hat{m}_V \, \mc B(V \to \mu^+ \mu^-) \, \hat{\Gamma}_{\textrm{tot}}^V}{\hat{q}^2 - \hat{m}_V^2 + i \hat{m}_V \hat{\Gamma}^V_{\textrm{tot}}}~.
\ee
In this relation, $\bar C = C_1 + C_{2}/3 + C_3 + C_{4}/3 + C_5 + C_{6}/3$. In order to address the theory uncertainty due to this modelling, we subsequently scan simultaneously over the BW normalisation factors $|\eta_V| \in [1, 3]$ as well as on the phases $\delta_V \in [0, 2\pi)$ with independent uniform distributions.\footnote{On the other hand, in the discussion of any $\ADGmmy$ error component {\em other} than broad-charmonium pollution, we will assume $|\eta_V| = 1$ and take, for the rest of the parameters, the numerical values in table \ref{tab:input}.} Such procedure is expected to provide a conservative way to measure the deviation from naive factorisation ($|\eta_V| = 1$ and $\delta_V = 0$).\footnote{As a matter of fact, it was found that $|\eta_V| \simeq 2.5$ and $\delta_V \simeq \pi$ gives a good description of $B \to K \mu^+ \mu^-$ data \cite{Lyon:2014hpa,Aaij:2013pta}.}
We perform the above scan for $\sh_{\min} \in [0.50, 0.70]$, and for all the NP scenarios in eq. (\ref{eq:NPbenchmarks}), as well as for the SM case.
The result is shown in fig. \ref{fig:ADG_cc_ff} as yellow bands.
For each given value of $\sh_{\min}$ the central value and the values in the upper and lower bands are calculated as, respectively, the mean and $\pm 1\sigma$ of the distribution obtained through the corresponding scan.

The panels in fig.~\ref{fig:ADG_cc_ff} show that the broad-charmonium modelling uncertainty due to eq.~(\ref{eq:Vcc_shift}) (yellow `bands') has no appreciable impact on the prediction of $\ADGmmy$, compared with the uncertainty (blue bands) induced by f.f.'s, to be discussed in Sec. \ref{sec:ff_high_q2}. Since the shift induced by eq. (\ref{eq:Vcc_shift}) is typically of O(5\%), and its phase is completely misaligned with the $C_{9 {\rm SM}}$ phase, it is not obvious why most of this uncertainty cancels out in $\ADGmmy$. On closer inspection, one can understand this cancellation as the result of specific features, in particular the complete dominance, in this kinematic region, of quadratic contributions in $C_9$ and $C_{10}$;
the fact that the multiplying form factors $F_V$ and $F_A$ are real; the fact that the broad-charmonium shift in eq. (\ref{eq:Vcc_shift}) can be treated as a `small' modification of the numerically large SM value for $C_9$. These features ease cancellations between the numerator and the denominator of $\ADGmmy$---suggesting that, with respect to this class of long-distance contributions, $\ADGmmy$ behaves well like a ratio observable. We present a more detailed analytic argument in Appendix \ref{app:why_cc_small}.

Because of the above conclusions, the $C_{9}$ and $C_{10}$ dominance in this region may be regarded as one key advantage with respect to the low-$q^2$ region, where the $C_7$ contribution becomes important, without being dominant in the full kinematic range. It is quite encouraging that, as a result, $\ADGmmy$ is an observable largely immune to broad-charmonium uncertainties, given that they escape a rigorous description. The control of the theory prediction rests thus entirely in the f.f. error, to which we turn next.

\subsubsection{Impact of form-factor modelling}\label{sec:ff_high_q2}

\begin{figure}[h!]
  \centering
  \includegraphics[width=.55\textwidth]{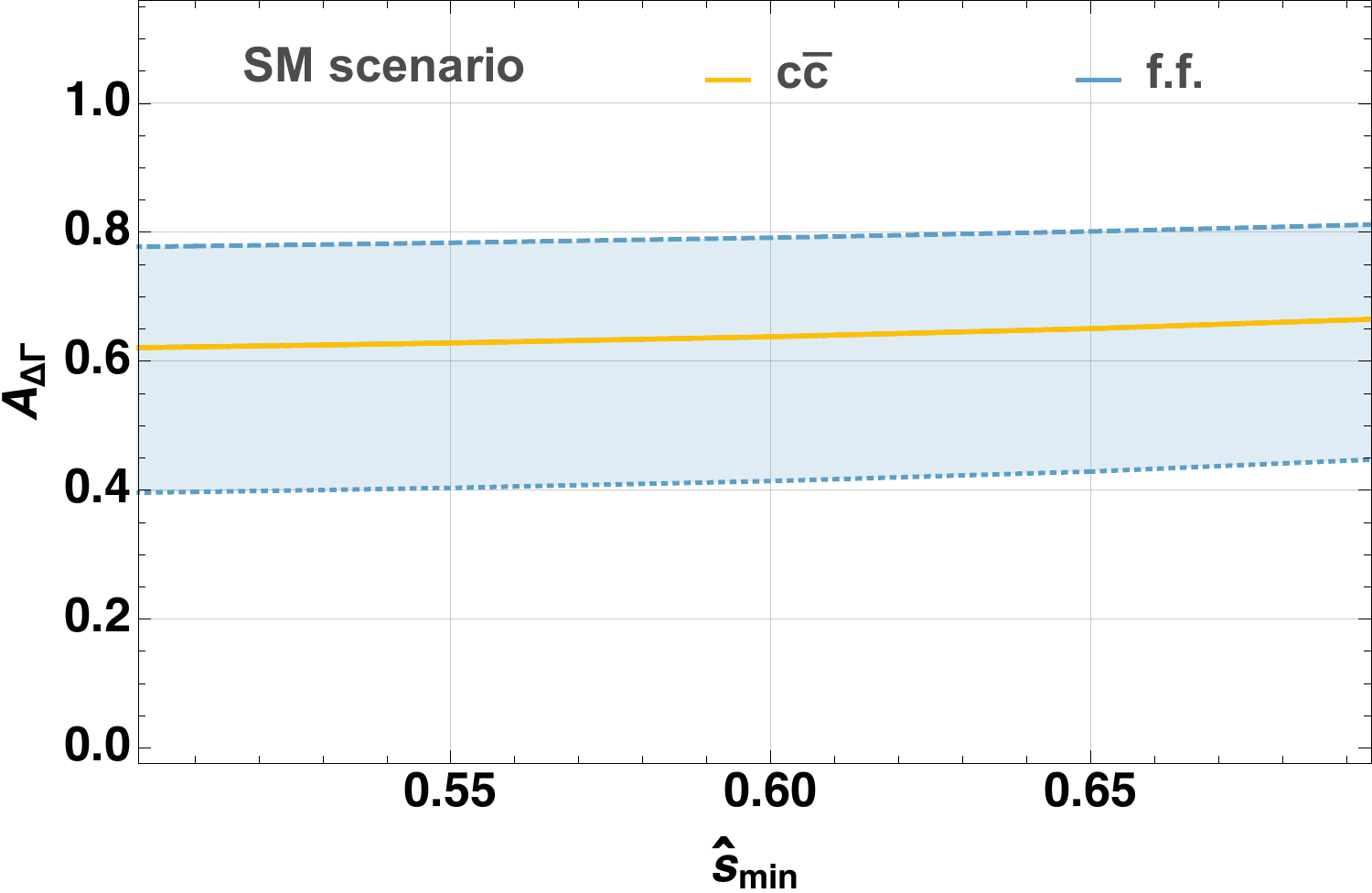}  
  \hfill
  \includegraphics[width=.485\textwidth]{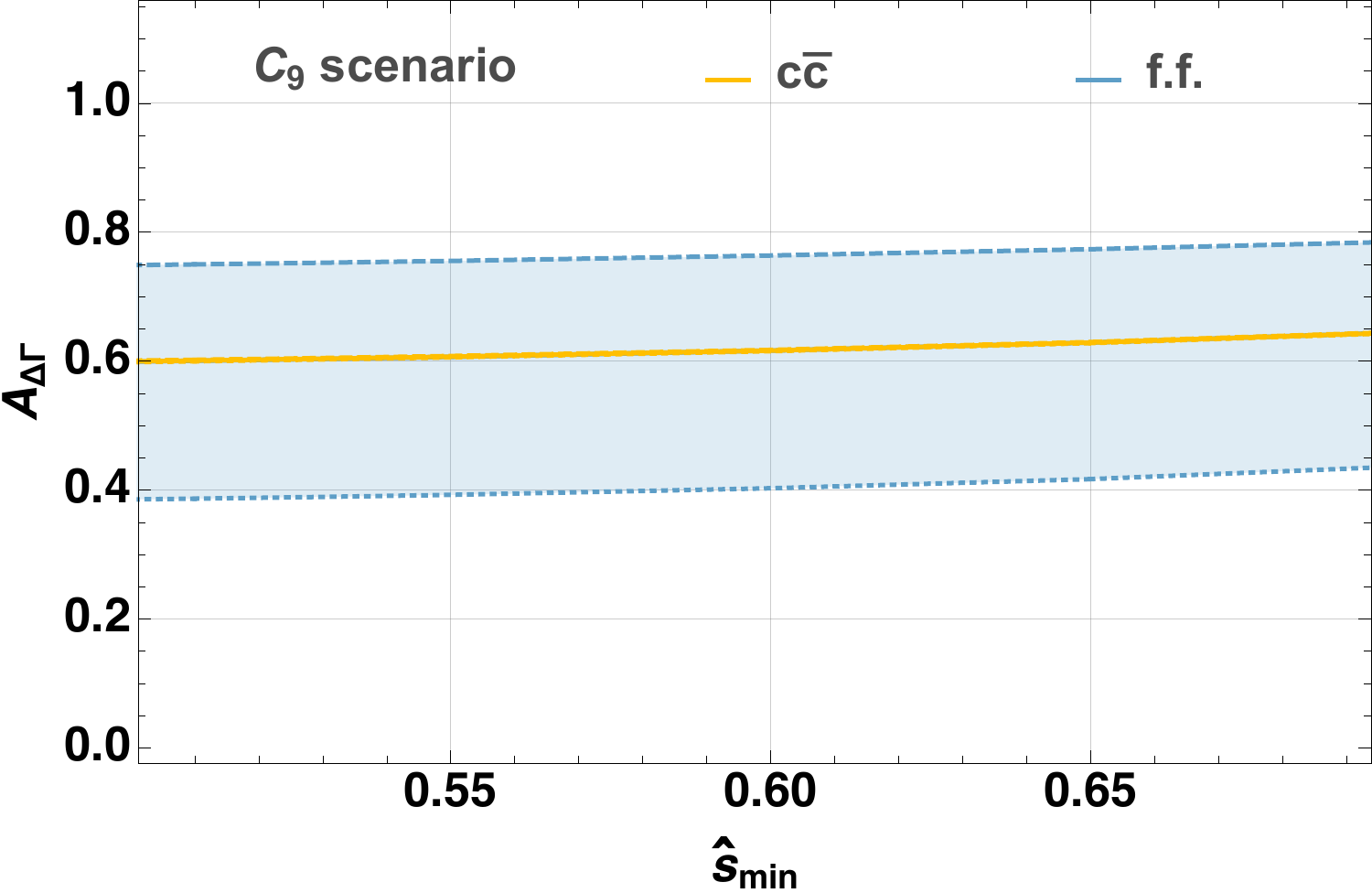}
  \hfill
  \includegraphics[width=.485\textwidth]{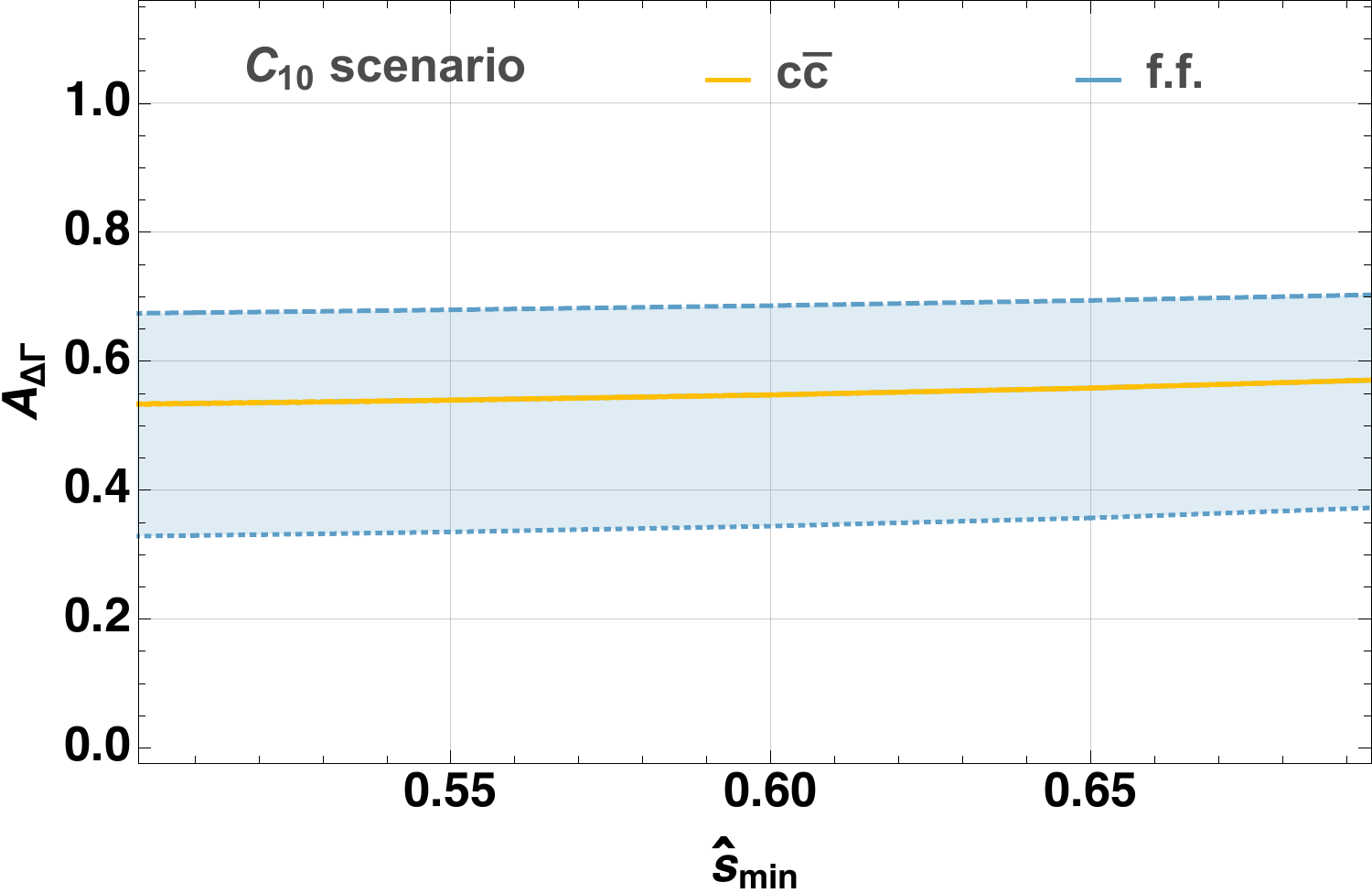}
  \hfill
  \includegraphics[width=.485\textwidth]{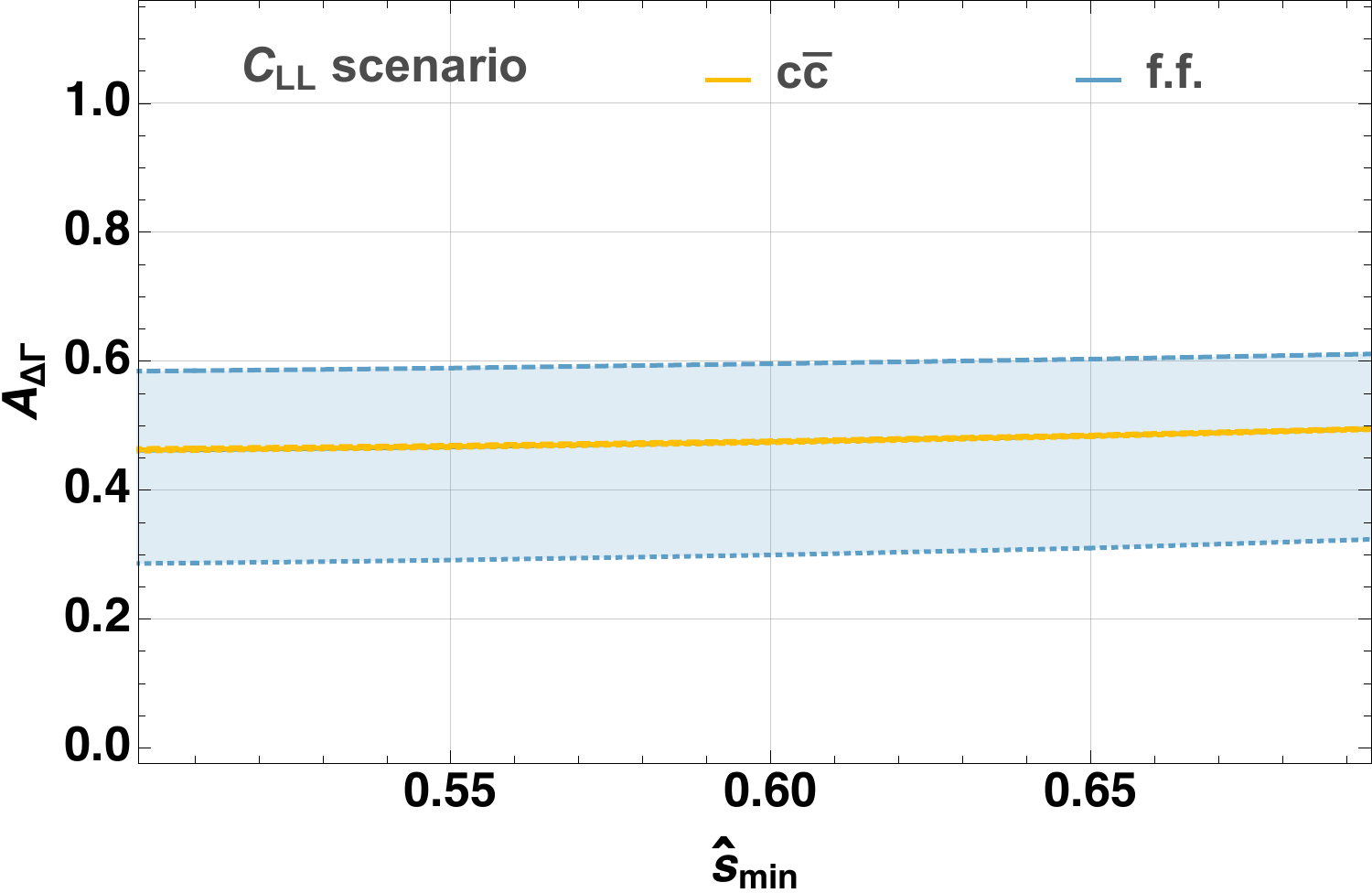}
  \hfill
  \includegraphics[width=.485\textwidth]{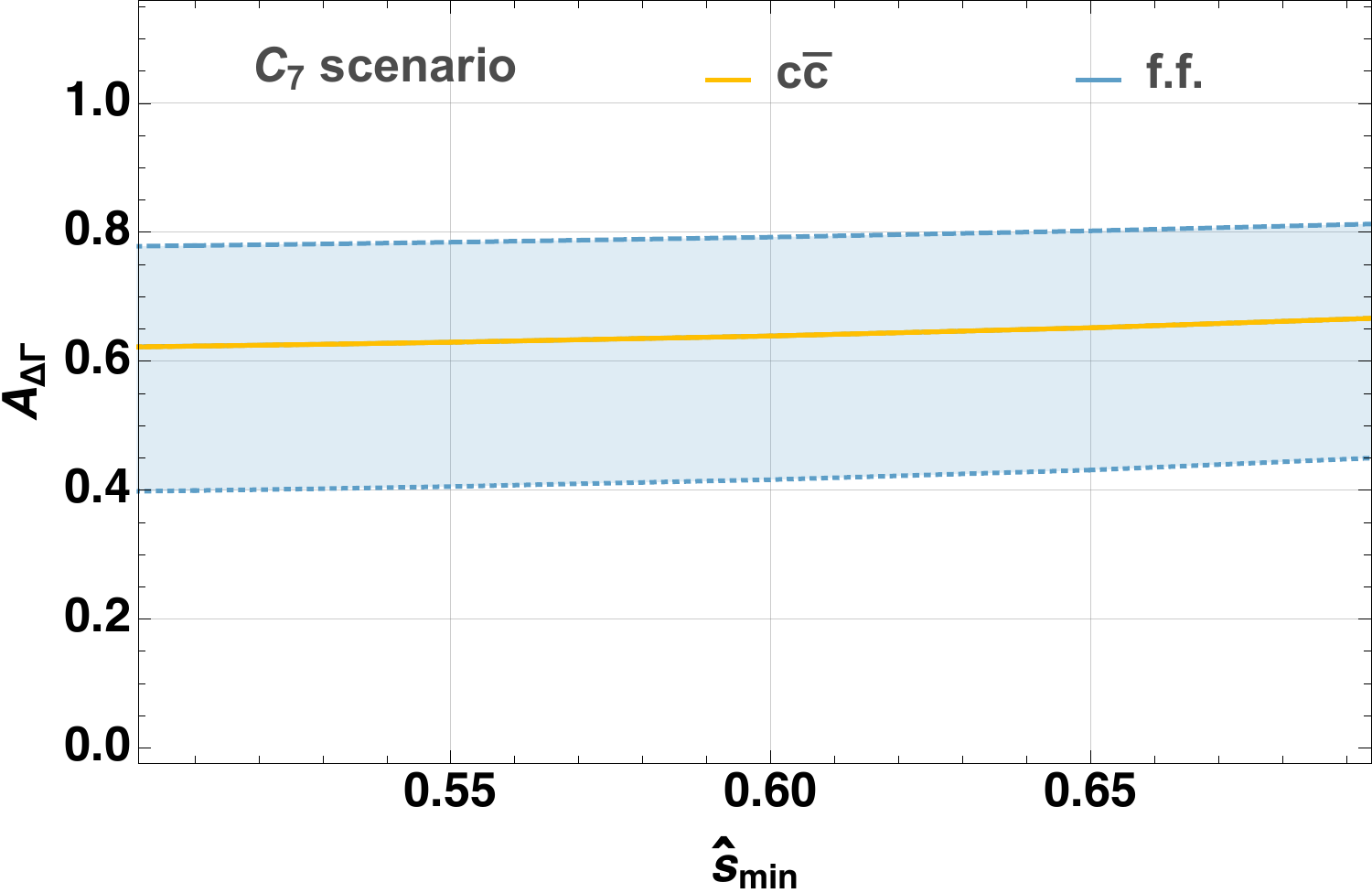}
  \caption{$\ADGmmy$ prediction in the kinematic interval $[\sh_{\min}, 1]$. The blue vs. yellow bands refer to the f.f. error, and respectively on the uncertainty associated to the modelling of broad-charmonium resonances. See text for details.}
  \label{fig:ADG_cc_ff}
\end{figure}

A further source of uncertainty that needs be addressed is the error attached to the f.f.'s $F_{V,A,TV,TA}$ that parameterise the $\bBs \to \gamma$ amplitudes, see eq. (\ref{eq:FVA_FTVTA}). As discussed in Sec. \ref{sec:low_q2}, the JPZ parameterization includes an estimate of the error on each of the $F_{V,A,TV,TA}$ f.f.'s \cite{Janowski:2021yvz}.\footnote{For recent progress on radiative leptonic decays in lattice QCD in this kinematic region, see \cite{Kane:2019jtj}.} We accordingly vary these f.f.'s within uncorrelated normal distributions around their respective errors. The impact of these errors on $\ADGmmy$ is shown in fig. \ref{fig:ADG_cc_ff}, as blue bands.
In calculating these bands, the narrow-charmonium contributions are modelled with eq. (\ref{eq:Vcc_shift}), with $|\eta_V| = 1$ and the $\delta_V$ phases as in table \ref{tab:input}, although we verified that a simultaneous variation of broad-charmonium and form-factor parameters leads to tiny differences.

Fig. \ref{fig:ADG_cc_ff} allows to visually compare the impact of the two sources of uncertainty, broad charmonium vs. f.f.'s, for each theory scenario.

We see that, in the whole $\sh_{\min}$ range considered and for any theory scenario, the f.f. uncertainty is always much larger than the uncertainty induced by broad-charmonium modelling.
The basic difference between the two sources of error may be appreciated as follows. On the one hand, the broad-charmonium correction enters `only' as a shift to $C_9$, hence efficient cancellations can take place, as discussed in detail in Appendix \ref{app:why_cc_small}. On the other hand, f.f.'s enter in different ways, all numerically relevant, for the different Wilson-coefficients combinations, as already mentioned at the end of Sec. \ref{sec:ADG_calculation}. As a consequence, cancellations of f.f. effects are much less efficient. The $\ADGmmy$ error induced by f.f.'s is still too important to allow to tell apart the chosen theory scenarios from each other. However, it is noteworthy that, in this kinematic region, the dominance of, jointly, $C_9$ and $C_{10}$ gives rise to a particularly high sensitivity to the $C_{LL}$ scenario, which could be resolvable with an improvement on f.f.'s uncertainties.

\section{Experimental Outlook and Conclusions} \label{sec:conclusions}

In this paper, we have discussed the theoretical interest of pursuing effective-lifetime measurements for $\Bsmumugamma$. With the method of Ref. \cite{Dettori:2016zff} these measurements, in the high-$q^2$ part of the spectrum, can be performed on the very same event sample as for $\Bsmumu$, i.e. di-muon events with no detected photon. The case of $\Bsmumugamma$ is actually one of many possible examples. In fact, the underlying idea---the idea that if one can estimate all other $\Bsmumu$ backgrounds as $q^2$ departs downwards from the $\MB^2$ peak, one may extract $\Bsmumugamma$ `parasitically' as {\em one particular} background---is in principle applicable to any event of the kind $\Bs \to \mumu X$. Such decay allows to reconstruct a primary and a secondary vertex (the latter through the di-muon), and thereby to perform time-dependent measurements, in turn usable to reconstruct the corresponding effective lifetime.\footnote{If one sums over $X$, one may even think of an `inclusive' effective lifetime. However, the definition of the corresponding theoretical observable would require accurate knowledge of the yield for each $X$. Besides, the constraining power of such observable would be diluted by the proliferation of channels, implying a proliferation of couplings.}

The specific task of measuring the effective lifetime for $\Bsmumugamma$ introduces a number of daunting challenges---although not insurmountable. 
The measurement of the effective lifetime of a \Bs decay requires a clean signal sample, a task that for \bsmumugamma decays is non-trivial, and a well measured secondary vertex, which is typically the case for displaced di-muons at LHC experiments. 
At present, two possible, different approaches exist to reconstruct these decays: 
a full reconstruction, more suited for the low-$q^2$ part of the spectrum, where the photon energy is larger and, as such, easier to reconstruct; 
and a method relying on partial reconstruction~\cite{Dettori:2016zff}, mostly relevant for the high-$q^2$ range, where the photon energy is small, making photon reconstruction inefficient. The two methods present different challenges as concerns possible backgrounds.

In the high-\qsquare region, the partial-reconstruction method allows high efficiency, comparable to the \bsmumu decay for which, as mentioned, a first measurement 
of the effective lifetime has been performed by the LHCb~\cite{Aaij:2017vad} and CMS~\cite{Sirunyan:2019xdu} experiments, combined in Ref.~\cite{LHCb-CONF-2020-002}.\footnote{The LHCb measurement has been recently updated in Refs.~\cite{LHCb:2021vsc,LHCb:2021awg} but the combination has not been updated yet.}
Assuming this method, the closer the di-muon mass 
to the \Bs mass the more similar the experimental performances in \bsmumugamma will be to the \bsmumu case. 
Alas, the backgrounds are considerably different, due to the absence of a clear narrow peak on which 
to perform a statistical background subtraction from the invariant mass sidebands. 
In particular the semileptonic $B\to h \mu \nu$, with the hadron $h = \pi, K$ misidentified as a muon, constitute a large background, 
together with the rarer $B\to h \mu \mu$ decays, where the additional hadron is not reconstructed. 
The latter in particular will constitute almost an irreducible background, although one possible handle to tame it may be the shape of the di-muon mass distribution. 
A measurement of the \bsmumugamma effective lifetime thus requires both a precise knowledge of the yield and of the di-muon mass distribution of the listed backgrounds. This is not unrealistic, for example a large sample of \bskmunu decays has been recently seen at LHCb~\cite{LHCb-PAPER-2020-038} and can be used to constrain its equivalent misidentified yield.
The lifetime of this decay, being a flavour-specific decay, is precisely known. 
Similar backgrounds from $B^0$ and $B^+$ decays will also have precisely known lifetimes. 
In this partially reconstructed method, a small uncertainty can arise from the absence of the photon momentum in the full $B$ reconstructed momentum used to calculate the boost (and hence the lifetime from the decay distance): the higher the $q^2$ considered the smaller is this uncertainty. This would add a small component to the experimental uncertainty on the lifetime on an event-by-event basis, which will average out with statistics. In addition, kinematic fits exploiting the known flight direction of the $B$ can further improve this resolution and relative measurements can cancel out these effects, and indeed partially reconstructed semileptonic decays are used for precision lifetime measurements~\cite{LHCb:2017knt}.

Conversely, at low di-muon mass, in order to distinguish \bsmumugamma signal decays from background it will be crucial to efficiently reconstruct the photon. 
While this reduces the overall signal efficiency, it allows to statistically separate it from background through the invariant mass peak at the \Bs mass. 
In this region, performances can be benchmarked against the $B^0_s \to \phi \gamma$ decay, that can also be reconstructed in the muonic final state, and that is well studied in its kaonic final state. 

Clearly, a quantitative study is needed to translate these considerations into a luminosity vs. $\ADGmmy$-accuracy plot.
In particular, in either of the above two cases detailed studies of the backgrounds are required to assess the performance on this new observable. Such survey requires a full experimental simulation and is thus beyond the scope of this paper. But it is clear that the measurement of this observable will require high statistics and is thus feasible in the upgrade phase of the LHC experiments.

In synthesis, the present paper introduces the $\Bsmumugamma$ effective lifetime.
This observable offers the possibility of probing all Wilson coefficients that are currently of interest in view of the $B$ anomalies, and, in particular, of novel CP-violating effects. Interestingly, scenarios with new weak phases not aligned with the corresponding ones in the SM contributions are under-constrained. Besides, the relevant CP-sensitive quantity---$\ADGmmy$---encoded in the effective lifetime is per definition a ratio-of-amplitudes (squared) observable. One can therefore expect a partial cancellation of the hadronic uncertainties inherent in the $\Bsmumugamma$ amplitude. We find that this cancellation is surprisingly efficient at high $q^2$ for broad-charmonium-modeling uncertainties, whereas errors induced by form factors are still sizeable and represent the main limitation at present. However, the high-$q^2$ region is especially favorable for non-perturbative, first-principle approaches to the calculation of the concerned form factors, e.g. within lattice QCD. Also interestingly, this is the same region where $\Bsmumugamma$ is already being accessed from the $\Bsmumu$ dataset.

\section*{Acknowledgments}

DG owes special thanks to Christoph Bobeth for many insights on Ref. \cite{Beneke:2020fot} as well as to Javi Virto in connection with Refs. \cite{Descotes-Genon:2015hea,Asatrian:2019kbk}; he also acknowledge Dmitri Melikhov for discussions, Francesco Polci for useful comments about Ref. \cite{Aaij:2021vac}, and Roman Zwicky for exchanges about Refs. \cite{Pullin:2021ebn,Janowski:2021yvz}. FD would like to thank Flavio Archilli, Matteo Rama and the whole LHCb \bsmumu group for useful discussions. We acknowledge Peter Stangl for help on various aspects of {\tt flavio}. This work is supported by the ANR under contract n. 202650 (PRC `GammaRare').

\appendix

\section{\boldmath Amplitudes and form factors: basic notation} \label{app:ffs}

The dynamics of the $\bBs \to \mmy$ amplitude may be parameterised by the following $b \to s \ell \ell$ effective Hamiltonian \cite{Buchalla:1995vs,Buras:1994dj,Misiak:1992bc} 
\be
\label{eq:Heff}
\mathcal H_{\mathrm{eff}} = \frac{4 G_F}{\sqrt 2}\left( \sum_{i=1}^2 
(\lambda_u C_i \mathcal{O}_i^u + \lambda_c C_i \mathcal  \mathcal{O}_i^c )       -\lambda_t \sum_{i=3}^{6} C_i \mathcal{O}_i   - \lambda_t \sum_{i=7}^{10} (  C_i  \mathcal {O}_i + C_i'  \mathcal{O}_i')  
 \right)~, 
\ee
where $\lambda_i \equiv V_{is}^*V_{ib}$ and $V$ the CKM matrix. The operators relevant to our discussion read
\begin{alignat}{4}
\label{eq:Oi}
& \mathcal O_1^q &\;=\;&  (\bar s^{\alpha} \gamma_\mu q_{L}^{\beta}) (\bar q^{\beta} \gamma^\mu b_{L}^{\alpha})~,  \qquad  \qquad 
& & \mathcal O_2^q &\;=\;&  (\bar s^{\alpha}  \gamma_\mu q_{L}^{\alpha}) (\bar q^{\beta} \gamma^\mu b_{L}^{\beta})~, \nonumber \\[0.1cm]
& \mathcal O_{7} &\;=\;& \frac{e}{16\pi^2} m_b \bar s \sigma_{\mu \nu} F^{\mu \nu} b_R~,
\qquad  \qquad 
& & \mathcal O_{8} &\;=\;& \frac{g_s}{16\pi^2} m_b\bar s \sigma_{\mu \nu} G^{\mu \nu} b_R~, 
\\[0.1cm]
& \mathcal O_{9} &\;=\;& \frac{e^2}{16\pi^2}(\bar s \gamma_\mu  b_L) (\bar \ell\gamma^\mu \ell) ~,
\qquad  \qquad
& & \mathcal O_{10} &\;=\;& \frac{e^2}{16\pi^2}(\bar s \gamma_\mu  b_L) (\bar \ell\gamma^\mu \gamma_5 \ell) ~, \nn
\end{alignat}
with $\alpha, \beta$ color indices and the primed operators obtained from eq. (\ref{eq:Oi}) by the replacements \{$L \to R$, $m_b \to m_s$\}. We use the covariant-derivative definition $D_\mu = \partial_\mu + i e Q_f A_\mu + i g_s G_\mu$ (where e.g. $Q_e = -1$). This definition, and the chosen normalisations of $\mc O_{7,8}$ yield $C^{\SM}_{7,8} < 0$.

With the above definitions, the amplitudes $\parenbar{\mc A}_{\mmy} \equiv \mc A (\parenbar{B}_s\hspace{-0.33cm}\phantom{1}^0 \to \mmy)$ entering eq. (\ref{eq:ADGmmy_app}) may be expressed as the sum of a `direct-emission' (DE) and a bremsstrahlung component \cite{Melikhov:2004mk,Kozachuk:2017mdk}, which read
\bea
\label{eq:A_DE}
&&\hspace{-1cm}\mc A_{\DE}(\bBs \to \mmy) ~=~ \frac{G_F}{\sqrt2} V_{tb} V_{ts}^* \frac{\alpha}{2\pi} \times \nn \\ 
&&\hspace{2.5cm} \Bigl \{ - \frac{2im_b C_7}{q^2} \, 
\< \ga(k,\eps)| \bar s \sigma_{\mu \nu} (1+\ga_5) q^\nu b | \bBs(p) \> \,
\bar u(p_2) \ga^\mu v(p_1) \nn \\
&& \hspace{2.7cm} + C_9 \, 
\< \ga(k,\eps)| \bar s \ga_\mu (1-\ga_5) b | \bBs(p) \> \,
\bar u(p_2) \ga^\mu v(p_1) \nn \\
&& \hspace{2.7cm} + C_{10} \,
\< \ga(k,\eps)| \bar s \ga_\mu (1-\ga_5) b | \bBs(p) \> \,
\bar u(p_2) \ga^\mu \ga_5 v(p_1) \Bigl \}\,,
\eea
\bea
\label{eq:A_Brems}
&&\mc A_{\Brems} ~=~ + i \frac{G_F}{\sqrt2} V_{tb} V_{ts}^* \frac{\alpha}{2\pi} e ~ X_{f} f_{B_s} ~ 2 m_{\mu} \, C_{10} \, \Bigl \{ \bar u(p_2) \left( \frac{\slashed \eps^* \slashed p}{t - m_\mu^2} - \frac{\slashed p \slashed \eps^*}{u - m_\mu^2} \right)  v(p_1) \Bigl \}\,.~~~~~
\eea
The $\mc O^{(')}_{7,9,10}$ matrix elements are parameterised by f.f.'s of dilepton momentum transfer $q^2$ \cite{Melikhov:2004mk,Kozachuk:2017mdk,Kruger:1996cv}\footnote{Here we assume the convention $\epsilon^{0123} = -1$.}
\bea
\label{eq:FVA_FTVTA}
\< \gamma(k,\la)|\bar s \gamma^\mu b | \bBs(q+k) \> &=& 
e \, \eps^{\mu \la^* q k} \frac{\FV(q^2)}{\MB}\,, \nn \\
\< \gamma(k,\la)|\bar s \gamma^\mu \gamma_5 b | \bBs(q+k) \> &=& 
i e \, (\la^{*\mu} \, q k - k^\mu \, \la^* q) \frac{\FA(q^2)}{\MB}\,, \nn \\
\< \gamma(k,\la)|\bar s \sigma^{\mu \nu} b q_\nu| \bBs(q+k) \> &=& 
i e \, \eps^{\mu \la^* q k} \FTV(q^2,0)\,, \nn \\
[0.2cm]
\< \gamma(k,\la)|\bar s \sigma^{\mu \nu} \gamma_5 b q_\nu| \bBs(q+k) \> &=& 
e \, (\la^{*\mu} \, q k - k^\mu \, \la^* q) \FTA(q^2,0)\,,
\eea
and $f_{B_s}$ is defined through
\be
\label{eq:fBs}
\< 0 | \bar s \ga^\mu \ga_5 b | \bBs(p) \> = i p^\mu f_{B_s} X_{f}~.
\ee
The amplitudes in eqs. (\ref{eq:A_DE})-(\ref{eq:A_Brems}) can be generalized to include primed coefficients through the substitutions
\be
C_7 \to C_7 \pm \frac{m_s}{m_b} C_7^\prime~, ~~~~C_{9,10} \to C_{9,10} \pm C_{9,10}^\prime~, 
\ee
where the $+$($-$) refers to terms proportional to $F_{V,TV}$ ($F_{A,TA}$ or $f_{\Bs}$).

Eq. (\ref{eq:A_DE}) exactly reproduces the result in \cite{Melikhov:2004mk}; eq. (\ref{eq:A_Brems}) matches the result in \cite{Melikhov:2004mk} provided $X_f = +1$. We note however that the PDG \cite{Zyla:2020zbs} and FLAG \cite{Aoki:2019cca} imply $X_f = -1$. This point is accounted for in Refs. \cite{Guadagnoli:2016erb,Kozachuk:2017mdk}.

The calculation of $\ADGmmy$ involves also the amplitudes for an initial $\Bs$. The $\Bs \to \ga$ hadronic matrix elements are related to the $\bBs \to \ga$ ones through a CP transformation, hence they depend on the phase $\phi_{\CP}$ appearing in eq. (\ref{eq:CPBs}). 
The amplitude for $\Bs \to \mmy$ is related to the amplitude for $\bBs \to \mmy$ by replacing CKM matrix entries and Wilson coefficients with their complex-conjugates, and by the following substitutions
\be
\label{eq:CPff}
\mc F_{V}^{(\bBs)} \rightarrow - e^{-i \phi_{\CP}} \mc F_{V}^{(\Bs)}~,~~~~
\mc F_{A}^{(\bBs)} \rightarrow + e^{-i \phi_{\CP}} \mc F_{A}^{(\Bs)}~,
\ee
where $\mc F_{V}^{(\bBs)}$ denote any of $F_V$ or $F_{TV}$, whereas $\mc F_{A}^{(\bBs)}$ denote any of $F_A$, $F_{TA}$, and $f_{B_s}$ in eqs. (\ref{eq:FVA_FTVTA})-(\ref{eq:fBs}), whereas $\mc F_{V,A}^{(\Bs)}$ denote the f.f.'s and decay constant for $\Bs$ matrix elements.\footnote{The analogous relations in the $A_\parallel,\perp$ notation can be found in \cite{Beneke:2020fot,Janowski:2021yvz}.  As discussed in the main text, the $\phi_\CP$ dependence cancels in $\ADGmmy$. For the translation of $A_\parallel,\perp$ amplitudes into the notation of eq. (\ref{eq:FVA_FTVTA}), see e.g. \cite{Guadagnoli:2017quo}.}

As concerns the dependence on weak phases, we note that a $\bBs$ ($\Bs$) initial state corresponds to a $b \to s$ ($\bar b \to \bar s$) Hamiltonian, with Wilson coefficients $C_{7, 9, 10}$ ($C_{7, 9, 10}^{*}$). As a consequence, both of $\bar{\mc A}$ and $\mc A^*$ are proportional to $C_{7, 9, 10}$ and the numerator of $\ADGmmy$ depends on $C_i^2$. Hence the dependence on the Wilson-coefficients' phases does not cancel in $\ADGmmy$ as it would in, e.g., $|\mc A|^2$.
However, and as remarked in the main text, the CKM-phase dependence in $\bar{\mc A}_{\mmy} \mc A_{\mmy}^*$ cancels exactly with the analogous dependence in $q/p$, so that the r.h.s. of eq. (\ref{eq:AbAs}) is proportional to $|\mc N|^2$. As a consequence, $\ADGmmy$ is sensitive to CP-violating phases coming from Wilson coefficients only.

\section{BBW vs. JPZ form factors}\label{app:BBW}

For the hadronic matrix elements relevant to our study we use two alternative approaches. The first one relies on parameterising the required matrix elements as in eq. (\ref{eq:FVA_FTVTA}). This parameterisation \cite{Melikhov:2004mk,Kozachuk:2017mdk,Kruger:1996cv} defines the f.f.'s $F_{V,A,TV,TA}$. These f.f.'s are subsequently estimated using a relativistic quark model (see \cite{Melikhov:1995xz,Melikhov:1997qk,Melikhov:2001zv} for details). We use the latest determination in \cite{Kozachuk:2017mdk} as reference for this approach. Importantly, $\FTV(q^2,0)$ and $\FTA(q^2,0)$ denote only the contributions where the e.m. penguin emits the di-lepton pair. In order to also include topologies where the e.m. penguin emits the real, final-state photon, as well as so-called weak-annihilation contributions \cite{Bosch:2002bv} one replaces
\be
\label{eq:barFTVTA}
F_{TV,TA}(q^2,0) \to \bar{F}_{TV,TA}(q^2)
\ee
in eqs. (\ref{eq:FVA_FTVTA}), with the $\bar{F}_{TV,TA}$ functions defined in Ref. \cite{Kozachuk:2017mdk}.

Recently, Beneke, Bobeth and Wang \cite{Beneke:2020fot} have performed the first evaluation of the $\bBsmmy$ amplitude at low $q^2 < 6$ \GeV$^2$ using rigorous factorisation methods. The amplitude is expressed as the sum of initial- plus final-state-radiation terms. The final-state-radiation terms correspond to $\bBs$-to-vacuum matrix elements proportional to $\mc O_{10}$, hence helicity-suppressed and completely negligible \cite{Dettori:2016zff,Beneke:2020fot,Aditya:2012im}. On the other hand, from the initial-state-radiation terms one can read off the `equivalent' of the $\FV, \FA, \bFTV$ and $\bFTA$ functions, to be denoted as $\FV^{\BBW}$, $\FA^{\BBW}$, $\FTV^{\BBW}$, $\FTA^{\BBW}$.
We find\footnote{Eqs. (\ref{eq:FVA_FTVTA}) use antisymmetric-tensor conventions as in \cite{Kozachuk:2017mdk}, which differ from Ref. \cite{Beneke:2020fot}'s. Below relations take this difference into account.}
\bea
\label{eq:FVABBW}
C_i \, F_{V,A}^{\rm BBW}(q^2) &=&
- Q_s F_{B_s} \frac{\MB}{2 E_\gamma} \, V_{i}^{\eff}(q^2) \,
\int_0^\infty \frac{d \omega}{\omega} \phi_+(\omega) \, J(2 E_\gamma, 0, \omega) \nn \\
&&- C_i \left \{ \xi_{B_s} \pm \frac{f_{B_s} \MB}{2 E_\gamma} 
\left( \frac{Q_b}{m_b} + \frac{Q_s}{2 E_\gamma} \right) \right \}\,,
\eea
where $i=9,10$ and $C_i$ stands for $C_9^{\eff}(q^2)$ and $C_{10}$ \cite{Chetyrkin:1996vx,Bobeth:1999mk}, respectively, and
\bea
\label{eq:FTVTABBW}
C_7^{\eff} F_{TV,TA}^{\rm BBW}(q^2) &=& - Q_s F_{B_s} \Bigg \{ \frac{\MB}{2 E_\gamma} V_{7}^{\eff}(q^2) 
\int_0^\infty \frac{d \omega}{\omega} \phi_+(\omega) \, J(2 E_\gamma, 0, \omega) \nn \\
&&\phantom{riighttt} + V_7^{\eff}(0) \int_0^\infty d \omega \phi_+(\omega) \, \frac{J(\MB, q^2, \omega)}{\omega - q^2 / \MB - i 0^+} \Bigg \} \nn \\
&&- C_7^{\eff} \left \{ \left( \xi_{B_s} + \tilde{\xi}_{B_s} + Q_b \frac{f_{B_s}}{E_\gamma} \right)
\pm Q_s \frac{f_{B_s}}{\MB}\frac{q^2}{4 E_\gamma^2} \right \} 
\mp \frac{\MB}{m_b} \frac{f_{B_s}}{E_\gamma} f_{V,A}(1 - \sh)\,.\nn \\
\eea
In the above expressions:\footnote{DG warmly acknowledges Christoph Bobeth for enlightening discussions about many details of Ref. \cite{Beneke:2020fot}.}
\begin{itemize}
\item For the $V_{i}^{\eff}(q^2) = C_i \times \left(1 + O\left(\frac{\alpha_s(\mu_h)}{4\pi}\right)\right)$, where $C_i = C_7^{\eff}, C_9^{\eff}, C_{10}$, with $\mu_h$ the hard scale, we use expressions in \cite{Beneke:2020fot}. We take the necessary $F_i^{j (u)}$ functions from \cite{Seidel:2004jh}, the $F_i^{j (c)}$ functions from \cite{Greub:2008cy}(see also \cite{Asatrian:2001de,Asatryan:2001zw}), with both sets validated through \cite{Asatrian:2019kbk}.

\item We comply with \cite{Beneke:2020fot} also as concerns the values of the Wilson coefficients $C_i$, $i = 1, ..., 6,9,10$, $C_{7,8}^{\eff}$ at the SCET `hard' scale of 5 GeV. For $C_9^{\eff}(q^2)$ we use \cite{Chetyrkin:1996vx,Bobeth:1999mk}.\footnote{We use the same values for JPZ-case \cite{Janowski:2021yvz} numerics as well. In particular, the Wilson coefficients appearing in the amplitudes eqs. (\ref{eq:A_DE})-(\ref{eq:A_Brems}) are to be understood as $C_{7,8}^{\eff}$, $C_9^{\eff}(q^2)$, $C_{10}$.}

\item $F_{B_s}$ is the HQET $B_s$-meson decay constant, whose relation with $f_{B_s}$ may be found in \cite{Beneke:2020fot,Beneke:2011nf};

\item The hard-collinear matching function $J$ can be read off from \cite{Beneke:2020fot,Wang:2016qii}, with $\mu$ ($\mu_0$) the hard-collinear (collinear) scales.

\item For the $B$-meson light-cone distribution amplitude $\phi_+$ we use the exponential model suggested in Ref. \cite{Grozin:1996pq}. Within this model the evolution from $\mu_0$ to $\mu$ can be performed using \cite{Beneke:2011nf}, and the inverse ($\lambda_{B_s}(\mu)$), as well as all logarithmic moments $\sigma_n(\mu)$ are analytically calculable \cite{Beneke:2011nf}, as is the second convolution integral in eq. (\ref{eq:FTVTABBW}). Using $\lambda_{B_s}(\mu_0) = 0.40(15)$ \cite{Beneke:2020fot}, we find, as a cross-check, $\sigma_1(\mu_0) = 1.49$ and $\sigma_2(\mu_0) = 3.9$, in good agreement with the ranges $1.5 \pm 1.0$ and $3 \pm 2$ suggested in \cite{Beneke:2011nf}. We then use
\be
\label{eq:laBs_value}
\lambda_{B_s}(\mu) = 0.44^{+0.15}_{-0.16} ~~ \GeV
\ee
in our numerics. This parameter represents the main source of error, as well-known.

\item The other main source of error comes from the next-to-leading-power f.f.'s $\xi$ and $\tilde \xi$. Following \cite{Beneke:2018wjp}, Ref. \cite{Beneke:2020fot} includes in the $\xi$ and $\tilde \xi$ definition a term that subtracts the leading-power contribution in the tree approximation, multiplied by $r_{\LP} = 0.2 \pm 0.2$. The $r_{\LP}$ range yields the second largest source of error in our case.

\item For the $m_b$ and $m_c$ masses, we use the pole-mass values in table \ref{tab:input}.\footnote{Ref. \cite{Beneke:2020fot} suggests to use the $b$-mass obtained in the so-called potential-subtracted renormalisation scheme~\cite{Beneke:1998rk}. We find this choice to have no impact on our $\ADGmmy$ numerics.}
The distinction with respect to $\overline{\mathrm{MS}}$ masses is dropped in the next-to-leading-power terms in eqs. (\ref{eq:FVABBW})-(\ref{eq:FTVTABBW}).

\item Any other function or symbol not commented upon can be retrieved from \cite{Beneke:2020fot}.

\end{itemize}
\begin{figure}[!ht]
  \centering
  \hspace{-0.2cm}\includegraphics[scale=0.49]{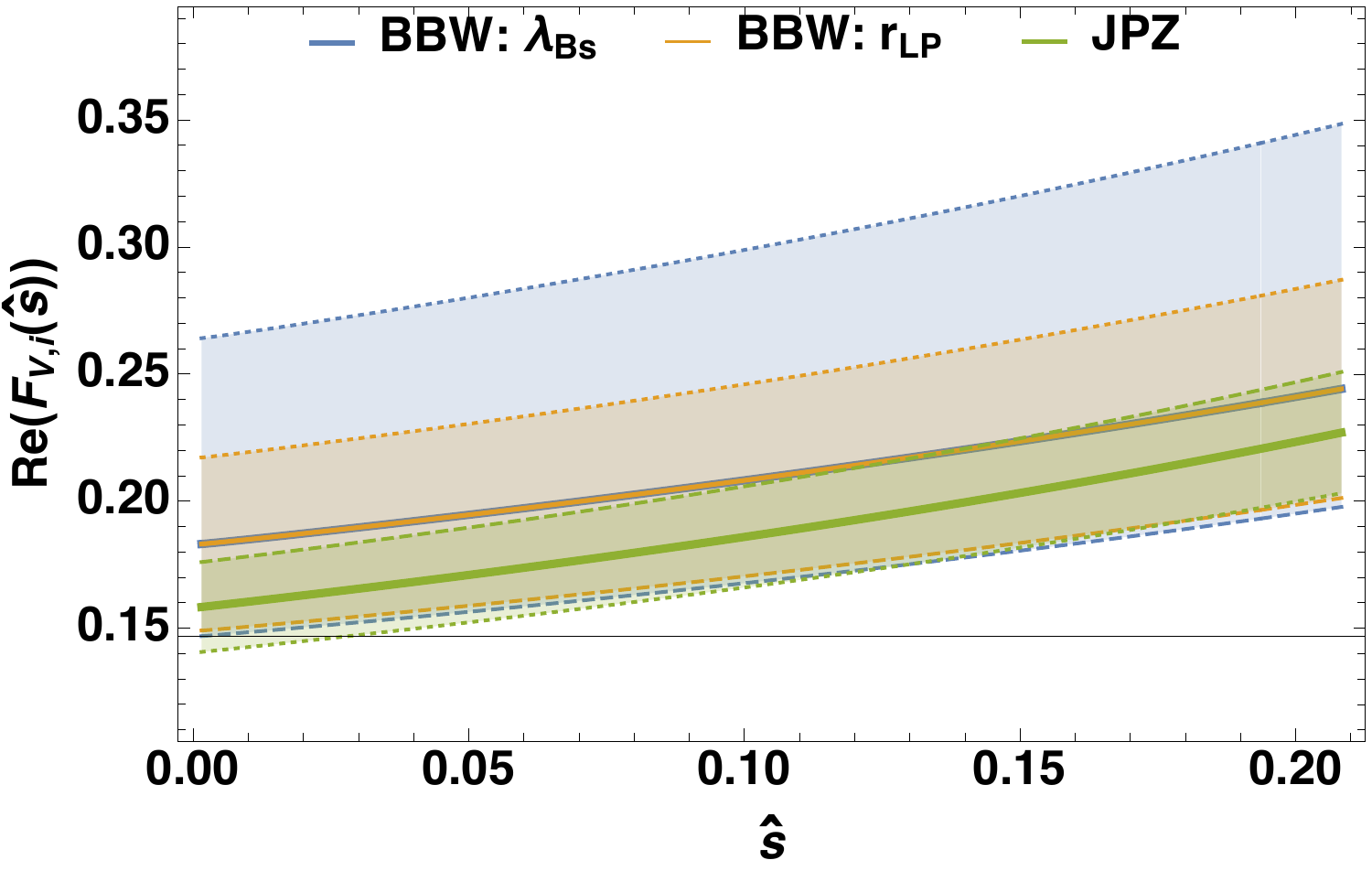}
  \hspace{0.2cm}\includegraphics[scale=0.49]{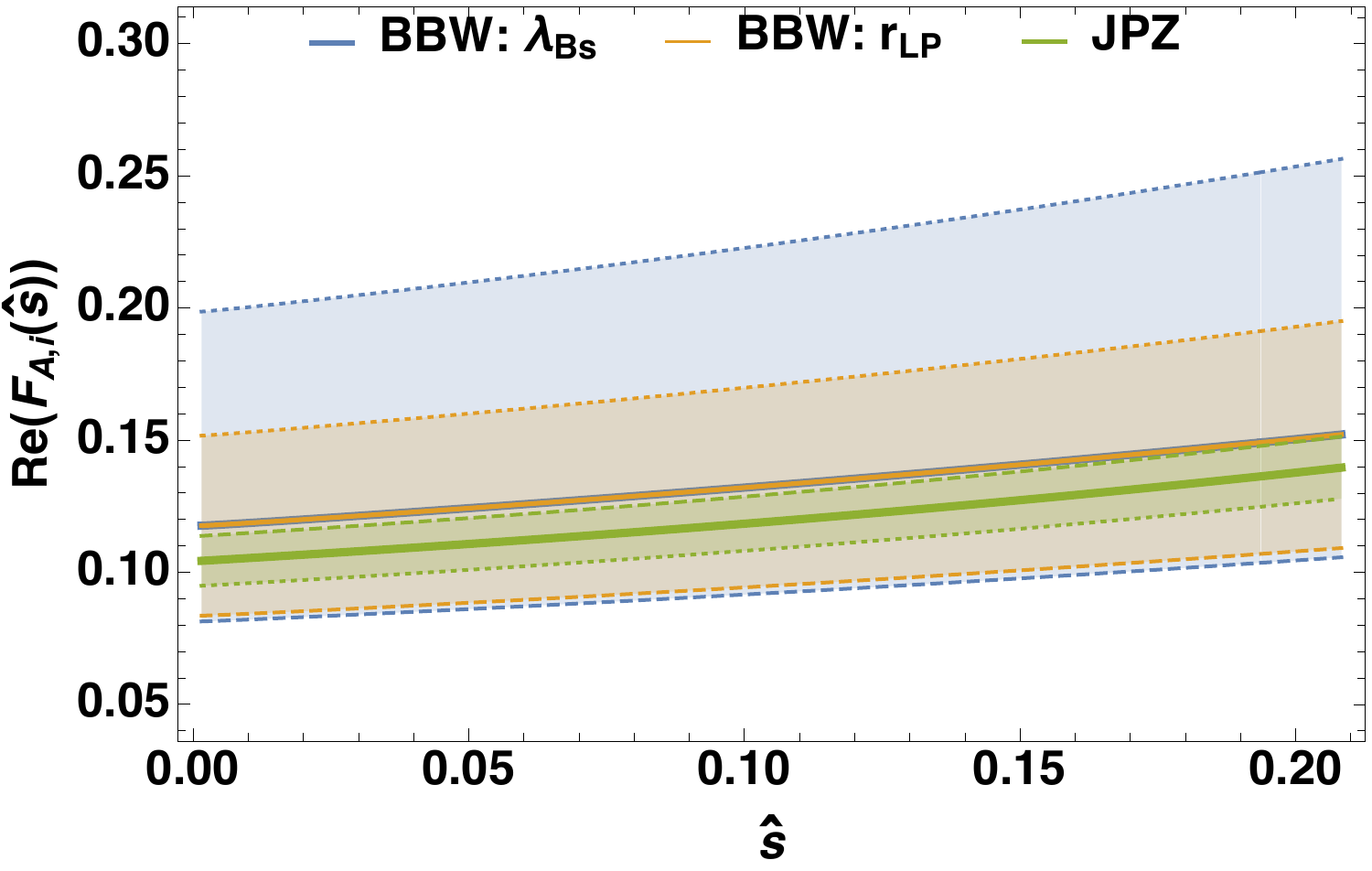}
  \includegraphics[scale=0.49]{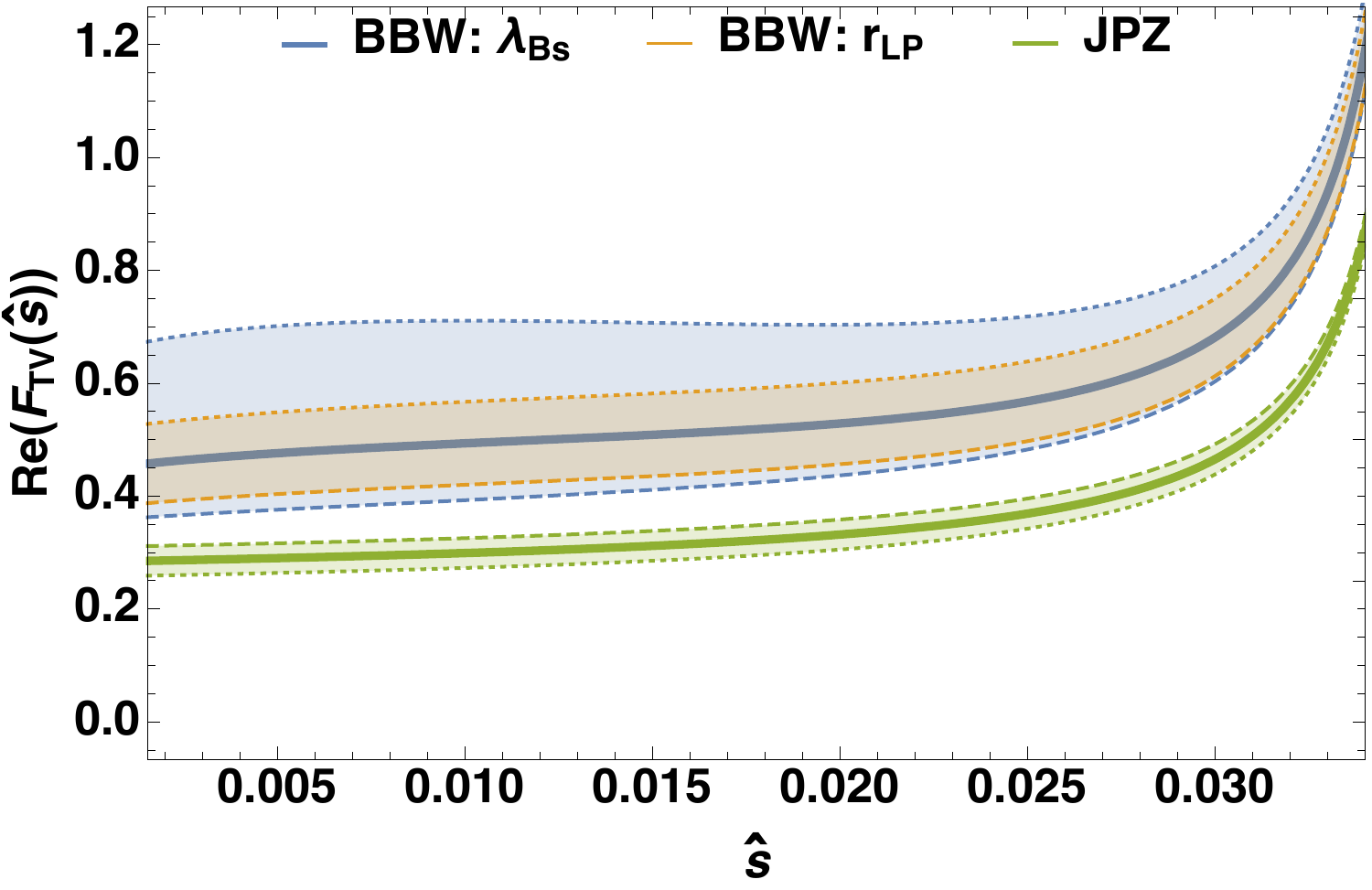}
  \hspace{0.2cm}
  \includegraphics[scale=0.49]{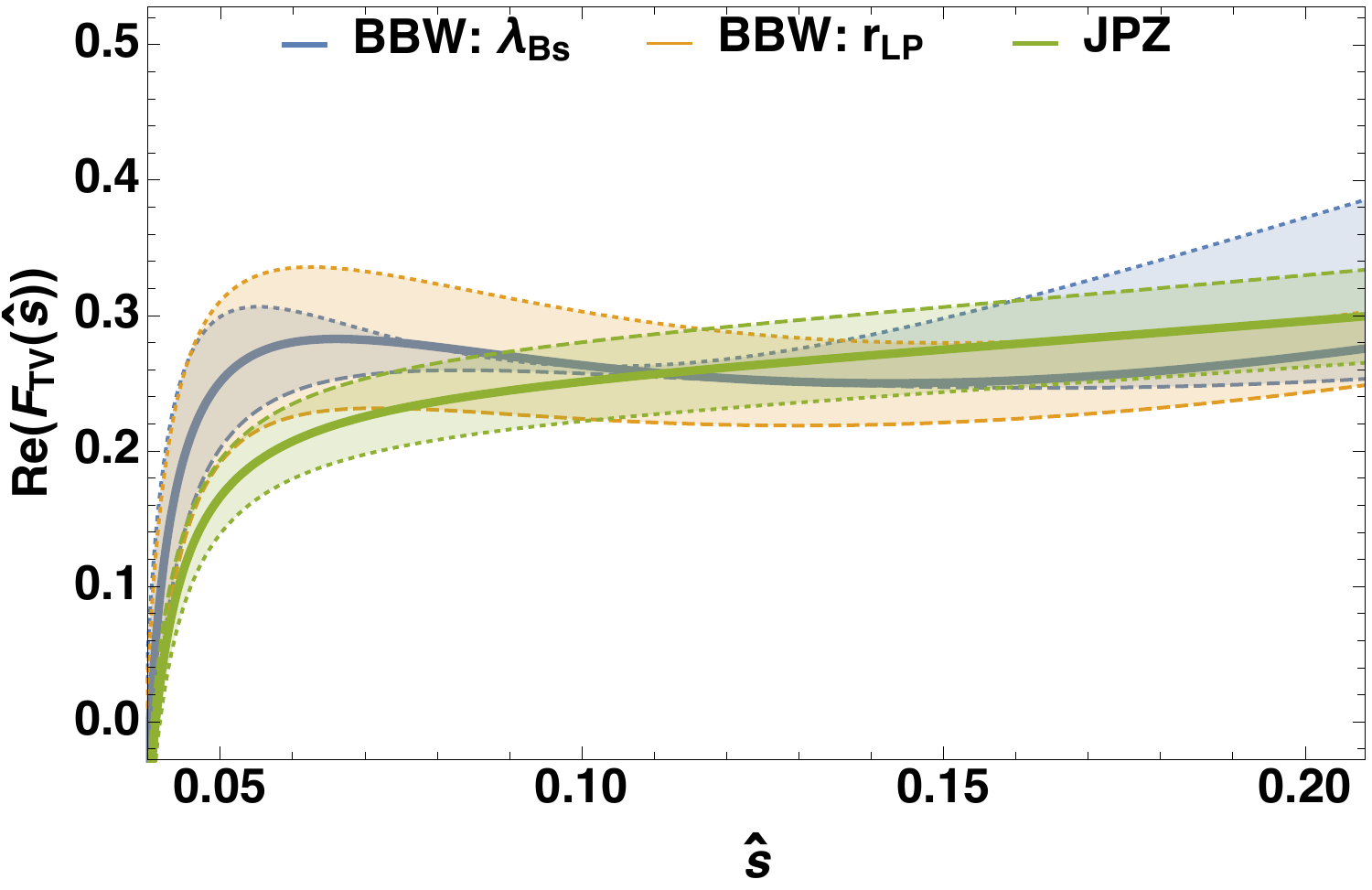}
  \includegraphics[scale=0.49]{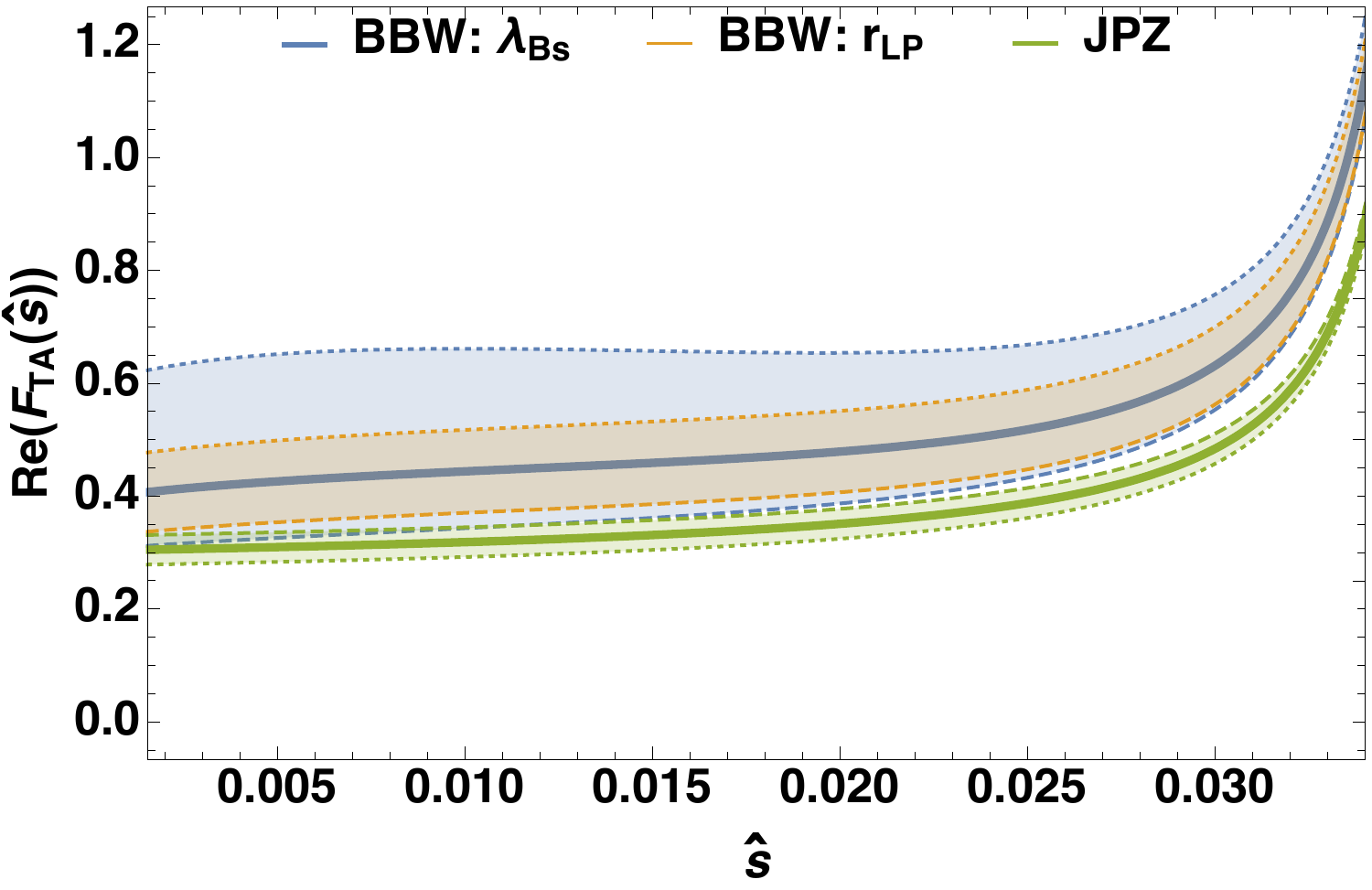}
  \hspace{0.2cm}
  \includegraphics[scale=0.49]{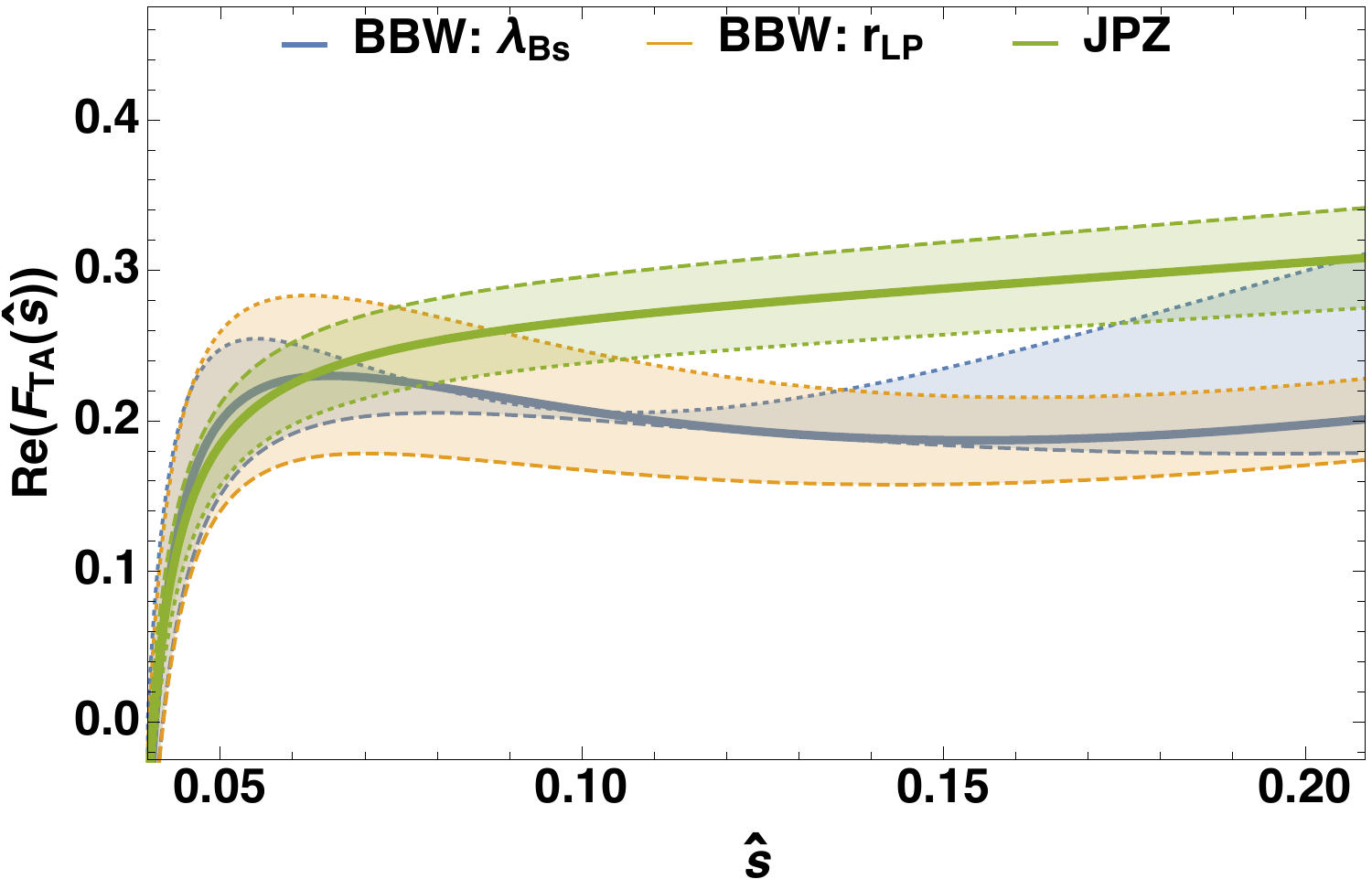}
  \vspace{-0.4cm}
  \caption{Comparison of the real parts of JPZ f.f.'s $F_{V}$, $F_{A}$ (first row left vs. right), $\bar F_{TV}$ (second row), $\bar F_{TA}$ (third row), with the BBW `effective' f.f. counterparts in eqs. (\ref{eq:FVABBW})-(\ref{eq:FTVTABBW}). Note that the first row shows vector and axial f.f.'s with an index $i = 9, 10$, as in eqs. (\ref{eq:FVABBW}). The BBW-f.f. range of variation due to the $\lambda_{B_s}$ and to the $r_{\LP}$ uncertainties, both discussed in the text, are shown separately (see legend entries). For rendering reasons, in the instance of tensor f.f.'s we show separately the cases of $\sh$ below (left panels) and respectively above (right panels) the $\phi$ peak.}
  \label{fig:Reff_BBW_vs_JPZ}
\end{figure}

\begin{figure}[!ht]
  \centering
  \hspace{-0.2cm}\includegraphics[scale=0.49]{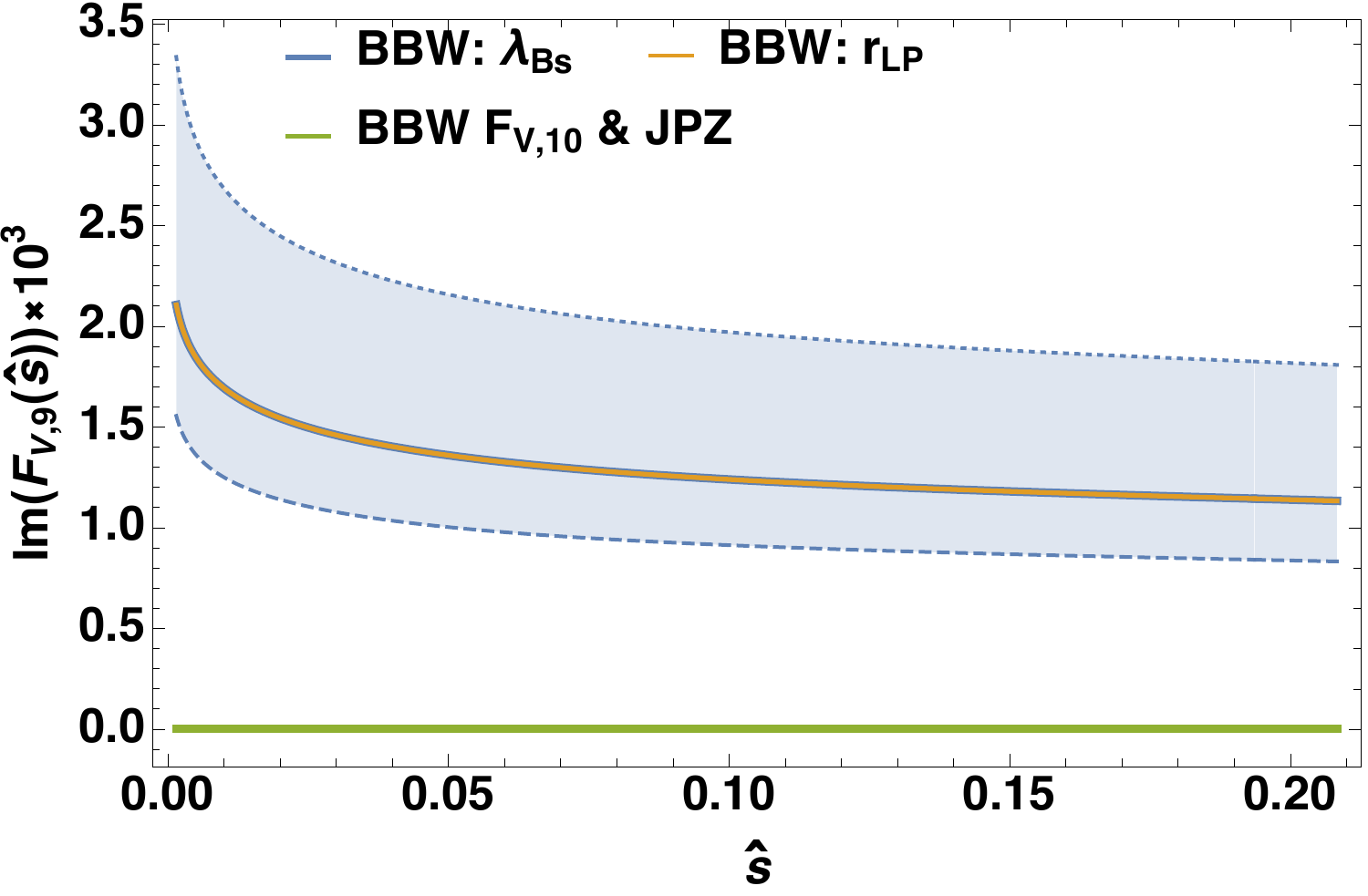}
  \hspace{0.2cm}\includegraphics[scale=0.49]{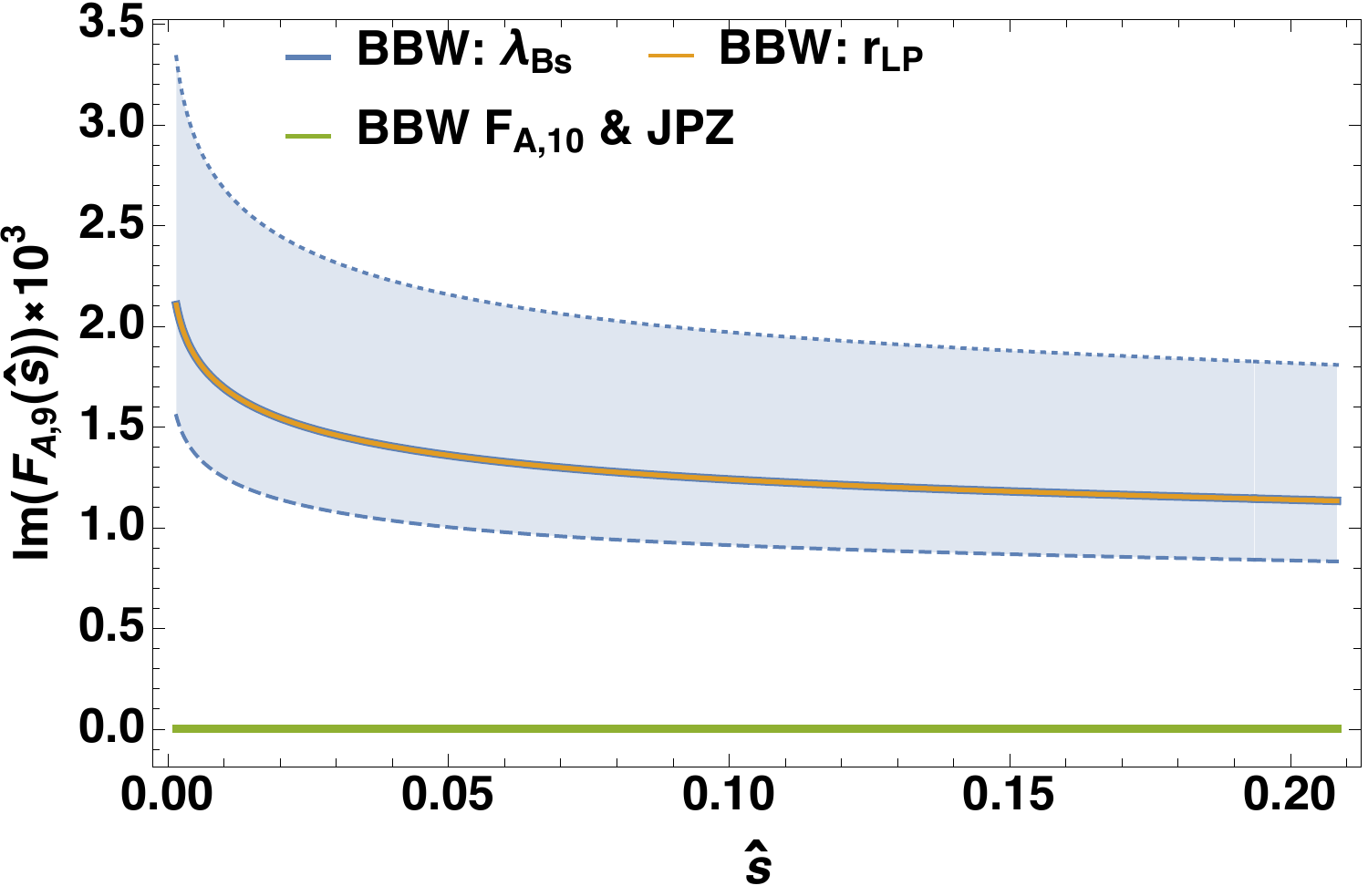}
  \includegraphics[scale=0.49]{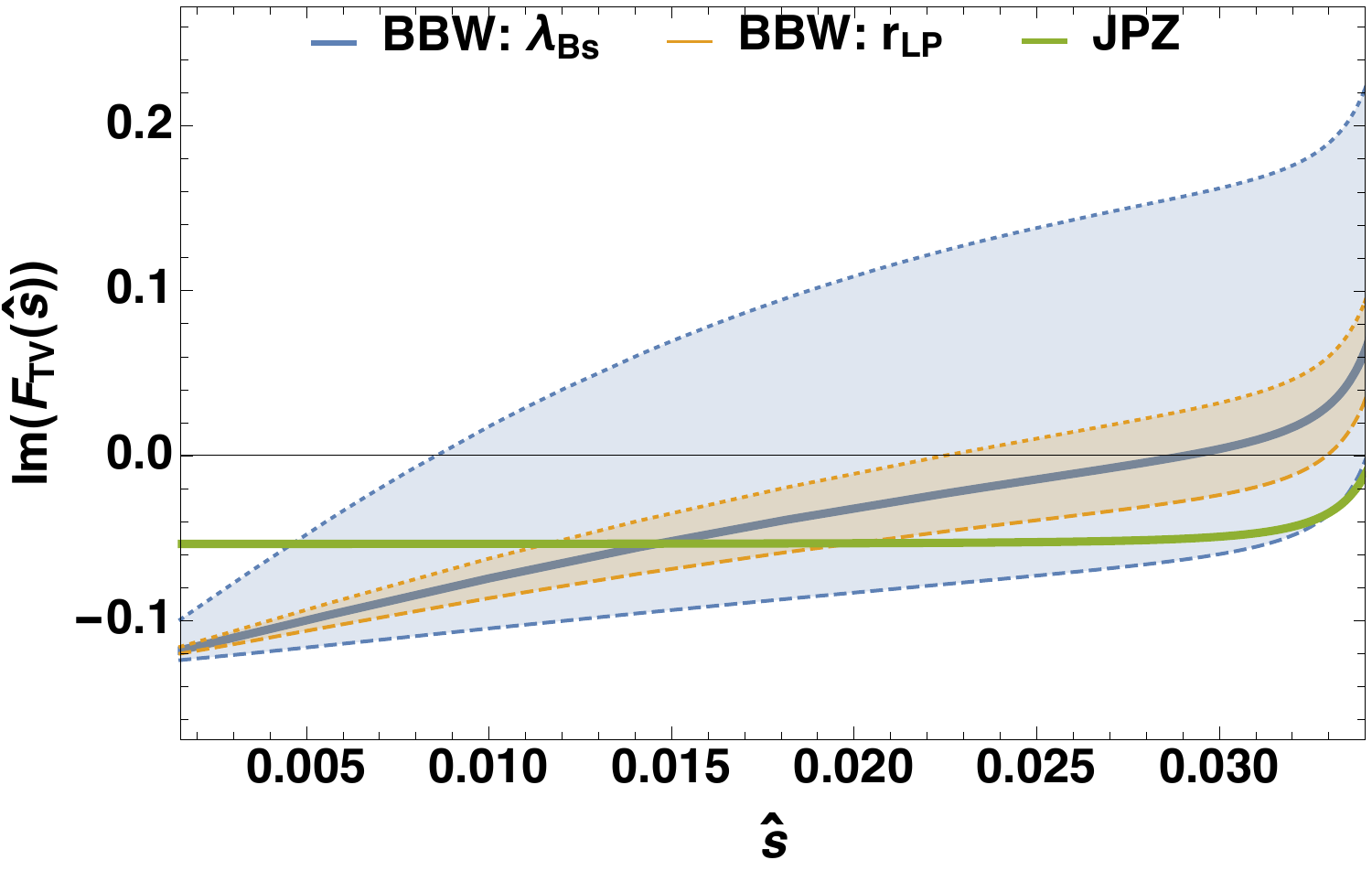}
  \hspace{0.2cm}
  \includegraphics[scale=0.49]{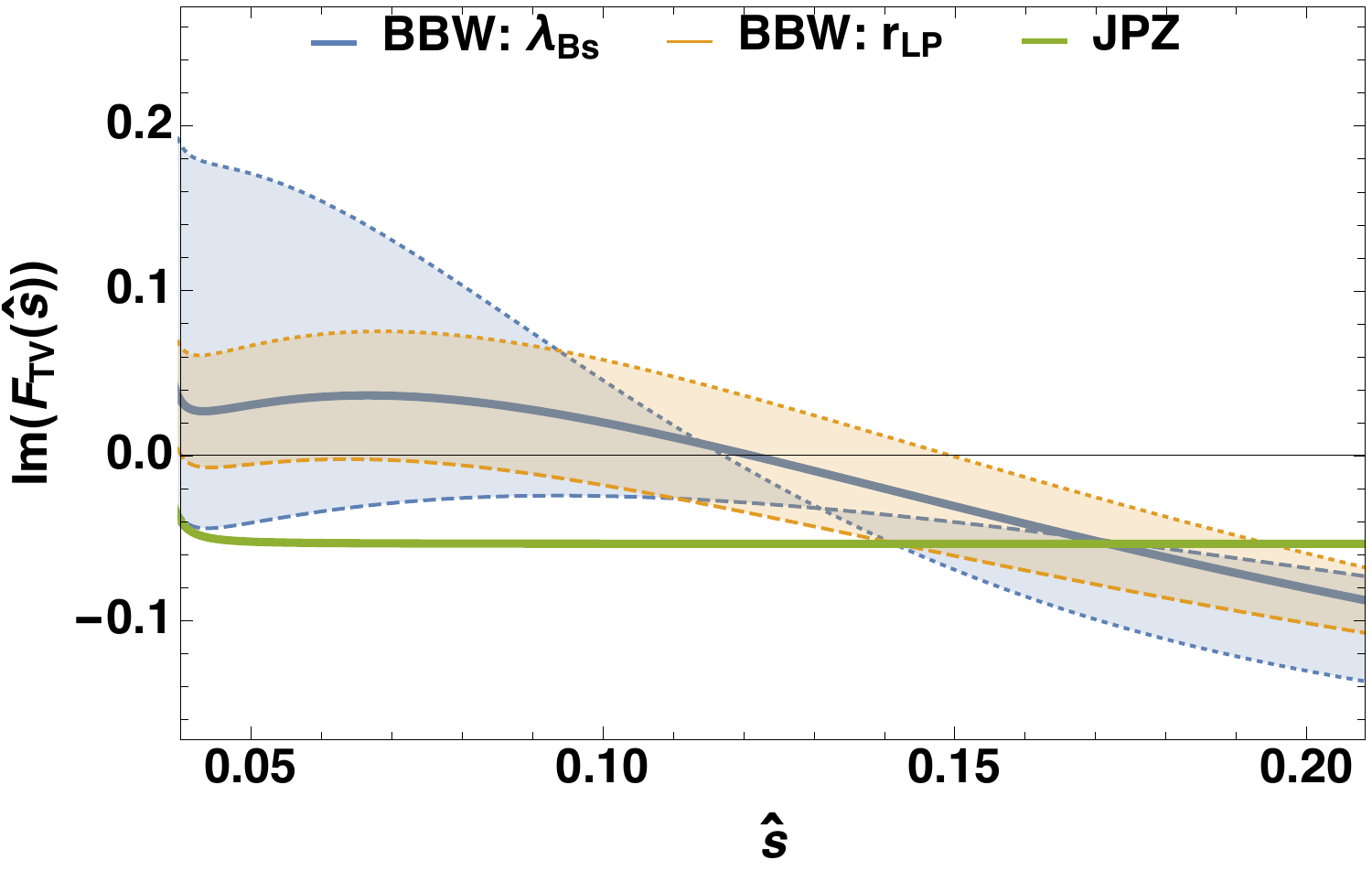}
  \includegraphics[scale=0.49]{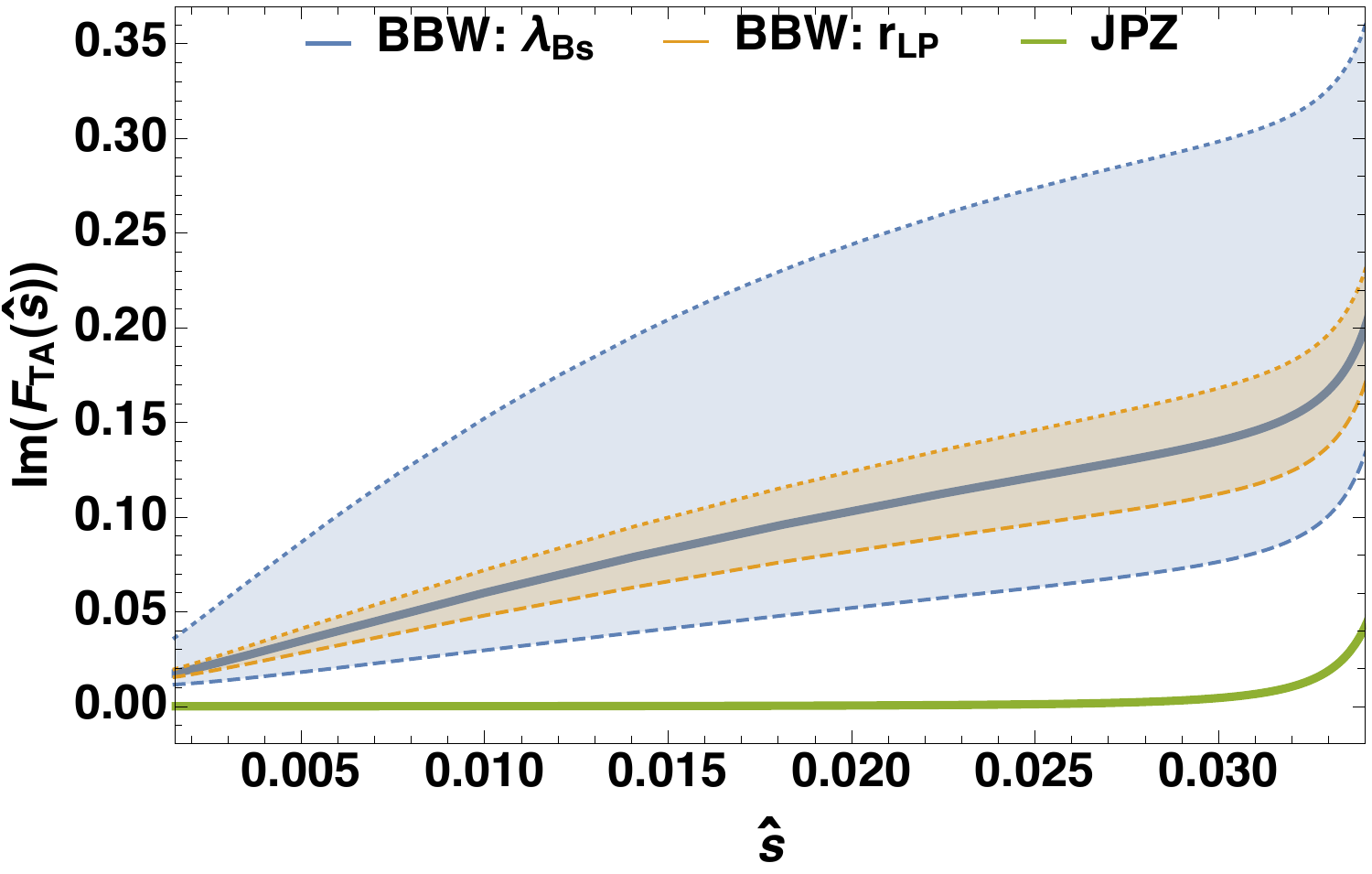}
  \hspace{0.2cm}
  \includegraphics[scale=0.49]{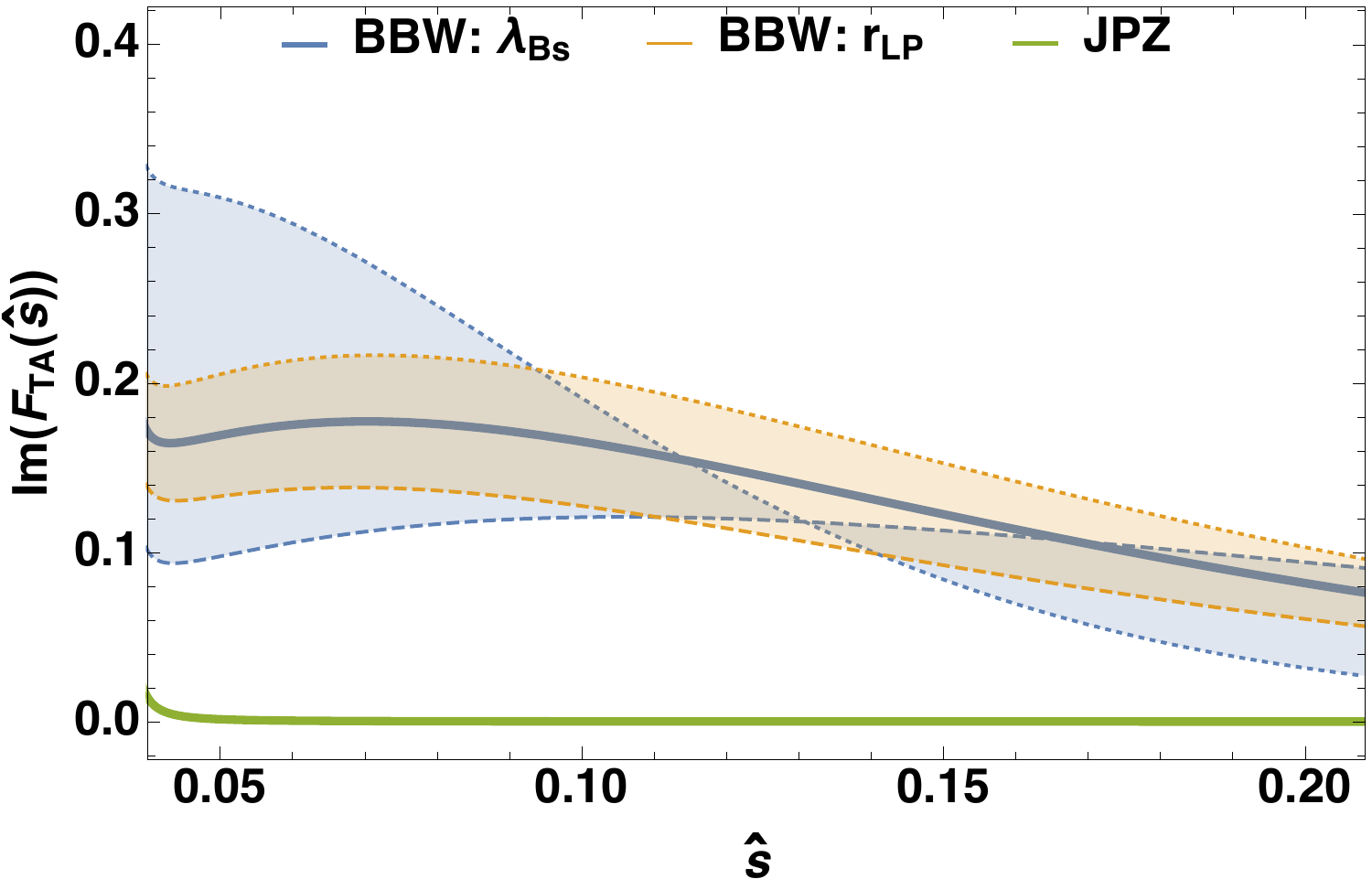}
  \vspace{-0.4cm}
  \caption{Same as fig. \ref{fig:Reff_BBW_vs_JPZ}, but for the imaginary parts of the BBW vs. JPZ f.f.'s.}
  \label{fig:Imff_BBW_vs_JPZ}
\end{figure}

It is useful to explicitly compare the JPZ f.f.'s in the $F_{V}$, $F_{A}$, $\bar F_{TV}$, $\bar F_{TA}$ nomenclature, with the BBW `effective' f.f.'s in eqs. (\ref{eq:FVABBW})-(\ref{eq:FTVTABBW}) in the low-$q^2$ region. We do so in figs. \ref{fig:Reff_BBW_vs_JPZ} and \ref{fig:Imff_BBW_vs_JPZ} for the real and imaginary parts, respectively. We show separately the BBW-f.f. range of variation due to the $\lambda_{B_s}$ and to the $r_{\LP}$ uncertainties discussed in the previous paragraph. 
Note that the JPZ parameterization in Ref. \cite{Janowski:2021yvz} yields real f.f.'s. The only source of an imaginary contribution is the shift due to the $\phi$ resonance, which affects only the tensor f.f.'s.
The `dictionary', provided in Ref. \cite{Janowski:2021yvz}, between JPZ and KMN f.f.'s allows to include such shift in a way akin to the KMN parameterization. We include this contribution only for the sake of the indicative comparison in fig. \ref{fig:Imff_BBW_vs_JPZ}, keeping in mind that the $\phi$ region is basically entirely excluded in the considered kinematic range.
The figures show that the overall agreement between the JPZ and BBW f.f.'s is mostly acceptable within the quite large errors inherent especially to the BBW parameterization.

\section{\boldmath Explicit formula for $\ADGmmy$} \label{app:ADG_formula}

Starting from eq. (\ref{eq:ADGmmy})
\be
\label{eq:ADGmmy_app}
\ADGmmy ~=~ 
- \frac{
\int d\hat{s} \, d \cos \theta \, \, f(\hat s, {\hat m}^2_\mu) \,\, \re \left( q / p \, \bar{\mc A}_{\mmy} \mc A_{\mmy}^* \right)
}{
\int d\hat{s} \, d \cos \theta \, \, f(\hat s, {\hat m}^2_\mu) \,\,
\left \vert \mc A_{\mmy} \right \vert^2
}\,,
\ee
we can perform the integration $\int_{-1}^{1} d \cos \theta$. We obtain
\bea
\label{eq:AbAs}
\int d \cos \theta ~ \re \left( q / p \, \bar{\mc A}_{\mmy} \mc A_{\mmy}^* \right) ~=~
|\mc N|^2 \sum_{i , j}^{7,9,10} \re (f_{ij} \, \C_i \C_j)
\eea
and 
\bea
\label{eq:A2}
\int d \cos \theta ~ \left \vert \mc A_{\mmy} \right \vert^2 ~=~
|\mc N|^2 \sum_{i , j}^{7,9,10} \re ( \bar f_{ij} C_i C_j^* )\,,
\eea
where the functions $\parenbar{f}_{ij}$ are given by
\begin{flalign}
\label{eq:f77}
f_{7 \, 7} ~=~ \frac{16}{3} \MB^4 \mbh^2 \, x^2 \, \frac{2 \mlh^2 + \sh}{\sh^2} (\bar F_{TA}^2 - \bar F_{TV}^2)\,, &&
\end{flalign}
\begin{flalign}
\label{eq:bf77}
\bar f_{7 \, 7} ~=~ \frac{16}{3} \MB^4 \mbh^2 \, x^2 \, \frac{2 \mlh^2 + \sh}{\sh^2} (|\bar F_{TA}|^2 + |\bar F_{TV}|^2)\,, &&
\end{flalign}
\begin{flalign}
\label{eq:f99}
f_{9 \, 9} ~=~ \frac{4}{3} \MB^4 x^2 (2 \mlh^2 + \sh) (\Fa{9}^2 - \Fv{9}^2)\,, &&
\end{flalign}
\begin{flalign}
\label{eq:bf99}
\bar f_{9 \, 9} ~=~ \frac{4}{3} \MB^4 x^2 (2 \mlh^2 + \sh) (|\Fa{9}|^2 + |\Fv{9}|^2)\,, &&
\end{flalign}
\begin{flalign}
\label{eq:f1010}
f_{10 \, 10} ~=~ - \frac{4 \MB^4 x^2}{3} 
(4 \mlh^2 - \sh)(\Fa{10}^2 - \Fv{10}^2) \,, &&
\end{flalign}
\begin{flalign}
\label{eq:bf1010}
\bar f_{10 \, 10} ~=~ - \frac{4 \MB^4 x^2}{3} 
(4 \mlh^2 - \sh)(|\Fa{10}|^2 + |\Fv{10}|^2) \,, &&
\end{flalign}
\begin{flalign}
\label{eq:f79}
f_{7 \, 9} + f_{9 \, 7} ~=~ \frac{16}{3} \MB^4 \mbh \, x^2 \, \frac{2 \mlh^2 + \sh}{\sh} (\bar F_{TA} \Fa{9} - \bar F_{TV} \Fv{9})\,, &&
\end{flalign}
\begin{flalign}
\label{eq:bf79}
\bar f_{7 \, 9} \, C_7 C_9^* + \bar f_{9 \, 7} \, C_9 C_7^* ~=~ \frac{16}{3} \MB^4 \mbh \, x^2 \, \frac{2 \mlh^2 + \sh}{\sh} \re\left[\bar F_{TA} C_7 \, (\Fa{9} C_9)^* + \bar F_{TV} C_7 \, (\Fv{9} C_9)^*\right]\,, &&
\end{flalign}
\begin{flalign}
\label{eq:f710}
f_{7 \, 10} = f_{10 \, 7} ~=~ 0 &&
\end{flalign}
\begin{flalign}
\label{eq:bf710}
\bar f_{7 \, 10} \, C_7 C^*_{10} + \bar f_{10 \, 7} \, C_{10} C^*_{7} ~=~ 128 \, \fB \MB^3 \, \mbh \mlh^2 \, \frac{x}{\sh} \, \frac{\atan(z)}{z} \re \left[ \bar F_{TV} C_7 \, C_{10}^* \right]\,, &&
\end{flalign}
\begin{flalign}
\label{eq:f910}
f_{9 \, 10} = f_{10 \, 9} ~=~ 0\,, &&
\end{flalign}
\begin{flalign}
\label{eq:bf910}
\bar f_{9 \, 10} \, C_9 C_{10}^* + \bar f_{10 \, 9} \, C_{10} C_{9}^* ~=~ 64 \, \fB \MB^3 \mlh^2 \, x \frac{\atan(z)}{z} \re \left[ \Fv{9} C_9 \, C_{10}^* \right] \,, &&
\end{flalign}%
with $x = 1 - \sh$ and $z = \sqrt{1 - 4 \mlh^2 / \sh}$. In the above formulae we abbreviated the $F_{V(A)}^{\rm BBW}$ f.f.'s appearing in eq. (\ref{eq:FVABBW}) with $F_{V(A), i}$, with $i = 9$ or 10. For JPZ f.f.'s it is understood that the cases $i = 9$ or 10 coincide.

We finally note that the above formul{\ae} include terms linear in $f_{B_s}$, but not quadratic ones. In fact, and as already mentioned in Sec. \ref{sec:high_q2}, we consider the $\Bsmmy$ spectrum with the bremsstrahlung contribution set to zero, whereas we keep interference terms, that the MonteCarlo does not subtract. These linear terms have actually little impact on the numerics, as we verified explicitly.

\section{\boldmath On the near-cancellation of broad-charmonium uncertainties}\label{app:why_cc_small}

From Sec. \ref{sec:broad-c} and fig. \ref{fig:ADG_cc_ff} we concluded that broad-charmonium modeling turns out to have an immaterial effect on the $\ADGmmy$ prediction. This finding is far from trivial, and calls for a closer inspection of the underlying mechanism. In this appendix we show by an analytic argument that the above finding is the result of well-defined cancellation patterns in the $\ADGmmy$ expression.

\begin{figure}[h]
  \centering
  \includegraphics[scale=0.56]{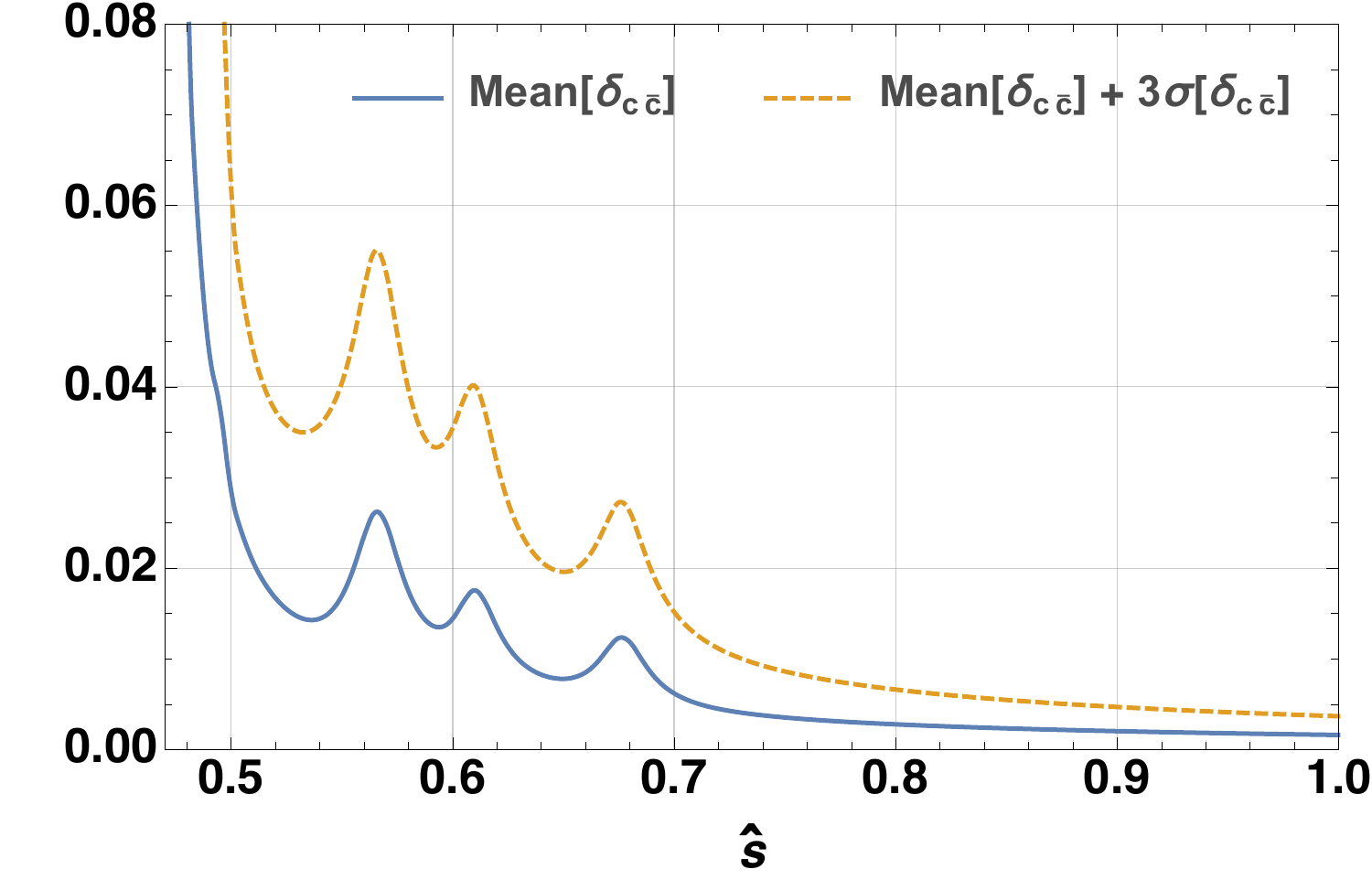}
  \caption{Blue solid line: mean of $|\delta_{c \bar c}|/C_{9, \rm SD}$ as a function of $\sh$, obtained from a uniform scan to $|\eta_V| \in [1,3]$ and $\delta_V \in [0, 2 \pi)$. Dashed orange line: mean plus 3 times the standard deviation.}
  \label{fig:RMSdelta}
\end{figure}

Plugging eqs. (\ref{eq:AbAs}) and (\ref{eq:A2}) into eq. (\ref{eq:ADGmmy_app}), we can write (see also definition (\ref{eq:PS_def}))
\be
\label{eq:ADGmmy_start}
\ADGmmy ~=~ 
- \frac{
\int_{\PS} \, \re \left( q / p \, \bar{\mc A}_{\mmy} \mc A_{\mmy}^* \right)
}{
\int_{\PS} \, \left \vert \mc A_{\mmy} \right \vert^2
} = 
\frac{
- \int_{\PS} \, \left[ \re (f_{99} C_9^2) \right] + N_{\rm rest}
}{
\int_{\PS} \, \left[ \re (\bar f_{99} |C_9|^2) \right] + D_{\rm rest}
}\,,
\ee
where $N_{\rm rest}$ and $D_{\rm rest}$ denote terms other than quadratic or sesquilinear in $C_9$. Broad-charmonium contributions are modelled through eq. (\ref{eq:Vcc_shift}). We accordingly write $C_9 = \bar C_{9} + \delta_{c \bar c}$, where $\delta_{c \bar c}$ denotes the last term (i.e. the full BW sum) in eq. (\ref{eq:Vcc_shift}), whereas $\bar C_9 = C_9^{\rm eff}(q^2) + C_{9, \rm NP}$ bundles the rest of the SM contribution as well as the NP shift.\footnote{\label{foot:C9_C10}More precisely \cite{Bobeth:1999mk,Beneke:2001at}
\be
C_{9}^{\rm eff}(q^{2}) = C_{9, \rm SD} + Y(q^{2}) - \frac{V_{ub} V_{us}^*}{V_{tb} V_{ts}^*}\left(\frac{4}{3} C_{1}+C_{2}\right) \left[h\left(q^{2}, m_u \right)-h\left(q^{2}, m_{c}\right) \right]\,,
\ee
with $C_{9, \rm SD} = 4.327$ the Wilson coefficient calculated at the EW matching scale, and RGE-evolved \cite{Bobeth:2003at,Huber:2005ig} to the $\mu_b = 5$ GeV scale, that in \cite{Beneke:2020fot} is identified with the hard factorization scale in the $N_f = 5$ theory. The difference between $C_9^{\rm eff}(q^2)$ and $C_{9, \rm SD}$ is a complex correction whose real part smoothly decreases from $4 \cdot 10^{-4}$ to $-1\%$ in the range of interest for this discussion, $\sh \in [0.5, 1]$, whereas its imaginary part is $+5\%$ and basically constant over the same range. For later reference, we also have $C_{10} = -4.262$, whose magnitude differs from $C_{9, \rm SD}$ by less than 2\%, and $C_7 = -0.303$.}
We estimate the complex shift $\delta_{c \bar c}$ from a simultaneous scan to the ten resonance parameters $|\eta_V| \in [1, 3]$ and $\delta_V \in [0, 2\pi)$  (see Sec. \ref{sec:broad-c} for details). For the sake of the present discussion, we need an absolute upper bound on the size of this shift. To this end we choose $\sh = 0.49$, which yields the $1\sigma$-range
\be
\label{eq:delta_cc}
|\delta_{c\bar c}| \in (4.3 \pm 1.6)\% \times C_{9, \rm SD}\, ,
\ee
with an arbitrary phase. In the discussion to follow, we will conservatively assume eq. (\ref{eq:delta_cc}) as the size of the $\delta_{c \bar c}$ complex correction {\em throughout the integration range}, $\sh \in [0.5, 1]$, considered for high $q^2$. Such assumption will prove to be enough for the sake of our argument. We note, however, that this is a truly conservative assumption. In fact, fig. \ref{fig:RMSdelta} displays the mean over the scan of $|\delta_{c \bar c}|/C_{9, \rm SD}$ as a function of $\sh$ (blue solid line) as well as the mean plus 3 times the standard deviation (dashed orange line).
The figure shows clearly that the size of the correction steadily decreases for $\sh > 0.49$, and that the correction is below 1\% for $\sh > 0.75$, i.e. in half of the actual integration range.

With this in mind, we will consider $\delta_{c\bar c}$ as constant over this range, and given, in size, by eq. (\ref{eq:delta_cc}). We can then write
\be
\label{eq:ADGmmy_Re(d/C)}
\ADGmmy ~\simeq~ 
\frac{
- \int_{\PS} f_{99} \bar C_{9}^2 \left[ 1 + 2 \, \re \left( \delta_{c \bar c} / \bar{C}_9 \right) \right] + N_{\rm rest}
}{
\int_{\PS} \bar f_{99} | \bar C_{9} |^2 \left[ 1 + 2 \, \re \left( \delta_{c \bar c}^* / \bar{C}_9^* \right) \right] + D_{\rm rest}
}
\ee
where, in view of eq. (\ref{eq:delta_cc}), we neglect terms quadratic in the $\delta_{c \bar c}$ shift, and we use the fact that $f_{99}$ (see eq. (\ref{eq:f99}) and fig. \ref{fig:Imff_BBW_vs_JPZ}) is real. We note that the $\delta_{c \bar c}$ shift also affects $C_7 \times C_9$ (eqs. (\ref{eq:f79}) and (\ref{eq:bf79})) as well as $C_9 \times C_{10}$ (eq. (\ref{eq:bf910})) interference terms, that are bundled in $N_{\rm rest}$ and $D_{\rm rest}$. We will come back to these terms later on.

\newcommand{\doC}{\epsilon}
Denoting $\re \left( \delta_{c \bar c} / \bar{C}_9 \right) = \doC/2$, and treating $\bar C_{9}$ and $\delta_{c \bar c}$ as constant over the $q^2$ range of interest (see footnote \ref{foot:C9_C10}), we can rewrite eq. (\ref{eq:ADGmmy_Re(d/C)}) as
\be
\label{eq:ADGmmy_(1+doC)}
\ADGmmy ~\simeq~ 
\frac{
N_{99} \, \bar C_{9}^2 (1+\doC) + N_{\rm rest}
}{
D_{99} \, |\bar C_{9}|^2 (1+\doC) + D_{\rm rest}
}~.
\ee
Because of eq. (\ref{eq:delta_cc}), we can expand eq. (\ref{eq:ADGmmy_(1+doC)}) for small $\doC$, obtaining
\bea
\label{eq:ADGmmy_[1+]}
\ADGmmy &\simeq&
\frac{
N_{99} \, \bar C_{9}^2 (1+\doC) + N_{\rm rest}
}{
D_{99} \, |\bar C_{9}|^2 (1+\doC) + D_{\rm rest}
} \nn \\
[0.2cm]
&\simeq&
\frac{N_{99} \, \bar C_{9}^2 + N_{\rm rest}}{D_{99} \, |\bar C_{9}|^2+ D_{\rm rest}} \left[ 1 + \left( \frac{D_{\rm rest}}{D_{99} \, |\bar C_{9}|^2 + D_{\rm rest}} - \frac{N_{\rm rest}}{N_{99} \, \bar C_{9}^2 + N_{\rm rest}} \right) \doC \right]~.
\eea
The $N$ and $D$ coefficients in the above expression can now be read off from relations (\ref{eq:f77})-(\ref{eq:bf910}). In particular, for $N_{99}$ and $D_{99}$ we have
\bea
\label{eq:N99}
N_{99} &=& - \frac{4}{3} \MB^4 \int d \sh d \cos \theta \, f(\sh, \mlh^2) \, x^2 (2 \mlh^2 + \sh) (F_A^2 - F_V^2)\,, \\
[0.4cm]
\label{eq:D99}
D_{99} &=& + \frac{4}{3} \MB^4 \int d \sh d \cos \theta \, f(\sh, \mlh^2) \, x^2 (2 \mlh^2 + \sh) (|F_A|^2 + |F_V|^2)\,.
\eea
Besides, we can express $N_{\rm rest}$ and $D_{\rm rest}$ as
\bea
\label{eq:Nrest}
N_{\rm rest} &=& + \frac{4}{3} \MB^4 C_{10}^2 \int d \sh d \cos \theta \, f(\sh, \mlh^2) \, x^2
(4 \mlh^2 - \sh)(F_A^2 - F_V^2) ~+~ \delta_N(\delta_{c \bar c})\,, \\
[0.4cm]
\label{eq:Drest}
D_{\rm rest} &=& - \frac{4}{3} \MB^4 |C_{10}|^2 \int d \sh d \cos \theta \, f(\sh, \mlh^2) \, x^2
(4 \mlh^2 - \sh)(|F_A|^2 + |F_V|^2) ~+~ \delta_D(\delta_{c \bar c})\,.~~
\eea
In the $N_{\rm rest}$ and $D_{\rm rest}$ relations we have not written explicitly terms that either involve $C_7$, or are bilinear in $C_9 C_{10}$, that are also concerned by the $\delta_{c \bar c}$ shift. These terms, indicated in eqs.~(\ref{eq:Nrest})-(\ref{eq:Drest}) by $\delta_{N,D} (\delta_{c \bar c})$, are suppressed by the small size of the $C_7$ coefficient in comparison with $|C_{9,10}|$ (see footnote \ref{foot:C9_C10}), or by the small ratio $f_{B_s} / \MB \simeq 0.04$. We find that the $\delta_{N,D} (\delta_{c \bar c})$ contributions are around 15\% of those explicitly written. Hence, recalling eq. (\ref{eq:delta_cc}), the broad-charmonium shift within these contributions may be estimated as a $\approx 4\% \times 15\%$ effect, i.e. well below 1\%.

We can further simplify eqs. (\ref{eq:N99})-(\ref{eq:Drest}) by noting that, for $\sh \in [0.5, 1]$, we can completely neglect $4 \mlh^2 = 1.6 \cdot 10^{-3}$ with respect to $\sh$. We have
\newcommand{\ibPS}{\int_{\overline \PS}}
\bea
\label{eq:N99_}
N_{99} &\simeq& + \ibPS (F_V^2 - F_A^2)\,, \\
[0.4cm]
\label{eq:D99_}
D_{99} &\simeq& + \ibPS (F_V^2 + F_A^2)\,, \\
[0.4cm]
\label{eq:Nrest_}
N_{\rm rest} &\simeq& + C_{10}^2 \ibPS (F_V^2 - F_A^2) ~+~ \delta_N(\delta_{c \bar c})\,, \\
[0.4cm]
\label{eq:Drest_}
D_{\rm rest} &\simeq& + C_{10}^2 \ibPS (F_V^2 + F_A^2) ~+~ \delta_D(\delta_{c \bar c})\,,
\eea
where we used the abbreviation
\be
\frac{4}{3} \MB^4 \int d \sh d \cos \theta \, f(\sh, \mlh^2) \, x^2 \sh ~\equiv~ \ibPS\, .
\ee
We see that, if we identify $\bar C_9$ with $C_{9, \rm SD}$---which holds to within 5\%, see footnote \ref{foot:C9_C10}---and we further use the (accidental) near-equality between $C_{9, \rm SD}^2$ and $C_{10}^2$---which holds to 3\%---we can rewrite the denominators in the rounded parenthesis of eq. (\ref{eq:ADGmmy_[1+]}) as
\bea
N_{99} \, \bar C_{9}^2 + N_{\rm rest} &\simeq& 2 C_{9, \rm SD}^2 \ibPS (F_V^2 - F_A^2) ~+~ \delta_N(\delta_{c \bar c}) \, ,\\
D_{99} \, |\bar C_{9}|^2 + D_{\rm rest} &\simeq& 2 C_{9, \rm SD}^2 \ibPS (F_V^2 + F_A^2) ~+~ \delta_D(\delta_{c \bar c}) \,.
\eea
In the same approximations, the term in the square brackets in eq. (\ref{eq:ADGmmy_[1+]}) reads
\bea
&&1 + \left( \frac{N_{99} \, \bar C_9^2}{N_{99} \, \bar C_{9}^2 + N_{\rm rest}} - \frac{D_{99} \, |\bar C_9|^2}{D_{99} \, |\bar C_{9}|^2 + D_{\rm rest}} \right) \doC \nn \\
[0.2cm]
\label{eq:deltacc_doC}
&&\simeq~ 1 + \frac{1}{2} \left( \frac{\delta_D(\delta_{c \bar c})}{2 C_9^2 \ibPS (F_V^2 + F_A^2)} - \frac{\delta_N(\delta_{c \bar c})}{2 C_9^2 \ibPS (F_V^2 - F_A^2)}\right) \doC \nn \\
[0.2cm]
\label{eq:deltacc_doC_}
&&\lesssim~ 1 \pm 0.075 \cdot \doC ~\lesssim~ 1 \pm 0.7 \%\, ,
\eea
where the last equality is obtained as follows. As discussed below eq. (\ref{eq:Drest}), each of the two terms in the rounded parenthesis in eq. (\ref{eq:deltacc_doC}) is around 15\%. Although the difference of these two terms would reveal further cancellations, we conservatively bound this difference with 15\% itself. We further assume $\doC = 9\%$ from eq. (\ref{eq:delta_cc})---which, as discussed below that equation is a very conservative assumption.

One final comment deserves the fraction multiplying the square bracket in eq. (\ref{eq:ADGmmy_[1+]}). Plugging in eqs. (\ref{eq:N99_})-(\ref{eq:Drest_}), we may expand around small $\delta_N$ and $\delta_D$. This would change the correction in eq. (\ref{eq:deltacc_doC_}) by further terms well below 1\%, see paragraph following eq. (\ref{eq:Drest}). 

The above analytic argument is meant to make more transparent why broad-charmonium resonances lead to uncertainties safely below 1\% for high-$q^2$, i.e. negligible {\em vis-\`a-vis} f.f. uncertainties. We emphasize that the above argument results from a series of very conservative assumptions, and that the less intuitive but more accurate numerical analysis shows cancellations down to the per-mil level due to {\em (i)} the mean value and standard deviation of $\delta_{c\bar c}$ being $\hat s$-dependent and generally much smaller (see fig. \ref{fig:RMSdelta}) than the values in eq. (\ref{eq:delta_cc}), which hold at the $\sh=0.49$ threshold, {\em (ii)} the ratio-property of $\ADGmmy$ as seen from the minus sign in eq.~(\ref{eq:deltacc_doC}) and {\em (iii)} the importance of $C_{10}^{2}$ terms, unaffected by charmonium resonances and of the same order of magnitude as the $C_9^2$ term. An important remark is that, on the other hand, this argument does {\em not} hold for values of $\hat s$ lower than the mentioned $\sh = 0.49$ threshold; e.g., for $\hat s=0.46$, the $\Psi(2S)$ resonance yields much large contributions than eq.~(\ref{eq:delta_cc}), and an expansion is no more justified.

\section{Table of inputs}\label{app:input_table}

We collect in table \ref{tab:input} all our inputs. We only quote central values, i.e. we do not quote errors, as none of the parameters in the table contributes a significant part of the theory uncertainties we compute. Any omitted parameter is taken from Ref. \cite{Beneke:2020fot}. As concerns the CKM input, we take the latest `New-Physics fit' available from \cite{UTfit-website,Ciuchini:2000de}. Similar results may be obtained from \cite{CKMfitter-website,Charles:2004jd}.
\noindent \begin{table*}[h]
\def\arraystretch{1.4}
\begin{center}
\begin{tabular}{|c|c|c||c|c|c|}
\hline
Parameter & Value & Ref. & Parameter & Value & Ref. \\ 
\hline
\hline
$M_{Bs}$ & $ 5.36688~\GeV$  & \cite{Zyla:2020zbs} & $\alpha_s (m_Z)$ & $0.1179$ & \multirow{2}{*}{\cite{Zyla:2020zbs}} \\
$f_{Bs}$ & $ 0.2303~\GeV$  & \cite{Aoki:2019cca} & $\alpha_{\rm e.m.}(m_b)$ & $1/133$ & \\ \hline
$m_b(m_b)$ & $ 4.18~\GeV$ & \multirow{4}{*}{\cite{Zyla:2020zbs}} & $\lambda$ & $ 0.2255$ & \multirow{4}{*}{\cite{Zyla:2020zbs,UTfit-website,CKMfitter-website}} \\
$m_c(m_c)$ & $ 1.27~\GeV$ & & $A$ & $0.785$ & \\
$m_{b}^{\rm pole}$ & $ 4.78~\GeV$ & & $\bar{\rho}$ & $0.147$ & \\
$m_c^{\rm pole}$ & $ 1.67~\GeV$ & & $\bar{\eta}$ & $0.377$ & \\ \hline
$m_{\psi(2S)} $ & $3.686~\GeV$ & \multirow{5}{*}{\cite{Zyla:2020zbs}} & $\Gamma_{\psi(2S)}$ & $ 0.294 \times 10^{-3}~\GeV$ & \multirow{5}{*}{\cite{Zyla:2020zbs}} \\
$m_{\psi(3770)} $ & $3.774~\GeV$ & & $\Gamma_{\psi(3770)}$ & $ 27.2 \times 10^{-3}~\GeV$ & \\
$m_{\psi(4040)} $ & $4.039~\GeV$ & & $\Gamma_{\psi(4040)}$ & $ 80 \times 10^{-3}~\GeV$ & \\
$m_{\psi(4160)} $ & $4.191~\GeV$ & & $\Gamma_{\psi(4160)}$ & $ 70 \times 10^{-3}~\GeV$ & \\
$m_{\psi(4415)} $ & $4.421~\GeV$ & & $\Gamma_{\psi(4415)}$ & $ 62 \times 10^{-3}~\GeV$ & \\ \hline
$\mathcal{B}(\psi(2S) \to \ell \ell)$ & $ 8.0\times 10^{-3} $ & \multirow{5}{*}{\cite{Zyla:2020zbs}} & $\delta_{\psi(2S)}$ & $0 $  & \multirow{5}{*}{\cite{Ablikim:2007gd}} \\
$\mathcal{B}(\psi(3770) \to \ell \ell)$ & $ 9.6 \times 10^{-6} $ & & $\delta_{\psi(3770)}$ & $0 $  & \\
$\mathcal{B}(\psi(4040) \to \ell \ell)$ & $ 10.7 \times 10^{-6} $ & & $\delta_{\psi(4040)}$ & $133 \times \pi/180 $  & \\
$\mathcal{B}(\psi(4160) \to \ell \ell)$ & $ 6.9 \times 10^{-6} $ & & $\delta_{\psi(4160)}$ & $301 \times \pi/180 $  & \\
$\mathcal{B}(\psi(4415) \to \ell \ell)$ & $ 9.4 \times 10^{-6} $ & & $\delta_{\psi(4415)}$ & $246 \times \pi/180 $  & \\ \hline
\end{tabular}
\caption{List of input parameters.}
\label{tab:input}
\end{center}
\def\arraystretch{1.4}
\end{table*}


\bibliographystyle{JHEP_patched.bst}
\bibliography{bibliography}

\end{document}